\documentstyle{article}
\newlength{\widthbigcirc}
\newlength{\widthoplus}
\newlength{\widthoplusbigcirc}
\newlength{\widthbox}
\newlength{\widthplus}
\newlength{\widthplusbox}
\newlength{\widthtimes}
\newlength{\widthtimesbox}
\newcommand{\scriptbf}[1]{\mbox{\scriptsize${\bf #1}$}}

\newtheorem{Fact}{Fact}
\newtheorem{Facts}{Facts}
\newtheorem{Proposition}{Proposition}
\newtheorem{Lemma}{Lemma}
\newtheorem{Tlemma}{Technical Lemma}
\newtheorem{Theorem}{Theorem}

\newenvironment{proof}
   {\begin{trivlist} \item[] {\bf Proof.}}{\rule{.3em}{2ex} \end{trivlist}}

\newcommand{\type}{ {\bf :} }

\newcommand{\morph}{{:}}
\newcommand{\memberof}{{\in}}
\newcommand{\define}{\stackrel{{\rm df}}{=}}

\newcommand{\tprod}{\otimes}
\newcommand{\bsum}{\oplus}

\newcommand{\bprod}{\triangle}

\newcommand{\imply}{{\Rightarrow}}

\newcommand{\tensorimplysource}
   {\! \mbox{\hspace{.21em}--\hspace{-.21em}} \backslash }
\newcommand{\tensorimplytarget}
   { / \mbox{\hspace{-.23em}--\hspace{.23em}} \!}

\newcommand{\typesum}{\, \mbox{${\bigcirc} \hspace{-\widthoplusbigcirc} {\oplus}$} \:}
\newcommand{\scriptsizetypesum}{{\bigcirc} \hspace{-\widthoplusbigcirc} \, {\oplus}}
\newcommand{\sumtype}{{\bigcirc} \hspace{-\widthoplusbigcirc} {\oplus}}
\newcommand{\scriptsizesumtype}{{\bigcirc} \hspace{-\widthoplusbigcirc} \, {\oplus}}
\newcommand{\bsumvert}{\, \mbox{${\Box} \hspace{-\widthplusbox} {+}$} \:}
\newcommand{\tprodvert}{\, \mbox{${\Box} \hspace{-\widthtimesbox} {\times}$} \:}

\newcommand{\complement}[2]{{#2}^{\neg}_{#1}}
\newcommand{\boolneg}[1]{{\sim}{#1}}

\newcommand{\abov}[1]{{\uparrow}{#1}}
\newcommand{\below}[1]{{\downarrow}{#1}}

\newcommand{\singleton}[1]{\{#1\}}

\newcommand{\disjoint}[1]{{\smile_{\!#1}}}

\newcommand{\total}[1]{{#1}^{\dagger}}

\newcommand{\principalfilter}[1]{{\uparrow}(#1)}

\newcommand{\powerof}[1]{{\cal P}(#1)}

\newcommand{\mat}[1]{{\cal M}(#1)}

\newcommand{\distrib}[1]{{\cal D}(#1)}

\newcommand{\Hoare}[1]{{\cal H}(#1)}
\newcommand{\dualHoare}[1]{{\cal H}^{\rm co}(#1)}

\newcommand{\monoid}[3]{\mho_{#1}^{#3}{#2}}

\newcommand{\obj}[3]{\Theta_{#1}^{#3}{#2}}
\newcommand{\yoneda}[3]{{\rm y}_{#1}^{#3}{#2}}
\newcommand{\comonoid}[3]{\Omega_{#1}^{#3}{#2}}

\newcommand{\regular}[3]{\Re_{#1}^{#3}(#2)}

\newcommand{\filter}[1]{{\cal F}(#1)}
\newcommand{\filtersource}[1]{{\cal F}_0(#1)}
\newcommand{\filtertarget}[1]{{\cal F}_1(#1)}
\newcommand{\ideal}[1]{{\cal I}(#1)}
\newcommand{\idealsource}[1]{{\cal I}_0(#1)}
\newcommand{\idealtarget}[1]{{\cal I}_1(#1)}
\newcommand{\uterm}[2]{\flat_{{#1}{#2}}}
\newcommand{\spano}[2]{\sharp_{{#1}{#2}}}
\newcommand{\utermspano}[2]{\natural_{{#1}{#2}}}
\newcommand{\square}[1]{\Box{#1}}
\newcommand{\directsquare}[2]{\Box^{#1}{#2}}
\newcommand{\inversesquare}[2]{\Box_{#1}{#2}}
\renewcommand{\diamond}[1]{\Diamond{#1}}
\newcommand{\directdiamond}[2]{\Diamond^{#1}{#2}}
\newcommand{\inversediamond}[2]{\Diamond_{#1}{#2}}
\newcommand{\interior}[1]{{#1}^\circ}
\newcommand{\closure}[1]{{#1}^\bullet}

\newcommand{\involution}[1]{{#1}^{\!\propto}}

\newcommand{\product}[2]{ {#1} {\times} {#2} }
\newcommand{\power}[2]{ {#2}^{#1} }
\newcommand{\copower}[2]{ {#1} {\bullet} {#2} }
\newcommand{\triproduct}[3]{\mbox{$ #1 {\times} #2 {\times} #3 $}}

\newcommand{\pair}[2]{\langle #1,#2 \rangle}
\newcommand{\relcopair}[4]{[{#1},{#2}]^{#3}_{#4}}
\newcommand{\relpair}[4]{\langle{#1},{#2}\rangle^{#3}_{#4}}
\newcommand{\domain}[1]{\partial_0{#1}}
\newcommand{\range}[1]{\partial_1{#1}}
\newcommand{\kernel}[1]{\emptyset_0{#1}}
\newcommand{\cokernel}[1]{\emptyset_1{#1}}
\newcommand{\source}[2]{{]{#1}[_{#2}}}
\newcommand{\target}[2]{{\rangle{#1}\langle^{#2}}}
\newcommand{\cotuple}[2]{{[{#1}]_{#2}}}
\newcommand{\tuple}[2]{{\langle{#1}\rangle^{#2}}}
\newcommand{\decomposition}[3]{({#1})_{#2}^{#3}}
\newcommand{\triple}[3]{\mbox{$ \langle #1,#2,#3 \rangle $}}
\newcommand{\quadruple}[4]{\mbox{$ \langle #1,#2,#3,#4 \rangle $}}
\newcommand{\quintuple}[5]{\mbox{$ \langle #1,#2,#3,#4,#5 \rangle $}}
\newcommand{\sextuple}[6]{\mbox{$ \langle #1,#2,#3,#4,#5,#6 \rangle $}}

\newcommand{\term}[3]{{#1} \stackrel{{#2}}{\rightharpoondown} {#3}}
\newcommand{\onespan}[3]{{#1} {\widehat{\;{#2}\;}} {#3}}
\newcommand{\twospan}[3]{{#1} \! \stackrel{{#2}}{\Box} \! {#3}}

\newcommand{\opterm}[3]{#1 \stackrel{#2}{\leftharpoondown} #3}

\newcommand{\yiya}[1]
   {\setlength{\unitlength}{#1pt}
    \begin{picture}(1,1)(-.5,-.5)
       \put(0,0){\circle{1}}
       \put(.25,0){\oval(.5,.5)[b]}
       \put(-.25,0){\oval(.5,.5)[t]}
    \end{picture}
    \setlength{\unitlength}{1pt}}
\newcommand{\yinyang}{\yiya{8}}

\begin{document}
\title{The Standard Aspect \\
       of \\
       Dialectical Logic \\
        \mbox{\footnotesize
              $\pi \alpha \lambda \iota \nu \tau o \nu o \zeta$      $\alpha \rho \mu o \nu \iota \eta$
            - $\pi \alpha \lambda \iota \nu \tau \rho o \pi o \zeta$ $\alpha \rho \mu o \nu \iota \eta$
              \normalsize}}
\author{Robert E. Kent}
\date{May 1989}
\maketitle

   \begin{abstract}
{\em Dialectical logic\/} is the logic of dialectical processes. 
The goal of dialectical logic is to introduce dynamic notions
into logical computational systems.
The fundamental notions of {\em proposition\/} and {\em truth-value\/} in standard logic
are subsumed by the notions of {\em process\/} and {\em flow\/} in dialectical logic.
Dialectical logic has a standard aspect,
which can be defined in terms of the ``local cartesian closure'' of subtypes.
The standard aspect of dialectical logic provides a natural program semantics
which incorporates Hoare's precondition/postcondition semantics
and extends the standard Kripke semantics of dynamic logic.
The goal of the standard aspect of dialectical logic is
to unify the logic of small-scale and large-scale programming.
\end{abstract}

\section*{Introduction}

Dynamic logic \cite{Kozen} seeks to bring dynamic notions into logic and program semantics
by basing this semantics and logic on the notion of ``predicate transformer''.
The alternate program semantics of Hoare-style ``precondition/postcondition assertions'' 
is usually viewed as a special case of dynamic logic.
Dialectical logic \cite{Kent88} seeks to bring dynamic notions into logic
by basing logic \cite{Lawvere69} on the notion of ``dialectical contradiction'' or {\em adjoint pair\/}.
How do these three logics connect together?
This paper will show that dynamic logic and Hoare-style precondition/postcondition assertional semantics are exactly equivalent,
and that dialectical logic subsumes both in the sense that
``dynamic logic is the standard aspect of dialectical logic''.
More particularly,
I show in this paper that the axioms of dynamic logic
(or alternatively, precondition/postcondition assertional axioms)
characterize precisely the dialectical logic notion of {\em dialectical flow category\/}
(or alternatively, {\em assertional category\/}, a notion related but not equivalent to Manes's assertional category \cite{Manes}). 
A dialectical flow category is a kind of {\em indexed adjointness\/} or {\em dialectical base\/}
which itself is a dialectical enrichment of the notion of {\em indexed preorder\/} \cite{Hyland}.
In fact,
a dialectical flow category is an indexed adjointness of {\em subtypes\/} which is locally cartesian closed.
The indexing category here is the enriched notion of a {\em join bisemilattice\/} \cite{Kent88}.
Dialectical flow categories objectivize the intuitive idea of predicate transformation or the ``dialectical flow of predicates''.

\section{Base Structures}

The essence of dynamic systems is concentrated in the notion of ``change''.
The changing aspect of dynamic systems is abstracted as ``state''.
Change is represented mathematically by the idea of ``arrow''
\begin{center}
   $\rightharpoondown$
\end{center}
The change that an arrow in a dynamic system symbolizes
is the change of an internal state.
This accords well with the fact the the notion of arrow is a {\em polar\/} notion.
An arrow has both a ``source'' and a ``target'' $y \stackrel{r}{\rightharpoondown} x$,
and these often specify ``current state'' and ``next state''
with an implied direction or polarity.
The source/target polarity is a binary dualistic typing of arrows.
This gives dynamic systems a logical type-theoretic flavor.
We further this merging of logic and dynamics
by identifying the logical notion of ``term'' with the dynamic notion of arrow.

The nondeterministic aspect of actual dynamic systems
prompts us to regard an term $\term{y}{r}{x}$ as a composite notion,
so that terms in dynamic systems with the same source/target typing
possess a relationship of ``nondeterminism''
\begin{center}
   $\preceq$
\end{center}
with $r \preceq_{yx} s$ representing the fact that ``$s$ is more nondeterministic than $r$''.
We assume that the nondeterministic order is a preorder.
In this paper we quotient out any nondeterministically equivalent terms,
$r \equiv s$ when $r \preceq s$ and $s \preceq r$,
and assume that nondeterminism is a partial order.

By itself,
the basic notion of term is only a potentially dynamic notion.
The actual dynamics of terms is concentrated in the notion of ``interaction''
\begin{center}
   $\tprod$
\end{center}
In dynamic systems,
terms function as ``autonomous agents'' or ``processes''.
Such term-processes interact through their source and target types,
and hence types represent the notion of ``local ports'' in dynamic systems.
Two terms $\term{z}{s}{y}$ and $\term{y}{r}{x}$ can interact
when they have a port (type) in common of opposite polarity
through which the interaction is facilitated, is conducted and/or takes place.
This binary interaction is a tensor product
$\term{z}{{s \tprod r}}{x}$,
with the port which conducts the interaction being ``hidden'' in the resulting product term.
In dynamic systems,
interaction is interpreted to be sychronization/communication between term-processes.
We assume that process interaction is associative and respects the nondeterminism order.

The notion of term-process is the fundamental notion,
with the notion of type-port a derived notion and special case.
Type-ports are regarded, and explicitly rendered,
as special term-processes $\term{x}{x}{x}$
which are ``nops'' or identities in process interactions.
The type-ports identity processes are aggregated together as a collection
\begin{center}
   ${\rm Id}$
\end{center}


\subsection{Tensors}

\paragraph{Biposets.}
Terms in dynamic systems form a biposet.
A {\em biposet\/} is another name for an ordered category;
that is,
a category ${\bf P} = \quadruple{{\bf P}}{\preceq}{\tprod}{{\rm Id}}$
whose homsets are posets and whose composition is monotonic on left and right.
We prefer to view biposets as vertical structures, preorders with a tensor product,
rather than as horizontal structures, ordered categories.

In more detail,
a biposet {\bf P} consists of the following data and axioms.
There is a collection of {\bf P}-objects $x,y,z, \cdots$ called {\em types\/},
     and a collection of {\bf P}-arrows $r,s,t, \cdots$ called {\em terms\/}.
Each term $r$ has a unique source type $y$ and a unique target type $x$,
denoted by the relational notation $\term{y}{r}{x}$.
The collection of terms from source type $y$ to target type $x$
is ordered by a binary relation $\preceq_{y,x}$ called {\em term entailment\/},
which is
transitive,
if $r \preceq s$ and $s \preceq t$ then $r \preceq t$,
reflexive
$r \preceq r$, and
antisymmetric,
$r \equiv s$ implies $r = s$,
where $r \equiv s$ when $r \preceq s$ and $s \preceq r$.
Dialectical logic entailment $\preceq_{y,x}$ between terms generalizes
standard logic entailment $\vdash$ between propositions.
For any two terms $\term{z}{s}{y}$ and $\term{y}{r}{x}$ with matching types
(target type of $s$ = source type of $r$) 
there is a composite term $\term{z}{s \tprod r}{x}$,
where $\tprod$ is a binary operation called {\em tensor product\/},
which is associative
$t \tprod (s \tprod r) = (t \tprod s) \tprod r$,
and monotonic on left and right,
$s \preceq s'$ and $r \preceq r'$ imply ${s \tprod r} \preceq {s' \tprod r'}$.
Tensor product represents the ``interaction'' of the two term-processes $s$ and $r$.
It allows each term $\term{y}{r}{x}$ to specify a {\em right direct flow\/}
${\bf P}[z,y] \stackrel{\tprod r}{\rightarrow} {\bf P}[z,x]$
and a {\em left direct flow\/}
${\bf P}[x,z] \stackrel{r \tprod}{\rightarrow} {\bf P}[y,z]$
for each type $z$.
Any type $x$ is a term $\term{x}{x}{x}$,
which is an identity $s \tprod x = s$ and $x \tprod r = r$,
specifying identity direct/inverse flow.

Any category {\bf C} is a biposet with the trivial identity homset order $\preceq_{y,x} = {=_{y,x}}$.
The category {\bf Rel} (also denoted {\bf Mfn}) of sets and binary relations (multivalued functions) is a biposet.
A biposet with one object (universal type) is called a {\em monoidal poset\/}.
For each {\bf P}-type $x$,
the collection ${\bf P}[x,x]$ of endoterms at $x$ is a monoidal poset.
Any commutative monoid $\triple{M}{\circ}{e}$ is a monoidal poset with the ``part (prefix) order''
$m \preceq n$ when there is a $p \memberof M$ such that $m \circ p = n$.
Given an alphabet $A$,
the monoid of formal $A$-languages $\powerof{A^\ast}$ is a monoidal poset,
whose terms are formal languages,
whose tensor product is language concatenation,
and whose identity is singleton empty string $\{\varepsilon\}$.
If {\bf P} is a biposet,
then the op-dual or opposite biposet ${\bf P}^{\rm op}$ is the opposite category with the same homset order as {\bf P},
and the co-dual biposet ${\bf P}^{\rm co}$ is (the same category) {\bf P} with the opposite homset order.
A {\em morphism of biposets\/} ${\bf P} \stackrel{H}{\rightarrow} {\bf Q}$
is a functor which preserves homset order.

\paragraph{Adjoint Pairs.}
For any opposed pair of ordinary relations $\term{Y}{R}{X}$ versus $\opterm{Y}{S}{X}$
the ``unit inequality'' $Y \subseteq R \tprod S$ and
the ``counit inequality'' $S \tprod R \subseteq X$
{\em taken together\/} are equivalent to the facts
that $R$ is the graph $R = \yoneda{}{(f)}{1}$ of a function $Y \stackrel{f}{\rightarrow} X$
and that $S$ is the transpose $S = R^{\rm op} = {\yoneda{}{(f)}{1}}^{\rm op} = \yoneda{}{(f)}{0}$.
On the other hand,
the graph $\term{Y}{\yoneda{}{(f)}{1}}{X}$ of any function $Y \stackrel{f}{\rightarrow} X$ 
and its transpose $\yoneda{}{(f)}{0} = (\yoneda{}{(f)}{1})^{\rm op}$
satisfy the unit and counit inequalities.
So these conditions describe functionality in the biposet {\bf Rel},
and can be used as a way of axiomatizing functionality in general biposets.
But they are also the defining conditions for internal adjoint pairs.

Two opposed terms $\term{y}{r}{x}$ and $\opterm{y}{s}{x}$ form an {\em adjoint pair\/} of terms or an {\em adjunction\/},
denoted by $\term{y}{r \dashv s}{x}$,
when they satisfy the ``unit inequality'' $y \preceq r \tprod s$ and the ``counit inequality'' $s \tprod r \preceq x$.
This axiomatizes ``functionality'' of {\bf P}-terms. 
The term $r$ is called the {\em left adjoint\/} and the term $s$ is called the {\em right adjoint\/} in the adjunction $r \dashv s$.
It is easy to show that right adjoints (and left adjoints) are unique,
when they exist:
if $\term{y}{r \dashv s_1}{x}$ and $\term{y}{r \dashv s_2}{x}$ then $s_1 = s_2$.
Denote the unique right adjoint of $\term{y}{r}{x}$ by $\opterm{y}{r^{\rm op}}{x}$. 
A {\em functional {\bf P}-term\/} is a {\bf P}-term with a right adjoint.
We usually use the notation $\term{y}{f \dashv f^{\rm op}}{x}$ for functional terms.
For any adjoint pair $\term{y}{f \dashv f^{\rm op}}{x}$:
when the unit is equality $y = f \tprod f^{\rm op}$ they are a {\em coreflective pair\/};
when the counit is equality $f^{\rm op} \tprod f = x$ they are a {\em reflective pair\/}; and
when both unit and counit are equalities they are an {\em inverse pair\/}.
For any functional term $\term{y}{f \dashv f^{\rm op}}{x}$:
the adjunction $f \dashv f^{\rm op}$ is a coreflection
   iff $f$ is an monomorphism (iff $f^{\rm op}$ is an epimorphism);
the adjunction is a reflection
   iff $f$ is an epimorphism (iff $f^{\rm op}$ is an monomorphism); and
the adjunction is a inversion
   iff $f$ is an isomorphism (iff $f^{\rm op}$ is an isomorphism),
   iff $f^{\rm op} = f^{-1}$ is the two-sided inverse of $f$.
An coreflective pair $\term{y}{i \dashv p}{x}$ is also called a {\em subtype\/} of $x$.
Adjoint pairs compose in the obvious way:
$(g \dashv g^{\rm op}) \tprod (f \dashv f^{\rm op}) = (g \tprod f) \dashv (f^{\rm op} \tprod g^{\rm op})$,
and $(x \dashv x)$ is the identity adjoint pair at $x$.
So each biposet {\bf P} has an associated adjoint pair category ${\bf P}^\dashv$,
whose objects are {\bf P}-types and whose arrows are {\bf P}-adjunctions.
Equivalently,
${\bf P}^\dashv$-arrows are just functional {\bf P}-terms.
There is an inclusion functor ${\bf P}^\dashv \stackrel{{\rm Inc}}{\rightarrow} {\bf P}$.
The construction ${(\;)}^\dashv$ can be described as
either ``internal dialecticality'' or ``functionality''.

\paragraph{Comonoids/Affirmation.}
For any type $x$ in a biposet {\bf P} a {\em comonoid\/} $u$ at $x$,
denoted by $u \type x$,
is an endoterm $\term{x}{u}{x}$
which satisfies the ``part'' axiom (coreflexivity) $u \preceq_{x,x} x$,
stating that $u$ is a part of the type (identity term) $x$,
and the ``idempotency'' axiom (cotransitivity) $u \preceq_{x,x} u \tprod u$.
Since $u \tprod u \preceq x \tprod u = u$,
we can replace the inequality in the idempotency axiom with the equality $u \tprod u = u$.
Comonoids are generalized subtypes.
Comonoids of type $x$ are ordered by entailment $\preceq_x \define \preceq_{x,x}$.
The bottom endoterm $0_x \define 0_{x,x}$ and the identity endoterm $x$
are the smallest and largest comonoids of type $x$, respectively.
For a functional term $\term{y}{f \dashv f^{\rm op}}{x}$ 
the composite endoterm $f^{\rm op} \tprod f$ is a comonoid of type $x$ associated with $f$.
This associated comonoid is the top comonoid $f^{\rm op} \tprod f = x$
iff $f$ is an epimorphism iff $f \dashv f^{\rm op}$ is a reflective pair.
Denote the poset of comonoids of type $x$ by $\comonoid{}{(x)}{}$.
We can interpret the poset $\comonoid{}{(x)}{}$ as a ``state-set'' indexed by the type $x$,
with a comonoid $u \memberof \comonoid{}{(x)}{}$ being a ``state'' of a system.
The state $u \memberof \comonoid{}{(x)}{}$ has internal structure
and is a composite object sharing an ordering of nondeterminism $\preceq_x$ with other states.
For any two comonoids $u, v \in \comonoid{}{(x)}{}$
the tensor product is a lower bound
$u \tprod v \preceq u$ and $u \tprod v \preceq v$
which is an upper bound for comonoids below $u$ and $v$:
if $w \preceq u$ and $w \preceq v$ then $w \preceq u \tprod v$.
If $u$ and $v$ commute $u \tprod v = v \tprod u$
then the tensor product $u \tprod v$ is a comonoid;
in which case it is the meet $u \tprod v = u \wedge v$ in $\comonoid{}{(x)}{}$.


In a complete Heyting category {\bf H} (see below) an endoterm $\term{x}{p}{x}$
contains a largest comonoid of the same type $x$,
called the {\em interior\/} of $p$ and denoted by $\interior{p}$.
The interior is defined as the join
$\interior{p} \define \bigsqcup \{ w \memberof \comonoid{}{(x)}{} \mid w \preceq_x p \}$,
and satisfies the condition
$w \preceq_x p$ iff $w \preceq_x \interior{p}$
for all comonoids $w \memberof \comonoid{}{(x)}{}$.
In an arbitrary biposet {\bf P},
we use this condition to define (and to assert the existence of) the interior of endoterms.
The interior $\interior{p}$, 
when it exists,
is the largest generalized {\bf P}-subtype inside $p$.
Any comonoid $w \memberof \comonoid{}{(x)}{}$ is its own interior $\interior{w} = w$.
The interior of the tensor product is the meet in $\comonoid{}{(x)}{}$:
$\interior{(u \tprod v)} = u \wedge v = \interior{(v \tprod u)}$.
The interior of endoterms models the ``affirmation modality'' of linear logic \cite{Girard}.

We are especially interested in biposets {\bf P}
for which any {\bf P}-endoterm has such an interior.
Such biposets can be called interior (or possibly affirmation) biposets.
A biposet {\bf P} is an {\em interior biposet\/}
when at each type $x$ the inclusion-of-comonoids monotonic function
$\comonoid{}{(x)}{} \stackrel{{\rm Inc}_x}{\longrightarrow} {\bf P}[x,x]$
has a right adjoint
${\bf P}[x,x] \stackrel{\interior{(\,)}}{\rightarrow} \comonoid{}{(x)}{}$
called {\em interior\/},
which with inclusion forms a coreflective pair of monotonic functions ${\rm Inc}_x \dashv \interior{(\,)}$.
Composition $\interior{(\,)} \cdot {\rm Inc}_x$ is an general interior operator on endoterms.
Any meets that exist in ${\bf P}[x,x]$
are preserved by interior
$\interior{(p \wedge q)} = \interior{(\interior{p} \tprod \interior{q})}$
for endoterms $p,q \memberof {\bf P}[x,x]$,
since interior is a right adjoint.
{\bf [Standardization property:]}
In interior biposets $\comonoid{}{(x)}{}$ a meet semilattice,
with the interior of tensor product $\interior{(u\tprod v)}$ of two comonoids $v,u \memberof \comonoid{}{(x)}{}$ being the meet in $\comonoid{}{(x)}{}$,
and the tensor product identity (or type) endoterm $x$ being the largest comonoid of type $x$.
This standardization property means that
the local contexts of comonoids $\{ \comonoid{}{(x)}{} \mid x \mbox{ a type} \}$ are standard contexts,
and shows why propositions (interpreted as comonoids) and programs (interpreted as terms) are subsumed by a single concept.
Any complete Heyting category is an interior biposet.

\paragraph{Monoids/Consideration.}
A monoid is the order-theoretic dual of a comonoid.
For any type $x$ in a biposet {\bf P} a {\em monoid\/} $m$ at $x$,
denoted by $m \type x$,
is an endoterm $\term{x}{m}{x}$
which satisfies the ``reflexivity'' axiom $x \preceq_{x,x} m$,
and the ``idempotency'' axiom (transitivity) $m \tprod m \preceq_{x,x} m$.
We can replace the inequality in the transitivity axiom with the equality $m \tprod m = m$.
Monoids of type $x$ are ordered by entailment $\preceq_x \define \preceq_{x,x}$.
The identity endoterm $x$ is the smallest monoid of type $x$.
There is no largest monoid, in general.
For a functional term $\term{y}{f \dashv f^{\rm op}}{x}$ 
the composite endoterm $f \tprod f^{\rm op}$ is a monoid of type $x$ associated with $f$
called the {\em kernel\/} of $f$.
The kernel is the bottom monoid $f \tprod f^{\rm op} = x$
iff $f$ is an monomorphism iff $f \dashv f^{\rm op}$ is a coreflective pair.
Denote the poset of monoids of type $x$ by $\monoid{}{(x)}{}$.
For any two monoids $m, n \in \monoid{}{(x)}{}$
the tensor product is an upper bound
$m \preceq m \tprod n$ and $n \preceq m \tprod n$
which is a lowwer bound for monoids above $u$ and $v$:
if $m \preceq k$ and $n \preceq k$ then $m \tprod n \preceq k$.
If $m$ and $n$ commute $m \tprod n = n \tprod m$
then the tensor product $m \tprod n$ is a comonoid;
in which case it is the join $m \tprod n = m \vee n$ in $\monoid{}{(x)}{}$.

In a complete Heyting category {\bf H} (see below) an endoterm $\term{x}{p}{x}$
is contained in a smallest monoid of the same type $x$,
called the {\em closure\/} of $p$ and denoted by $\closure{p}$.
The closure is defined as the meet
$\closure{p} \define \bigwedge \{ m \memberof \monoid{}{(x)}{} \mid p \preceq_x m \}$,
and satisfies the condition
$\closure{p} \preceq_x m$ iff $p \preceq_x m$
for all monoids $m \memberof \monoid{}{(x)}{}$.
In an arbitrary biposet {\bf P},
we use this condition to define (and to assert the existence of) the closure of endoterms.
The closure $\closure{p}$, 
when it exists,
is the smallest {\bf P}-monoid containing $p$.
Any monoid $m \memberof \monoid{}{(x)}{}$ is its own closure $\closure{m} = m$.
The closure of the tensor product is the join in $\monoid{}{(x)}{}$:
$\closure{(m \tprod n)} = m \vee n = \closure{(n \tprod m)}$.
The closure of endoterms models the ``consideration modality'' of linear logic \cite{Girard}.
A biposet {\bf P} is an {\em closure\/} (or {\em consideration\/}) {\em biposet\/}
when at each type $x$ the inclusion-of-monoids monotonic function
$\monoid{}{(x)}{} \stackrel{{\rm Inc}_x}{\longrightarrow} {\bf P}[x,x]$
has a left adjoint
${\bf P}[x,x] \stackrel{\closure{(\,)}}{\rightarrow} \monoid{}{(x)}{}$
called {\em closure\/},
which with inclusion forms a reflective pair of monotonic functions $\closure{(\,)} \dashv {\rm Inc}_x$.
Composition $\closure{(\,)} \cdot {\rm Inc}_x$ is an general closure operator on endoterms.
Any joins that exist in ${\bf P}[x,x]$
are preserved by closure
$\closure{(p \vee q)} = \closure{(\closure{p} \bsum \closure{q})}$
for endoterms $p,q \memberof {\bf P}[x,x]$,
since closure is a left adjoint.
{\bf [Standardization property:]}
In closure biposets $\monoid{}{(x)}{}$ a join semilattice,
with the closure of tensor product $\closure{(m \tprod n)}$ of two monoids $m,n \memberof \monoid{}{(x)}{}$ being the join in $\monoid{}{(x)}{}$,
and the tensor product identity (or type) endoterm $x$ being the smallest monoid of type $x$.
Any complete Heyting category is an closure biposet,
with the closure defined to be the denumerable join
$\closure{p} \define \bigvee_n p^n = x \vee p \vee p \tprod p \cdots $.

An {\em involutive biposet\/} ${\bf P} = \pair{{\bf P}}{\involution{(\;)}}$
is a biposet {\bf P} with a morphism of biposets
${\bf P}^{\rm op} \stackrel{\involution{(\;)}}{\longrightarrow} {\bf P}$
which is 
monotonic,
self-inverse ${\involution{(\;)}}^{\rm op} \cdot \involution{(\;)} = {\rm Id}_P$,
identity on subtypes,
and adjunctive on functional terms;
so that
$\involution{(s \tprod r)} = \involution{r} \tprod \involution{s}$ for composable terms $\term{z}{s}{y}$ and $\term{y}{r}{x}$,
$\involution{x} = x$ for all types $x$,
if $r \preceq s$ then $\involution{r} \preceq \involution{s}$ for parallel terms $\term{y}{r,s}{x}$,  
$\involution{\involution{r}} = r$,
$\involution{u} = u$ for all comonoids $u \memberof \comonoid{}{(x)}{}$, and
$\involution{r} = r^{\rm op}$ for functional terms $\term{y}{r \dashv r^{\rm op}}{x}$.
In an involutive interior biposet,
interior must commute with involution:
$\interior{(\involution{r})} = \involution{(\interior{r})}$ for all terms $\term{y}{r}{x}$.
Dually, in an involutive closure biposet,
closure must commute with involution:
$\closure{(\involution{r})} = \involution{(\closure{r})}$ for all terms $\term{y}{r}{x}$.
In an involutive biposet with the interpretation of term entailment as nondeterministic order, 
functional terms are minimally nondeterministic (maximally deterministic) terms:
if $\term{y}{f \dashv f^{\rm op}}{x}$ and $\term{y}{g \dashv g^{\rm op}}{x}$ 
are parallel functional terms satisfying $f \preceq g$,
then $f \dashv g^{\rm op}$ and hence $f = g$.

A {\em topological biposet\/} {\bf P} is a classical biposet
with both interior and closure
which satisfy
$\boolneg{(\interior{p})} = \closure{(\boolneg{p})}$.

\paragraph{Dialectical Flow.}
In objective dialectics,
since dialectical contradictions are represented by adjunctions,
systems of dialectical contradictions are represented by diagrams in (pseudofunctors into) the category {\bf Adj}
whose objects are small categories and whose morphisms are adjoint pairs of functors. 
We call such a (pseudo)functor ${\bf B} \stackrel{E}{\longrightarrow} {\bf Adj}$
a {\em dialectical base} or an {\em indexed adjointness\/},
and use the notation
$E(y \stackrel{r}{\rightarrow} x) = (E^r \dashv E_r) \morph E(y) \rightarrow E(x)$.
A dialectical base can be split
into its {\em direct flow aspect\/} ${\bf B} \stackrel{E^{(\;)}}{\longrightarrow} {\bf Cat}$
and its {\em inverse flow aspect\/} ${\bf B}^{\rm op} \stackrel{E_{(\;)}}{\longrightarrow} {\bf Cat}$.
In this paper we are chiefly concerned with dialectical bases of preorders
which transform predicates (subtypes);
in other words, dialectical predicate transformers.
Here a dialectical base ${\bf B} \stackrel{E}{\longrightarrow} {\bf adj}$
factors through the category {\bf adj} of preorders and adjoint pairs of monotonic functions,
and direct flow ${\bf B} \stackrel{E^{(\;)}}{\longrightarrow} {\bf PO}$
and inverse flow ${\bf B}^{\rm op} \stackrel{E_{(\;)}}{\longrightarrow} {\bf PO}$
map to preorders (and usually semilattices).

Let {\bf P} be any biposet.
Any functional term $\term{y}{f \dashv f^{\rm op}}{x}$ defines
a {\em direct image\/} monotonic function
$\scriptbf{P}[y,y] \stackrel{\scriptbf{P}^f}{\longrightarrow} {\bf P}[x,x]$
defined by ${\bf P}^f(q) \define f^{\rm op} \tprod q \tprod f$ for endoterms $\term{y}{q}{y}$,
a {\em inverse image\/} monotonic function
${\bf P}[y,y] \stackrel{\scriptbf{P}_f}{\longleftarrow} {\bf P}[x,x]$
defined by ${\bf P}_f(p) \define f \tprod p \tprod f^{\rm op}$ for endoterms $\term{x}{p}{x}$.
It is easy to check that direct and inverse image is an adjoint pair of monotonic functions
${\bf P}(\term{y}{f}{x}) = {\bf P}[y,y] \stackrel{\scriptbf{P}^f \dashv \scriptbf{P}_f}{\longrightarrow} {\bf P}[x,x]$
for each functional {\bf P}-term $\term{y}{f \dashv f^{\rm op}}{x}$.
Let {\bf adj} be the category of preorders and adjoint pairs of monotonic functions.
The construction ${\bf P}$,
mapping types to their poset of endoterms
${\bf P}(x) = {\bf P}[x,x]$ and
mapping functional {\bf P}-terms to their adjoint pair of direct/inverse image adjunction,
is a dialectical base (indexed adjointness) ${\bf P}^\dashv \stackrel{\scriptbf{P}}{\longrightarrow} {\bf adj}$,
mapping functional {\bf P}-terms into the subcategory of {\bf adj}
consisting of monoidal posets and adjoint pairs of monotonic functions.

For each functional term $\term{y}{f \dashv f^{\rm op}}{x}$
and each {\bf P}-comonoid $v \memberof \comonoid{}{(y)}{}$,
the endoterm $\term{x}{f^{\rm op} \tprod v \tprod f}{x}$ is a {\bf P}-comonoid
$(f^{\rm op} \tprod v \tprod f) \memberof \comonoid{}{(x)}{}$.
So the direct image monotonic function ${\bf P}^f$ restricts to {\bf P}-comonoids.
Denote this restriction by
$\comonoid{}{(y)}{} \stackrel{\comonoid{}{}{f}}{\longrightarrow} \comonoid{}{(x)}{}$
and call it the {\em direct image\/} also.
In an interior biposet {\bf P}
the direct image function has a right adjoint 
$\comonoid{}{(y)}{} \stackrel{\comonoid{f}{}{}}{\longleftarrow} \comonoid{}{(x)}{}$ 
called the {\em inverse image\/} monotonic function,
and defined to be the interior
$\comonoid{f}{(u)}{} \define \interior{(f \tprod u \tprod f^{\rm op})}$
for each {\bf P}-comonoid $u \memberof \comonoid{}{(x)}{}$.
We regard the direct and inverse image operators
as ``the vertical dialectical flow of predicates''.
If we denote this adjointness by
$\comonoid{}{(f)}{} \define (\comonoid{}{}{f} \dashv \comonoid{f}{}{})$,
then the (vertical flow) comonoid construction $\comonoid{}{}{}$ is an indexed adjointness
(dialectical base)
${\bf P}^\dashv \stackrel{\comonoid{}{}{}}{\longrightarrow} {\bf adj}$,
mapping functional {\bf P}-terms into the subcategory of {\bf adj}
consisting of meet semilattices and adjoint pairs of monotonic functions.

When {\bf P} is an affirmation bisemilattice,
the fact that the (vertical flow) comonoid construction $\comonoid{}{}{}$ is an indexed adjointness
(dialectical base)
${\bf P}^\dashv \stackrel{\comonoid{}{}{}}{\longrightarrow} {\bf adj}$
can equivalently be expressed by the fact that
there is an indexed category
$\comonoid{{\rm P}}{}{} \stackrel{T^{\dashv}}{\longrightarrow} {\bf P}^{\dashv}$:
objects of $\comonoid{{\rm P}}{}{}$ are pairs $u \type x$ or typed comonoids $u \memberof \comonoid{}{(x)}{}$;
arrows $\term{v \type y}{f}{u \type x}$ of $\comonoid{{\rm P}}{}{}$ are functional terms $\term{y}{f \dashv f^{\rm op}}{x}$ in {\bf P}
satisfying $\comonoid{}{(u)}{f} \preceq u$ iff $v \tprod f \preceq f \tprod u$ iff $v \preceq \comonoid{f}{(u)}{}$.
For a topological biposet {\bf P},
the indexed category $\comonoid{{\rm P}}{}{}$ is a weak internal model for Milner's calculus of concurrent processes
without synchronization (interaction).
Milner's calculus can be interpreted in topological biposets as follows.
\begin{center}
   \begin{tabular}{|lc||lc|}
      \hline
      \multicolumn{2}{|c||}{``Milner's Calculus''} & \multicolumn{2}{c|}{topological biposets} \\
      \hline
      \hline
      sets of actions                 & $A$            & types                               & $x$           \\
      \multicolumn{2}{|c||}{(synchronization labels)}  &                                     &               \\
      typed processes                 & $P \type A$    & typed comonoids                     & $u \type x$   \\
                                      &                & \multicolumn{2}{|c|}{(objects of $\comonoid{{\rm P}}{}{}$)}             \\
      \hline
      action application              & $a {\bf ;} P$  & tensor product                & $s \tprod r$  \\
      nondeterministic choice         & $P + Q$        & boolean sum                   & $s \bsum r$   \\
      synchronization product         & $P [] Q$       & \multicolumn{2}{|c|}{(not represented)}      \\
      \hline
      restriction                     & $P |_{B}$      & inverse image                 & $\comonoid{i}{(u)}{} \define \interior{(i \tprod u \tprod p)}$ \\
      \multicolumn{2}{|r||}{where $P \type A$ and $B \subseteq A$} &                   & for subtype $\term{y}{{i \dashv p}}{x}$                        \\
      morphism                        & $Q[\phi]$      & direct image                  & $\comonoid{}{(v)}{f} \define (f^{\rm op} \tprod v \tprod f)$   \\
      \multicolumn{2}{|r||}{where $Q \type B$ and $B \stackrel{\phi}{\rightarrow} A$}  &                              & for functional term $\term{y}{{f \dashv f^{\rm op}}}{x}$       \\
      \hline
      recursion                       & $\vec{x} = \vec{P}(\vec{x})$     & monoidal closure              & $\closure{u}$ \\
      \multicolumn{2}{|c||}{with fixpoint solution ${\bf fix} \; \vec{x}\vec{P}(\vec{x})$} & \multicolumn{2}{c|}{(consideration modality)}                            \\
      \hline
   \end{tabular}
\end{center}
Although only a weak internal model,
this gives the correct orientation for a more adequate model discussed later.

For any pair of functional {\bf P}-terms
$\term{y_0}{f_0 \dashv f_0^{\rm op}}{x_0}$ and $\term{y_1}{f_1 \dashv f_1^{\rm op}}{x_1}$
there is a (vertical) adjoint pair of monotonic functions
${\bf P}[y_0,y_1] \stackrel{{\rm P}^{[f_0,f_1]} \dashv {\rm P}_{[f_0,f_1]}}{\longrightarrow} {\bf P}[x_0,x_1]$
called the direct/inverse image of {\bf P}-terms,
with the direct image map defined by
${\bf P}^{[f_0,f_1]}(\term{y_0}{y}{y_1}) \define f_0^{\rm op} \tprod y \tprod f_1$
and the inverse image map defined by
${\bf P}_{[f_0,f_1]}(\term{x_0}{x}{x_1}) \define f_0 \tprod x \tprod f_1^{\rm op}$.
By defining  ${\bf P}[y_0,y_1]$ to be the {\bf P}-homset
and ${\bf P}[f_0,f_1]$ to be the adjointness
${\bf P}[f_0,f_1] \define ({\bf P}^{[f_0,f_1]} \dashv {\bf P}_{[f_0,f_1]})$,
we form a base
$\power{2}{({\bf P}^\dashv)} \stackrel{{\rm P}}{\longrightarrow} {\bf adj}$
for the (vertical) dialectical flow of arbitrary {\bf P}-terms,
where $\power{2}{({\bf P}^\dashv)} \define \product{{\bf P}^\dashv}{{\bf P}^\dashv}$
is the 2nd power of the category of functional terms ${\bf P}^\dashv$.
This extends the vertical dialectical flow of endoterms
${\bf P}(f) = ({\bf P}^f \dashv {\bf P}_f)$
and the dialectical base
${\bf P}^\dashv \stackrel{{\rm P}}{\longrightarrow} {\bf adj}$
defined previously,
since the diagram
\begin{center}
   $\begin{array}{rcccl}
       {\bf P}^\dashv &          & \stackrel{\Delta_{{\rm P}^\dashv}}{\longrightarrow} &          & \power{2}{({\bf P}^\dashv)} \\
       & \stackrel{{\rm P}}{\searrow} &                                  & \stackrel{{\rm P}}{\swarrow} & \\
       && {\bf adj} &&
    \end{array}$
\end{center}
commutes,
where $\Delta_{{\rm P}^\dashv}$ is the usual diagonal functor,
the product-pairing of the identity functor on ${\rm P}^\dashv$ with itself;
that is,
${\bf P}(f) = {\bf P}[f,f]$.

Equivalent to the dialectical base
$\power{2}{({\bf P}^\dashv)} \stackrel{{\rm P}}{\longrightarrow} {\bf adj}$
is the indexed category $\stackrel{\scriptscriptstyle \rightharpoondown}{\bf P}$
called the {\em vertical category of {\bf P}-terms\/}
with source-target indexing functor
$\stackrel{\scriptscriptstyle \rightharpoondown}{\bf P} \stackrel{\partial_{\rm P}}{\longrightarrow} \power{2}{({\bf P}^\dashv)}$.
An object of $\stackrel{\scriptscriptstyle \rightharpoondown}{\bf P}$
is a {\bf P}-term $x = (\term{x_0}{x}{x_1})$.
A morphism of $\stackrel{\scriptscriptstyle \rightharpoondown}{\bf P}$,
denoted by $y \stackrel{F}{\Rightarrow} x$ and called a {\em vertical arrow\/} of {\bf P}-terms,
from a source {\bf P}-term $y = (\term{y_0}{y}{y_1})$ to a target {\bf P}-term $x = (\term{x_0}{x}{x_1})$,
consists of a pair of functional {\bf P}-terms
$\term{y_0}{f_0 \dashv f_0^{\rm op}}{x_0}$
and $\term{y_1}{f_1 \dashv f_1^{\rm op}}{x_1}$
which satisfy the direct image inequality
${\bf P}^{[f_0,f_1]}(y) \preceq_{x_0,x_1} x$,
or equivalently the inverse image inequality
$y \preceq_{y_0,y_1} {\bf P}_{[f_0,f_1]}(x)$.
Note that vertical arrow $F$ consists of the pair $\pair{f_0}{f_1}$
{\em plus\/} either of the above inequality constraints.
The indexing functor
$\stackrel{\scriptscriptstyle \rightharpoondown}{\bf P} \stackrel{\partial_{\rm P}}{\longrightarrow} \power{2}{({\bf P}^\dashv)}$
is the product-pairing of
the source functor
$\stackrel{\scriptscriptstyle \rightharpoondown}{\bf P} \stackrel{\partial_0^{\rm P}}{\longrightarrow} {\bf P}^\dashv$,
where $\partial_0^{\rm P}(\term{x_0}{x}{x_1}) = x_0$
and $\partial_0^{\rm P}(F) = f_0$,
and
the target functor
$\stackrel{\scriptscriptstyle \rightharpoondown}{\bf P} \stackrel{\partial_0^{\rm P}}{\longrightarrow} {\bf P}^\dashv$
defined similarly.

Equivalent to the above commuting square of dialectical bases,
is the following commuting square of indexed categories
\begin{center}
   $\begin{array}{ccc}
       \dot{\bf P}
       & 
       \makebox{\rule{.1in}{0in}}
       \makebox[0in]{\raisebox{.12in}[0in][0in]{\scriptsize$\Delta_{\rm P}$}} \makebox[0in]{$\longrightarrow$}
       \makebox{\rule{.1in}{0in}}
       &
       \stackrel{\scriptscriptstyle \rightharpoondown}{\bf P}
       \\
       \makebox[0in][r]{\scriptsize${\rm T}_{\rm P}$} \downarrow
       &
       & \downarrow \makebox[0in][l]{\scriptsize$\partial_{\rm P}$}
       \\
       {\bf P}^\dashv
       &
       \makebox[0in]{\raisebox{-.12in}[0in][0in]{\scriptsize$\Delta_{{\rm P}^\dashv}$}} \makebox[0in]{$\longrightarrow$}
       &
       \power{2}{({\bf P}^\dashv)}
    \end{array}$
\end{center}
Equivalent to this commuting square are the two identities
$\Delta_{\rm P} \cdot \partial_0^{\rm P} = {\rm T}_{\rm P} = \Delta_{\rm P} \cdot \partial_1^{\rm P}$.
The combination of both the horizontal and vertical aspects of
$\stackrel{\scriptscriptstyle \rightharpoondown}{\bf P}$
forms a double category,
with 
zero-cells (objects) being {\bf P}-types $x$,
one-cells  (arrows)  being {\bf P}-terms $\term{x_0}{x}{x_1}$, and
two-cells  (squares) being vertical arrows of {\bf P}-terms $y \stackrel{F}{\Rightarrow} x$.
Although we emphasize the horizontal/internal aspect of biposets in this paper,
in the section on spannable biposets we briefly discuss
the vertical/external aspect with respect to interaction of concurrent processes.

\paragraph{Dialectical Systems.}
At the base level,
a term-process $r$ can interact (communicate) with itself
iff it is an endoterm $\term{x}{r}{x}$.
By identifying self-reference (recursion) with growing dynamic systems,
the importance of endoterms is emphasized.
But there is another way in which an arbitrary term-process $\term{y}{r}{x}$
can manifest self-interaction.
It can pair itself with another term of the same type:
two term-processes of the same type $\term{y}{s,r}{x}$
can organize themselves into a complementary whole.
When viewed as a relational graph 
\begin{center}
   $x \stackrel{s}{\leftharpoondown} y \stackrel{r}{\rightharpoondown} x$
\end{center}
(a special relational span)
the pair has the potential for self-interaction. 
This self-interaction is often expressed as a composite dialectical motion \cite{Kent87}
consisting of inverse flow along one term and direct flow along the other.
Such a complementary pair (two working together as one) is called a {\em dialectical system\/}.

In an arbitrary dialectical base ${\bf B} \stackrel{{\rm E}}{\longrightarrow} {\bf Adj}$
a {\em dialectical system\/} ${\sf S} = (y \stackrel{\iota,o}{\longrightarrow} x)$ is a graph in {\bf B},
with inverse flow specifier $\iota$ and direct flow specifier $o$.
The dialectical interaction (complementary union)
of the component terms of a dialectical system
occurs through both source and target type-ports.
The notion of {\em reproduction\/} in a dialectical system is specified by the dialectical flow (fixpoint operator)
$\yinyang_\iota^o \define {\bf E}_\iota \cdot {\bf E}^o$.
This reproduction operator can be interpreted as the ``polar-turning structure'' of the preSocratic Greek philosopher Heraclitus \cite{Hussey},
and in Greek is rendered 
$\pi \alpha \lambda \iota \nu \tau \rho o \pi o \zeta$ $\alpha \rho \mu o \nu \iota \eta$.
An object $\phi \in {\bf P}(x)$ is {\em reproduced\/} when it satisfies the fixpoint equation $\yinyang_\iota^o(\phi) = \phi$.
[A philosophical note:
The notion of complementary union
(two working together in one)
is not that of ``synthesis''.
Neither of the opposites is ``transformed''.
Indeed,
with synthesis,
dialectical motion would cease! 
The notion of ``reproduction'' is one of equilibrium of motion,
not lack of motion.]
Here the yin-yang symbol $\yinyang_\iota^o$ is used as a reminder of ancient dialectics;
{\em yin\/} inverse flow along $\iota$ and {\em yang\/} direct flow along $o$.
Dialectical systems are the ``motors of nature'' specifying the dialectical motion of structured entities,
and a dialectical base provides the ``motive power'' for this motion
(from the dialectical point-of-view ``motion'' is synonymous with ``transformation'').

\paragraph{Spans.}
From the opposite standpoint to that above,
we might ask whether an arbitrary endoterm can be manifested as a dialectical system.
In dynamic sytems this question can be rephrased as
``How is {\em change\/} manifested as a dialectical phenomenon?''.
We give an answer to this question in the next few sections
by axiomatizing contexts where arbitrary terms can factor into functional spans.

For any category {\bf C} a {\bf C}-{\em span\/} $\onespan{y}{\rho}{x}$ from $y$ to $x$
is a pair of {\bf C}-arrows $\rho = (y \stackrel{r_0}{\leftarrow} r \stackrel{r_1}{\rightarrow} x)$
called the {\em legs\/} of the span,
with common source {\bf C}-object $r$ and target {\bf C}-objects $y$ and $x$.
Term entailment is defined in terms of morphisms of spans:
for any parallel pair of spans $\onespan{y}{\sigma,\rho}{x}$
where $\sigma = (y \stackrel{s_0}{\leftarrow} s \stackrel{s_1}{\rightarrow} x)$
and $\rho = (y \stackrel{r_0}{\leftarrow} r \stackrel{r_1}{\rightarrow} x)$,
the order $\sigma \preceq_{yx} \rho$ holds when there exists a {\bf C}-arrow $s \stackrel{h}{\rightarrow} r$
which commutes with the legs of the spans $h \cdot r_0 = s_0$ and $h \cdot r_1 = s_1$.
A span is potentially a specifier for a dialectically composite direct flow
(and a dialectically composite inverse flow in the exact case).
Term entailment $\preceq$ is only a preorder.
To get a poset we must quotient out in the usual fashion.
Alternatively,
we can discuss things in terms of equivalence,
where $\rho \equiv \sigma$ iff both $\rho \preceq \sigma$ and $\sigma \preceq \rho$,
instead of equality.
For any two {\bf C}-arrows
$y \stackrel{g}{\rightarrow} z$ and $x \stackrel{f}{\rightarrow} w$
there is a vertical direct image of {\bf C}-spans
$\widehat{{\bf C}}[y,x] \stackrel{\widehat{{\rm C}}^{[g,f]}}{\longrightarrow} \widehat{{\bf C}}[z,w]$,
which is defined by
$\widehat{{\bf C}}^{[g,f]}(y \stackrel{r_0}{\leftarrow} r \stackrel{r_1}{\rightarrow} x)
 \define
 (z \stackrel{r_0 \cdot g}{\leftarrow} r \stackrel{r_1 \cdot f}{\rightarrow} w)$.
Initially,
since we are not assuming the existence of any constraints on the underlying category ${\bf C}$ (exactness),
neither a tensor product in the horizontal aspect
nor an inverse image operator in the vertical aspect
necessarily exist.


An {\em exact category\/} {\bf C} is a category possessing canonical finite limits.
Let {\bf C} be any exact category,
with $1$ denoting the terminal object (empty product) in {\bf C};
so that there is a unique {\bf C}-arrow $x \stackrel{1_x}{\rightarrow} 1$ from any {\bf C}-object $x$ to $1$.
Horizontally the collection of {\bf C}-spans forms a biposet
(we ambiguously use the same notation $\widehat{\bf C}$ for this).
A type (an object of $\widehat{\bf C}$) is just a {\bf C}-object,
and a term $\onespan{y}{\rho}{x}$ (an arrow of $\widehat{\bf C}$) is a {\bf C}-span.
Tensor product is pullback-composition:
for terms $\onespan{z}{\sigma}{y}$ and $\onespan{y}{\rho}{x}$
where $\sigma = (z \stackrel{s_0}{\leftarrow} s \stackrel{s_1}{\rightarrow} y)$
and $\rho = (y \stackrel{r_0}{\leftarrow} r \stackrel{r_1}{\rightarrow} x)$,
the tensor product term $\onespan{z}{\sigma \tprod \rho}{x}$ is
$\sigma \tprod \rho = (z \stackrel{\hat{r}_0 \cdot s_0}{\longleftarrow} {s \tprod r} \stackrel{\hat{s}_1 \cdot r_1}{\longrightarrow} x)$
where $s \stackrel{\hat{r}_0}{\leftarrow} {s \tprod r} \stackrel{\hat{s}_1}{\rightarrow} r$
is the pullback of the opspan $s \stackrel{s_1}{\rightarrow} y \stackrel{r_0}{\leftarrow} r$.
The identity span at type $x$ is $x \stackrel{x}{\leftarrow} x \stackrel{x}{\rightarrow} x$.
There is an obvious involution on spans,
where $\involution{\rho} = (x \stackrel{r_1}{\leftarrow} r \stackrel{r_0}{\rightarrow} y)$
for any span
$\rho = (y \stackrel{r_0}{\leftarrow} r \stackrel{r_1}{\rightarrow} x)$.
The construction $\widehat{(\;)}$ can be described as
either ``external dialecticality'' or ``spanning''.
\begin{Fact}
   If {\bf C} is an exact category,
   then $\widehat{{\bf C}}$ is an involutive biposet.
\end{Fact}

As usual,
a span $\onespan{y}{\rho}{x}$ is a functional span
when $y \preceq \rho \tprod \involution{\rho}$ and $\involution{\rho} \tprod \rho \preceq x$.
If $\rho$ is a functional span and $\sigma \equiv \rho$,
then $\sigma$ is also a functional span.
In analogy with the relative Yoneda embeddings of enriched functors as bimodules in enriched category theory,
arrows of an exact category {\bf C} can be embedded as adjoint pairs of spans;
that is,
as functional spans.
There is a covariant embedding
${\bf C} \stackrel{\yoneda{{\rm C}}{}{1}}{\longrightarrow} \widehat{\bf C}$
of arrows as spans:
any {\bf C}-arrow $y \stackrel{f}{\rightarrow} x$ 
is a span $\onespan{y}{\yoneda{{\rm C}}{(f)}{1}}{x}$
where $\yoneda{{\rm C}}{(f)}{1} = (y \stackrel{y}{\leftarrow} y \stackrel{f}{\rightarrow} x)$.
There is a contravariant embedding
${\bf C}^{\rm op} \stackrel{\yoneda{{\rm C}}{}{0}}{\longrightarrow} \widehat{\bf C}$
of arrows as spans:
any {\bf C}-arrow $y \stackrel{f}{\rightarrow} x$ 
is a span $\onespan{x}{\yoneda{{\rm C}}{(f)}{0}}{y}$
where $\yoneda{{\rm C}}{(f)}{0} = (x \stackrel{f}{\leftarrow} y \stackrel{y}{\rightarrow} y)$.
These spans are adjoint 
$\onespan{y}{\yoneda{{\rm C}}{(f)}{1} \dashv \yoneda{{\rm C}}{(f)}{0}}{x}$
and hence mutually involutive
$\yoneda{{\rm C}}{(f)}{0} = \involution{\yoneda{{\rm C}}{(f)}{1}}$.
If we denote this adjunction by
$\yoneda{{\rm C}}{(f)}{} \define (\yoneda{{\rm C}}{(f)}{1} \dashv \yoneda{{\rm C}}{(f)}{0})$,
then
\begin{center}
   ${\bf C} \stackrel{\yoneda{{\rm C}}{}{}}{\longrightarrow} {\widehat{\bf C}}^\dashv$
\end{center}
is a functor called the {\em Yoneda embedding\/} of {\bf C} into ${\widehat{\bf C}}^\dashv$ 
the category of functional spans.  
Term entailment on functional spans of this form reduces to equality:
$\yoneda{{\rm C}}{(f)}{} \preceq \yoneda{{\rm C}}{(g)}{}$ iff $f = g$.
Also,
$\yoneda{{\rm C}}{(f)}{1} \preceq {\yoneda{{\rm C}}{(g)}{1}}^{\rm op} = \yoneda{{\rm C}}{(g)}{0}$ iff $f \tprod g = y$.
So $\yoneda{{\rm C}}{}{}$ is an embedding,
since it is bijective on objects and faithful.
\begin{Tlemma}
   Any functional span $\rho$ is equivalent to a Yoneda span $\rho \equiv \yoneda{{\rm C}}{(f)}{1}$.
   So for any exact category {\bf C},
   the Yoneda embedding $\yoneda{{\rm C}}{}{}$ is actually an isomorphism
   \begin{center}
      ${\bf C} \stackrel{\yoneda{{\rm C}}{}{}}{\cong} {\widehat{\bf C}}^\dashv$
   \end{center}
\end{Tlemma}
\begin{proof}
   Let $\onespan{y}{\rho}{x}$ be any functional span,
   where $\rho = (y \stackrel{r_0}{\leftarrow} r \stackrel{r_1}{\rightarrow} x)$.
   The tensor products $\involution{\rho} \tprod \rho$ and $\rho \tprod \involution{\rho}$
   are defined by means of the pullback spans
   $r \stackrel{s_0}{\leftarrow} s \stackrel{s_1}{\rightarrow} r$
   and
   $r \stackrel{t_0}{\leftarrow} t \stackrel{t_1}{\rightarrow} r$
   (these are also called the kernel pairs)
   of the opspans
   $r \stackrel{r_1}{\rightarrow} x \stackrel{r_1}{\leftarrow} r$
   and
   $r \stackrel{r_0}{\rightarrow} y \stackrel{r_0}{\leftarrow} r$,
   respectively.
   The counit inequality $\involution{\rho} \tprod \rho \preceq x$
   means that there is a {\bf C}-arrow $t \stackrel{h}{\rightarrow} x$ such that
   $t_0 \cdot r_1 = h = t_1 \cdot r_1$. 
   The unit inequality $y \preceq \rho \tprod \involution{\rho}$
   means that there is a {\bf C}-arrow $y \stackrel{k}{\rightarrow} s$ such that
   $k \cdot s_0 \cdot r_0 = y = k \cdot s_1 \cdot r_0$.
   By definition of pullback,
   there is a unique {\bf C}-arrow $y \stackrel{g}{\rightarrow} t$ such that
   $g \cdot t_0 = k \cdot s_0$ and $g \cdot t_1 = k \cdot s_1$.
   Since $r_0 \cdot k \cdot s_0 \cdot r_0 = r_0$,
   by definition of pullback,
   there is a unique {\bf C}-arrow $r \stackrel{p}{\rightarrow} t$ such that
   $p \cdot t_0 = r_0 \cdot k \cdot s_0$ and $p \cdot t_1 = r$.
   Define the {\bf C}-arrow $y \stackrel{f}{\rightarrow} x$ by
   $f \define g \cdot h$.
   Then
   $r_0 \cdot f
    = r_0 \cdot g \cdot h
    = r_0 \cdot g \cdot t_0 \cdot r_1
    = r_0 \cdot k \cdot s_0 \cdot r_1
    = p \cdot t_0 \cdot r_1
    = p \cdot h
    = p \cdot t_1 \cdot r_1
    = r_1$.
    The equality $r_0 \cdot f = r_1$ means the span inequality
    $\rho \preceq \yoneda{{\rm C}}{(f)}{1}$.
    Also,
    $g \cdot t_0 \cdot r_1 = g \cdot h = f$ 
    and
    $g \cdot t_0 \cdot r_0 = k \cdot s_0 \cdot r_0 = y$ 
    imply the span inequality
    $\yoneda{{\rm C}}{(f)}{1} \preceq \rho$.
\end{proof}
So there is precisely one Yoneda span in each equivalence class of functional spans.
We use this Yoneda span, or its {\bf C}-arrow, as the representative.
Moreover,
$\yoneda{{\rm C}}{}{1}
 = {\bf C} \stackrel{\yoneda{{\rm C}}{}{}}{\cong} {\widehat{\bf C}}^\dashv \stackrel{{\rm Inc}}{\longrightarrow} \widehat{{\bf C}}$
and
$\yoneda{{\rm C}}{}{0}
 = {\bf C}^{\rm op} \stackrel{(\yoneda{{\rm C}}{}{1})^{\rm op}}{\longrightarrow} {\widehat{\bf C}}^{\rm op} \stackrel{\involution{(\,)}}{\longrightarrow} \widehat{{\bf C}}$,
so that $\yoneda{{\rm C}}{}{1}$ is essentially the inclusion ${\widehat{\bf C}}^\dashv \stackrel{{\rm Inc}}{\longrightarrow} \widehat{{\bf C}}$
and $\yoneda{{\rm C}}{}{0}$, $\yoneda{{\rm C}}{}{1}$ and $\yoneda{{\rm C}}{}{}$ are interdefinable.
We refer to any of these equivalent constructions as the {\em Yoneda embedding\/}.

Any {\bf C}-arrow $y \stackrel{f}{\rightarrow} x$ having target $x$
has traditionally been called a ``subobject'' or ``generalized element'' of $x$
in the local cartesian context of topos theory \cite{Seeley}.
Let ${\bf C} {\downarrow} x$ denote subobjects of type $x$.
Any exact category {\bf C} naturally defines a dialectical flow of its own subobjects.
Each {\bf C}-arrow $y \stackrel{f}{\rightarrow} x$
specifies by composition a direct-image monotonic function
${{\bf C} {\downarrow} y} \stackrel{{\rm C}^f}{\longrightarrow} {{\bf C} {\downarrow} x}$
defined by ${\bf C}^f(w \stackrel{h}{\rightarrow} y) \define h \cdot f$,
and specifies by pullback an inverse-image monotonic function
${{\bf C} {\downarrow} y} \stackrel{{\rm C}_f}{\longleftarrow} {{\bf C} {\downarrow} x}$
defined by ${\bf C}^f(z \stackrel{g}{\rightarrow} y) \define \hat{z} \stackrel{\hat{g}}{\rightarrow} y$ 
where the {\bf C}-span
$y \stackrel{\hat{g}}{\leftarrow} \hat{z} \stackrel{\hat{f}}{\rightarrow} z$
is the pullback of the {\bf C}-opspan
$y \stackrel{f}{\rightarrow} x \stackrel{g}{\leftarrow} z$.
Defining 
${\bf C}(x) \define {{\bf C} {\downarrow} x}$ for each type $x$
and
${\bf C}(f) \define ({\bf C}^f \dashv {\bf C}_f)$
the direct/inverse image adjointness
for each {\bf C}-arrow $y \stackrel{f}{\rightarrow} x$,
makes the exact category ${\bf C}$ a dialectical base
${\bf C} \stackrel{{\rm C}}{\longrightarrow} {\bf adj}$.

Given any type $x$,
a span-endoterm of type $x$ is a {\bf C}-endospan $\onespan{x}{\pi}{x}$,
where $\pi = (x \stackrel{p_0}{\leftarrow} y \stackrel{p_1}{\rightarrow} x)
           = y \stackrel{p_0,p_1}{\rightarrow} x$.
A {\bf C}-endospan is also called a {\bf C}-{\em graph\/} with
``edges'' {\bf C}-object $y$,
``nodes'' {\bf C}-object $x$,
``source'' {\bf C}-arrow $p_0$ and
``target'' {\bf C}-arrow $p_1$.
A {\bf C}-endospan is potentially a specifier for dialectical flow.
An endospan $\onespan{x}{\phi}{x}$ of the form
$\phi = (x \stackrel{f}{\leftarrow} y \stackrel{f}{\rightarrow} x)$
is a ``diagonal span'',
a {\bf C}-graph with only self-loop edges $y$ on nodes $x$.
Since diagonal spans at $x$ satisfy both partiality $\phi \preceq x$ and idempotency $\phi = \phi \tprod \phi$, 
they are precisely the $x$-comonoids in the biposet $\widehat{{\bf C}}$.
Each endospan $\onespan{x}{\pi}{x}$,
where $\pi = (x \stackrel{p_0}{\leftarrow} y \stackrel{p_1}{\rightarrow} x)$,
has an interior $\onespan{x}{\interior{\pi}}{x}$ diagonal span (span-comonoid),
which is the equalizer of the {\bf C}-{\em graph\/} diagram
$y \stackrel{p_0,p_1}{\longrightarrow} x$,
and consists of the self-loop edges part of the graph.
So $\widehat{{\bf C}}$ is an interior biposet.
Recall that the usual direct/inverse image vertical flow of comonoids
in the biposet of {\bf C}-spans
forms a dialectical base
${\widehat{\bf C}}^\dashv \stackrel{\comonoid{}{}{}}{\longrightarrow} {\bf adj}$,
with $\comonoid{}{(x)}{} = \comonoid{\widehat{{\rm C}}}{(x)}{}$
 denoting span-comonoids at any {\bf C}-type $x$,
and  $\comonoid{}{(\rho)}{} = (\comonoid{}{}{\rho} \dashv \comonoid{\rho}{}{})$
 denoting the usual vertical flow of comonoids at any functional span $\onespan{y}{\rho}{x}$.

{\bf C}-subobjects $y \stackrel{f}{\rightarrow} x$ at $x$
can be identified with {\bf C}-comonoids ${\Delta_x(f)} \in \comonoid{\widehat{{\rm C}}}{(x)}{}$
where $\Delta_x(f) = (x \stackrel{f}{\leftarrow} y \stackrel{f}{\rightarrow} x)$.
The operator $\Delta_x$ is a generalized diagonal operator.
This diagonal operator is a bijection
${{\bf C} {\downarrow} x} \stackrel{\Delta_x}{\cong} \comonoid{\widehat{{\rm C}}}{(x)}{}$,
inducing an order on $x$-subobjects ${\bf C} {\downarrow} x$ which is the traditional order,
and making $\Delta_x$ an order isomorphism.
The direct/inverse image adjoint pairs and diagonal inverse pairs
form a commuting square
\begin{center}
   $\begin{array}{ccc}
       {{\bf C}(y)}
       & 
       \makebox{\rule{.1in}{0in}}
       \makebox[0in]{\raisebox{.12in}[0in][0in]{\scriptsize$\Delta_y$}} \makebox[0in]{$\cong$}
       \makebox{\rule{.1in}{0in}}
       &
       \comonoid{}{(y)}{}
       \\
       \makebox[0in][r]{\scriptsize${\rm C}(f)$} \downarrow
       &
       & \downarrow \makebox[0in][l]{\scriptsize$\comonoid{}{(\yoneda{}{(f)}{})}{}$}
       \\
       {{\bf C}(x)}
       &
       \makebox[0in]{\raisebox{-.12in}[0in][0in]{\scriptsize$\Delta_x$}} \makebox[0in]{$\cong$}
       &
       \comonoid{}{(x)}{}
    \end{array}$
\end{center}
for any {\bf C}-arrow $y \stackrel{f}{\rightarrow} x$,
so that the Yoneda/diagonal pair $\yoneda{{\rm C}}{}{}/\Delta_{\rm C}$
is an isomorphism of dialectical bases
\begin{center}
   $\begin{array}{rcccl}
       {\bf C} && \stackrel{\yoneda{{\rm C}}{}{}}{\cong} && {\widehat{\bf C}}^\dashv \\
       & \stackrel{{\rm C}}{\searrow} & \stackrel{\Delta_{\rm C}}{\cong} & \stackrel{\comonoid{}{}{}}{\swarrow} & \\
       && {\bf adj} &&
    \end{array}$
\end{center}
with the direct/inverse image flow
${\bf C} \stackrel{{\rm C}}{\longrightarrow} {\bf adj}$
isomorphic via diagonal
to the usual vertical flow of comonoids in the biposet of {\bf C}-spans
${\bf C} \stackrel{\yoneda{{\rm C}}{}{}}{\cong}
 {\widehat{\bf C}}^\dashv \stackrel{\comonoid{}{}{}}{\longrightarrow} {\bf adj}$.
This isomorphism of dialectical bases
effectively translates the local cartesian context of topos theory into dialectical logic,
and shows that dialectical logic subsumes topos theory in a very natural fashion.

Terms in the biposet of spans $\widehat{\bf C}$
have the following interesting horizontal decomposition property:
any span $\onespan{y}{\rho}{x}$ where $\rho = (y \stackrel{r_0}{\leftarrow} r \stackrel{r_1}{\rightarrow} x)$,
factors as the two leg span-objects
$\rho = r_0^{\rm op} \tprod r_1
      = \yoneda{{\rm C}}{(r_0)}{0} \tprod \yoneda{{\rm C}}{(r_1)}{1}$.
We now axiomatize this decomposition property.

\paragraph{Spannable Biposets.}
Change in dynamic systems is represented by terms in biposets and described as state-transition.
Now terms and state-transitions are dialectical notions:
to change state we must ``leave'' a current state and ``arrive'' at a next state.
This source/target, leaving/arriving, starting/finishing {\em transition dialectic\/},
which is concretely realized in {\bf C}-spans $\widehat{{\bf C}}$,
will be abstractly axiomatized here by the notion of spannable biposets.

Let {\bf P} be any biposet with category of functional terms ${\bf P}^\dashv$. 
For any pair of {\bf P}-types $y$ and $x$,
an element $\onespan{y}{\rho}{x}$ in the poset $\widehat{{\bf P}^\dashv}[y,x]$ 
is a span of functional terms
$\rho = (y \stackrel{r_0 \dashv r_0^{\rm op}}{\leftharpoondown} r \stackrel{r_1 \dashv r_1^{\rm op}}{\rightharpoondown} x)$,
which is also called a functional span.
There is a canonical {\em underlying {\bf P}-term\/} monotonic function
$\widehat{{\bf P}^\dashv}[y,x] \stackrel{\uterm{y}{x}}{\longrightarrow} {\bf P}[y,x]$,
which ``flattens'' a functional span to a term 
$\uterm{y}{x}(y \stackrel{r_0 \dashv r_0^{\rm op}}{\leftharpoondown} r \stackrel{r_1 \dashv r_1^{\rm op}}{\rightharpoondown} x)
 \define
 \term{y}{r_0^{\rm op} \tprod r_1}{x}$,
mapping a span of functional terms to the tensor product of its legs.
For the special case ${\bf P} = \widehat{{\bf C}}$ where {\bf C} is exact,
the underlying term functions are isomorphisms
$\widehat{{\widehat{{\bf C}}}^\dashv}[y,x] \stackrel{\uterm{y}{x}}{\cong} \widehat{{\bf C}}[y,x]$.
In the poset $\widehat{{\bf P}^\dashv}$,
the (vertical) direct image of functional spans
$\widehat{{\bf P}^\dashv}[y,x] \stackrel{\widehat{{\rm P}^\dashv}^{[g,f]}}{\longrightarrow} \widehat{{\bf P}^\dashv}[z,w]$
is defined by
$\widehat{{\bf P}^\dashv}^{[g,f]}(y \stackrel{r_0}{\leftharpoondown} r \stackrel{r_1}{\rightharpoondown} x)
 \define
 (z \stackrel{r_0 \tprod g}{\leftharpoondown} r \stackrel{r_1 \tprod f}{\rightharpoondown} w)$.
In arbitrary {\bf P},
since
$\uterm{z}{w}(\widehat{{\bf P}^\dashv}^{[g,f]}(\rho))
 = g^{\rm op} \tprod r_0^{\rm op} \tprod r_1 \tprod f
 = {\bf P}^{[g,f]}(\uterm{y}{x}(\rho))$
for any functional span
$\rho = (y \stackrel{r_0 \dashv r_0^{\rm op}}{\leftharpoondown} r \stackrel{r_1 \dashv r_1^{\rm op}}{\rightharpoondown} x)$,
we have the identity
   $\widehat{{\bf P}^\dashv}^{[g,f]} \cdot \uterm{z}{w} = \uterm{y}{x} \cdot {\bf P}^{[g,f]}$,
which asserts naturality of the underlying {\bf P}-term operator
$\widehat{{\bf P}^\dashv}^{[\,,\,]} \stackrel{\uterm{}{}}{\Rightarrow} {\bf P}^{[\,,\,]}$
between direct flows.
We will axiomatically extend this identity to a commuting square of adjoint pairs.
We do this in two steps.

First of all,
we assume the existence of right adjoints
$\widehat{{\bf P}^\dashv}[y,x] \stackrel{\spano{y}{x}}{\longleftarrow} {\bf P}[y,x]$
to the underlying {\bf P}-term functions.
So each {\bf P}-term $\term{y}{r}{x}$ has an associated {\em overlying span\/} of functional terms
$\spano{y}{x}(r)
 = (y \stackrel{r_0 \dashv r_0^{\rm op}}{\leftharpoondown} r \stackrel{r_1 \dashv r_1^{\rm op}}{\rightharpoondown} x)$,
with
$\uterm{y}{x}(\rho') \preceq r$ for each functional span
$\rho' = (y \stackrel{r'_0}{\leftharpoondown} r' \stackrel{r'_1}{\rightharpoondown} x)$ and each {\bf P}-term $\term{y}{r}{x}$
iff
$(r'_0)^{\rm op} \tprod r'_1 \preceq r$
iff
there exists a functional term $\term{r'}{h \dashv h'}{r}$ such that $h \tprod r_0 = r'_0$ and $h \tprod r_1 = r'_1$
iff
$\rho' \preceq \spano{y}{x}(r)$.
We assume that the {\bf P}-term underlying
$\spano{y}{x}(r)$ is $r$ itself;
that is,
each {\bf P}-term factors through its associated functional span
$r = r_0^{\rm op} \tprod r_1$.
So the adjoint pair $\utermspano{y}{x}$ defined by
$\utermspano{y}{x} \define (\uterm{y}{x} \dashv \spano{y}{x})$
is a reflective pair of monotonic functions with identity counit
$\spano{y}{x} \cdot \uterm{y}{x} = {\rm Id}$.
We also assume that the overlying span operator $\spano{}{}$ includes
as special cases the Yoneda functional term embeddings,
satisfying the ``axiom of functional extension''
\begin{center}
   $\begin{array}{r@{\;=\;}lcr@{\;=\;}l}
      \spano{y}{x}(f) & \yoneda{{\rm P}^\dashv}{(f)}{1}
      &&
      \spano{x}{y}(f^{\rm op}) & \yoneda{{\rm P}^\dashv}{(f)}{0}
    \end{array}$
\end{center}
for any functional term $\term{y}{f \dashv f^{\rm op}}{x}$;
that is,
that
$\spano{y}{x}(\term{y}{f}{x})
 = (y \stackrel{y \dashv y}{\leftharpoondown} y \stackrel{f \dashv f^{\rm op}}{\rightharpoondown} x)$
and
$\spano{x}{y}(\term{x}{f^{\rm op}}{y})
 = (x \stackrel{f \dashv f^{\rm op}}{\leftharpoondown} y \stackrel{y \dashv y}{\rightharpoondown} y)$.
More abstractly,
${\rm Inc}_{yx} \cdot \spano{y}{x}
 = \yoneda{{\rm P}^\dashv,yx}{}{1}$
and .
When ${\bf P} = \widehat{{\bf C}}$ for exact {\bf C}, 
the overlying span functions give the span decomposition that motivated this section
$\spano{{\rm C},}{yx}(\rho)
 = (y \stackrel{\yoneda{{\rm C}}{(r_0)}{0}}{\longleftarrow} r \stackrel{\yoneda{{\rm C}}{(r_1)}{1}}{\longrightarrow} x)$,
for any span $\term{y}{\rho}{x}$
where $\rho = (y \stackrel{r_0}{\leftarrow} r \stackrel{r_1}{\rightarrow} x)$;
so that
$\spano{{\rm C},}{yx} = \widehat{\yoneda{{\rm C}}{}{}}_{yx}$
the $yx$-th component
$\widehat{{\widehat{{\bf C}}}^\dashv}[y,x] \stackrel{\widehat{\yoneda{{\rm C}}{}{}}_{yx}}{\cong} \widehat{{\bf C}}[y,x]$
of the yoneda isomorphism.
If {\bf P} is an involutive biposet ${\bf P} = \pair{{\bf P}}{\involution{(\,)}}$,
then we assume $\spano{}{}$ commutes with involution:
$\involution{\spano{y}{x}(r)} \equiv \spano{x}{y}(\involution{r})$.

Secondly,
we assume the existence of a right adjoint $\widehat{{\bf P}^\dashv}_{[g,f]}$ to the direct image of spans $\widehat{{\bf P}^\dashv}^{[g,f]}$
called the {\em inverse image\/} of spans,
satisfying the ``adjointness'' axiom
$\widehat{{\bf P}^\dashv}^{[g,f]}(\rho) \preceq \sigma$
iff $\rho \preceq \widehat{{\bf P}^\dashv}_{[g,f]}(\sigma)$
iff there exists a functional term $\term{r}{h \dashv h'}{s}$ such that $h \tprod s_0 = r_0 \tprod g$ and $h \tprod s_1 = r_1 \tprod f$
for any two functional spans
$\term{y}{\rho}{x}   = (y \stackrel{r_0 \dashv r_0^{\rm op}}{\leftharpoondown} r \stackrel{r_1 \dashv r_1^{\rm op}}{\rightharpoondown} x)$
and
$\term{z}{\sigma}{w} = (z \stackrel{s_0 \dashv s_0^{\rm op}}{\leftharpoondown} s \stackrel{s_1 \dashv s_1^{\rm op}}{\rightharpoondown} w)$.
In particular,
$\widehat{{\bf P}^\dashv}_{[f,x]}(x) = 
 (y \stackrel{y \dashv y}{\leftharpoondown} y \stackrel{f \dashv f^{\rm op}}{\rightharpoondown} x)$
and
$\widehat{{\bf P}^\dashv}_{[x,f]}(x) = 
 (x \stackrel{f \dashv f^{\rm op}}{\leftharpoondown} y \stackrel{y \dashv y}{\rightharpoondown} y)$.
By defining $\widehat{{\bf P}^\dashv}[g,f]$ to be the adjointness
$\widehat{{\bf P}^\dashv}[g,f] \define (\widehat{{\bf P}^\dashv}^{[g,f]} \dashv \widehat{{\bf P}^\dashv}_{[g,f]})$
and $\widehat{{\bf P}^\dashv}[y,x]$ to be the homset,
we get a dialectical base
$\product{{\bf P}^\dashv}{{\bf P}^\dashv} \stackrel{\widehat{{\rm P}^\dashv}}{\longrightarrow} {\bf adj}$.
We further assume that these direct/inverse image and term/span adjoint pairs
form a commuting square
\begin{center}
   $\begin{array}{ccc}
       \widehat{{\bf P}^\dashv}[y,x]
       & 
       \makebox{\rule{.1in}{0in}}
       \makebox[0in]{\raisebox{.12in}[0in][0in]{\scriptsize$\utermspano{y}{x}$}} \makebox[0in]{$\longrightarrow$}
       \makebox{\rule{.1in}{0in}}
       &
       {\bf P}[y,x]
       \\
       \makebox[0in][r]{\scriptsize$\widehat{{\rm P}^\dashv}[g,f]$} \downarrow
       &
       & \downarrow \makebox[0in][l]{\scriptsize${\bf P}[g,f]$}
       \\
       \widehat{{\bf P}^\dashv}[z,w]
       &
       \makebox[0in]{\raisebox{-.12in}[0in][0in]{\scriptsize$\utermspano{z}{w}$}} \makebox[0in]{$\longrightarrow$}
       &
       {\bf P}[z,w]
    \end{array}$
\end{center}
for any two functional terms
$\term{y}{g \dashv g^{\rm op}}{z}$ and $\term{x}{f \dashv f^{\rm op}}{w}$,
so that
$\widehat{{\bf P}^\dashv} \stackrel{\utermspano{}{}}{\Rightarrow} {\bf P}$ 
is a morphism of bases of (vertical) dialectical flow.
The inverse aspect asserts the ``inverse flow axiom''
$\spano{z}{w} \cdot \widehat{{\bf P}^\dashv}_{[g,f]} \equiv {\bf P}_{[g,f]} \cdot \spano{y}{x}$;
that is, 
$\widehat{{\bf P}^\dashv}_{[g,f]}(\spano{z}{w}(s))
 \equiv \spano{y}{x}({\bf P}_{[g,f]}(s))
      = \spano{y}{x}(g \tprod s \tprod f^{\rm op})$
for any {\bf P}-term $\term{z}{s}{w}$.
The axiom of functional extension is equivalent to the axiom
that ``$\spano{}{}$ preserves identities''
$\spano{x}{x}(x)
 = (x \stackrel{x \dashv x}{\leftharpoondown} x \stackrel{x \dashv x}{\rightharpoondown} x)
 = x$,
since if this holds then
$\spano{y}{x}(f)
 = \widehat{{\bf P}^\dashv}_{[f,x]}(\spano{x}{x}(x))
 = \widehat{{\bf P}^\dashv}_{[f,x]}(x)
 = (y \stackrel{y \dashv y}{\leftharpoondown} y \stackrel{f \dashv f^{\rm op}}{\rightharpoondown} x)$
and
$\spano{x}{y}(f^{\rm op})
 = \widehat{{\bf P}^\dashv}_{[x,f]}(\spano{x}{x}(x))
 = \widehat{{\bf P}^\dashv}_{[x,f]}(x)
 = (x \stackrel{f \dashv f^{\rm op}}{\leftharpoondown} y \stackrel{y \dashv y}{\rightharpoondown} y)$.
If
$\sigma = (y \stackrel{s_0 \dashv s_0^{\rm op}}{\leftharpoondown} s \stackrel{s_1 \dashv s_1^{\rm op}}{\rightharpoondown} x)$  
is a functional span
and
$\tau   = (y \stackrel{t_0 \dashv t_0^{\rm op}}{\rightharpoondown} t \stackrel{t_1 \dashv t_1^{\rm op}}{\leftharpoondown} x)$  
is a functional opspan between $y$ and $x$,
then
$\sigma \preceq \spano{y}{x}(t_0 \tprod t_1^{\rm op})$
iff
$\sigma \preceq \widehat{{\bf P}^\dashv}_{[t_0,t_1]}(\spano{t}{t}(t))$
iff
$\sigma \preceq \widehat{{\bf P}^\dashv}_{[t_0,t_1]}(t)$
iff
$\widehat{{\bf P}^\dashv}^{[t_0,t_1]}(\sigma) \preceq t$
iff
$(t \stackrel{s_0 \tprod t_0}{\leftharpoondown} s \stackrel{s_1 \tprod t_1}{\rightharpoondown} t)
 \preceq (t \stackrel{t}{\leftharpoondown} t \stackrel{t}{\rightharpoondown} t)$
iff
there exists a functional term $\term{s}{h \dashv h'}{t}$ such that $h = s_0 \tprod t_0$ and $h = s_1 \tprod t_1$
iff
$s_0 \tprod t_0 = s_1 \tprod t_1$.
This states that the span overlying an opspan is the pullback of the opspan.

A biposet {\bf P} which satisfies the above axioms is said to be {\em spannable\/};
that is,
a {\em spannable biposet\/} {\bf P} is a biposet for which:
\begin{enumerate}
   \item there exist overlying span monotonic functions $\spano{y}{x}$
      which are right adjoint left inverse
      to the underlying {\bf P}-term monotonic functions $\uterm{y}{x}$
      and which extend the two Yoneda embeddings
      $\yoneda{{\rm P}^\dashv,yx}{}{0}$ and $\yoneda{{\rm P}^\dashv,yx}{}{1}$;
   \item there exist inverse image monotonic functions
      $\widehat{{\bf P}^\dashv}_{[g,f]}$
      for spans which are right adjoint to the direct image monotonic functions
      $\widehat{{\bf P}^\dashv}^{[g,f]}$; and
   \item the term/span reflective pairs $\utermspano{y}{x} \define (\uterm{y}{x} \dashv \spano{y}{x})$
      form a morphism of dialectical bases
      $\widehat{{\bf P}^\dashv} \stackrel{\utermspano{}{}}{\Rightarrow} {\bf P}$. 
\end{enumerate}
\begin{Proposition}
   \begin{enumerate}
      \item If {\bf C} is an exact category,
            then $\widehat{{\bf C}}$ is a spannable biposet.
      \item If {\bf P} is a spannable biposet,
            then ${\bf P}^\dashv$
            (the horizontal category of functional terms)
            is an exact category.
   \end{enumerate}
\end{Proposition}
So equivalently,
a {\em spannable biposet\/} {\bf P} is a biposet
(1) whose associated category of functional terms ${\bf P}^\dashv$ is exact, and
(2) whose underlying {\bf P}-term operators $\uterm{y}{x}$
    have right adjoints left inverses
    $\spano{y}{x}$ called overlying span operators,
    such that the reflective pairs of monotonic functions
    $\utermspano{y}{x} \define (\uterm{y}{x} \dashv \spano{y}{x})$
    form a morphism of bases of (vertical) dialectical flow
    $\widehat{{\bf P}^\dashv} \stackrel{\utermspano{}{}}{\Rightarrow} {\bf P}$.
Given two functional spans
$\sigma = (z \stackrel{s_0 \dashv s_0^{\rm op}}{\leftharpoondown} s \stackrel{s_1 \dashv s_1^{\rm op}}{\rightharpoondown} y)$ 
and
$\rho = (y \stackrel{r_0 \dashv r_0^{\rm op}}{\leftharpoondown} r \stackrel{r_1 \dashv r_1^{\rm op}}{\rightharpoondown} x)$,
the {\em tensor product\/} functional span is defined by
$\sigma \tprod \rho \define \widehat{{\bf P}^\dashv}^{[s_0,r_1]}(\spano{s}{r}(s_1 \tprod r_0^{\rm op}))$.
This span tensor product is the usual definition of tensor product of spans in exact categories,
and {\bf P}-types and spans of functional {\bf P}-terms form a double biposet $\widehat{{\bf P}^\dashv}$.

\begin{Proposition}
   If {\bf P} is a spannable biposet,
   then $\stackrel{\scriptscriptstyle \rightharpoondown}{\bf P}$
   (the vertical category of term arrows)
   is an exact category.
\end{Proposition} 
\begin{proof}
\end{proof}
So any biposet {\bf P} has an associated category
$\widehat{\stackrel{\scriptscriptstyle \rightharpoondown}{{\bf P}}}$
of spans of vertical arrows of {\bf P}-terms,
which is indexed by {\bf P}-terms.
A 2-{\em span\/} $\twospan{y}{R}{x}$,
from source {\bf P}-term $y = (\term{y_0}{y}{y_1})$
to target {\bf P}-term $x = (\term{x_0}{x}{x_1})$,
is a pair of vertical arrows of {\bf P}-terms
$R = (y \stackrel{G}{\Leftarrow} z \stackrel{F}{\Rightarrow} x)$
called the {\em legs\/} of the 2-span
with common source {\bf P}-term $z = (\term{z_0}{z}{z_1})$
and target {\bf P}-terms $y$ and $x$.
The vertical tensor product $S \tprodvert R$ of two composable spans
$\twospan{z}{S}{y}$ and $\twospan{y}{R}{x}$
is defined (as usual) by pullback then composition in
$\stackrel{\scriptscriptstyle \rightharpoondown}{\bf P}$.
For the special case ${\bf P} = \widehat{{\bf C}}$ where {\bf C} is exact,
a 2-span $\twospan{y}{R}{x}$,
from source {\bf C}-span
 $y = (b_0 \stackrel{y_0}{\leftarrow} b \stackrel{y_1}{\rightarrow} b_1)$
to target {\bf C}-span
 $x = (a_0 \stackrel{x_0}{\leftarrow} a \stackrel{x_1}{\rightarrow} a_1)$,
is a pair of vertical arrows of {\bf C}-spans
$R = (y \stackrel{G}{\Leftarrow} z \stackrel{F}{\Rightarrow} x)$,
where
$G$ is a pair of {\bf C}-arrows
 $c_0 \stackrel{g_0}{\rightarrow} b_0$ and  $c_1 \stackrel{g_1}{\rightarrow} b_1$
and
$F$ is a pair of {\bf C}-arrows
 $c_0 \stackrel{f_0}{\rightarrow} a_0$ and  $c_1 \stackrel{f_1}{\rightarrow} a_1$ 
such that there is a pair of {\bf C}-arrows
 $c \stackrel{g}{\rightarrow} b$ and  $c \stackrel{f}{\rightarrow} a$ 
which together form a commuting composite square of {\bf C}-arrows
\begin{center}
   $\begin{array}{ccccc}
       b_0
       & 
       \makebox{\rule{.1in}{0in}}
       \makebox[0in]{\raisebox{.12in}[0in][0in]{\scriptsize$y_0$}} \makebox[0in]{$\longleftarrow$}
       \makebox{\rule{.1in}{0in}}
       &
       b
       & 
       \makebox{\rule{.1in}{0in}}
       \makebox[0in]{\raisebox{.12in}[0in][0in]{\scriptsize$y_1$}} \makebox[0in]{$\longrightarrow$}
       \makebox{\rule{.1in}{0in}}
       &
       b_1
       \\
       \makebox[0in][r]{\scriptsize$g_0$} \uparrow
       &&
       \makebox[0in][r]{\scriptsize$g$} \uparrow
       &&
       \uparrow \makebox[0in][l]{\scriptsize$g_1$}
       \\
       c_0
       & 
       \makebox{\rule{.1in}{0in}}
       \makebox[0in]{\raisebox{.12in}[0in][0in]{\scriptsize$z_0$}} \makebox[0in]{$\longleftarrow$}
       \makebox{\rule{.1in}{0in}}
       & 
       c
       & 
       \makebox{\rule{.1in}{0in}}
       \makebox[0in]{\raisebox{.12in}[0in][0in]{\scriptsize$z_1$}} \makebox[0in]{$\longrightarrow$}
       \makebox{\rule{.1in}{0in}}
       &
       c_1
       \\
       \makebox[0in][r]{\scriptsize$f_0$} \downarrow
       &&
       \makebox[0in][r]{\scriptsize$f$} \downarrow
       &&
       \downarrow \makebox[0in][l]{\scriptsize$f_1$}
       \\
       a_0
       &
       \makebox[0in]{\raisebox{-.12in}[0in][0in]{\scriptsize$x_0$}} \makebox[0in]{$\longleftarrow$}
       &
       a
       &
       \makebox[0in]{\raisebox{-.12in}[0in][0in]{\scriptsize$x_1$}} \makebox[0in]{$\longrightarrow$}
       &
       a_1
    \end{array}$
\end{center}
The combination of both the horizontal and vertical aspects of 
$\widehat{\stackrel{\scriptscriptstyle \rightharpoondown}{{\bf P}}}$
forms a double biposet, with 
zero-cells (objects) being {\bf P}-types $x$,
one-cells  (arrows)  being {\bf P}-terms $\term{x_0}{x}{x_1}$, and
two-cells  (squares) being vertical spans of {\bf P}-terms $\twospan{y}{R}{x}$.

If $\term{z}{s}{y}$ and $\term{y}{r}{x}$ are any two composable {\bf P}-terms,
then from this definition of span tensor product we get
$\spano{z}{y}(s) \tprod \spano{y}{x}(r) \preceq \spano{z}{x}(s \tprod r)$,
which states that $\spano{}{}$ is lax functorial.
The underlying {\bf P}-term of the tensor product span is
$\uterm{z}{x}(\sigma \tprod \rho)
 = \uterm{z}{x}(\widehat{{\bf P}^\dashv}^{[s_0,r_1]}(\spano{s}{r}(s_1 \tprod r_0^{\rm op})))
 = {\bf P}^{[s_0,r_1]}(\uterm{s}{r}(\spano{s}{r}(s_1 \tprod r_0^{\rm op})))
 = {\bf P}^{[s_0,r_1]}(s_1 \tprod r_0^{\rm op})
 = s_0^{\rm op} \tprod s_1 \tprod r_0^{\rm op} \tprod r_1
 = \uterm{z}{y}(\sigma) \tprod \uterm{y}{x}(\rho)$.
So,
the underlying {\bf P}-term operator
\begin{center}
   $\widehat{{\bf P}^\dashv} \stackrel{\uterm{{\rm P}}{}}{\longrightarrow} {\bf P}$
\end{center}
on a spannable biposet {\bf P}
is (horizontally) functorial from functional spans to arbitrary terms.
It is a morphism of biposets,
since it preserves term joins.
It is a ``quotient'' morphism of biposets,
since it is a full functor and bijective on objects.

\begin{Theorem}
   The spanning construction $\widehat{(\,)}$
   is left adjoint to
   the functionality construction $(\,)^\dashv$
   \begin{center}
      \shortstack{Exact\\Categories} $\stackrel{\widehat{(\,)} \dashv (\,)^\dashv}{\longrightarrow}$ \shortstack{Spannable\\Biposets}
   \end{center}
   forming a coreflection,
   with $\widehat{(\,)}$ embedding exact categories into spannable biposets,
   $(\,)^\dashv$ coreflecting spannable biposets onto exact categories,
   unit components given by the Yoneda isomorphisms 
   ${\bf C} \stackrel{\yoneda{{\rm C}}{}{}}{\cong} {\widehat{\bf C}}^\dashv$
   and counit components given by the underlying term biposet morphisms
   $\widehat{{\bf P}^\dashv} \stackrel{\uterm{{\rm P}}{}}{\longrightarrow} {\bf P}$.
\end{Theorem}


\subsection{Sums}

\paragraph{Join Bisemilattices.}
The structural aspect of the semantics of dialectical logic is best defined in terms of bisemilattices.
A {\em join bisemilattice\/} or {\em semiexact biposet\/} is a biposet whose homsets are finitely complete (join-)semilattices
and whose composition is finitely (join-) continuous.
Horizontally the term ``semilattice-valued category'' might be indicated,
but vertically from a bicategorical viewpoint the term ``bisemilattice'' seems appropriate.

In more detail,
a join bisemilattice
${\bf P} = \triple{\quadruple{{\bf P}}{\preceq}{\tprod}{{\rm Id}}}{\bsum}{0}$
consists of the data and axioms of a biposet
${\bf P} = \quadruple{{\bf P}}{\preceq}{\tprod}{{\rm Id}}$
plus the following.
For any two parallel terms $\term{y}{s,r}{x}$
there is a join term $\term{y}{s \bsum r}{x}$,
where $\bsum$ is a binary operation called {\em (standard) boolean sum\/},
satisfying the usual adjointness condition
$s \bsum r \preceq_{y,x} t$ iff $s \preceq_{y,x} t$ and $r \preceq_{y,x} t$. 
Standard sum represents (a special case of) the ``parallel combination'' of the two processes $s$ and $r$.
Another, perhaps more interesting, parallel combination is derived from the notion of type sum (defined below).
For any pair of types $y$ and $x$ there is an {\em empty\/} (or {\em bottom\/}) term $\term{y}{0_{y,x}}{x}$
satisfying $0_{y,x} \preceq r$. 
The tensor product is finitely join-continuous
(distributive w.r.t. finite joins) on the right and the left,
$s \tprod (r_1 \bsum \cdots \bsum r_n) = (s \tprod r_1) \bsum \cdots \bsum (s \tprod r_n)$
and $(s_1 \bsum \cdots \bsum s_m) \tprod r = (s_1 \tprod r) \bsum \cdots \bsum (s_m \tprod r)$.
The join $v \bsum u$ of any two comonoids $v$ and $u$ of type $x$ is also a comonoid of type $x$.
{\bf [Standardization property:]}
In a join bisemilattice {\bf P},
the poset of comonoids $\comonoid{}{(x)}{}$ is a lattice
with meet being the tensor product $\tprod$ and join being the boolean sum $\bsum$.
Furthermore,
the meet distributes over the join,
so that $\comonoid{}{(x)}{}$ is a distributive lattice.
This standardization property means that
the local contexts (monoidal semilattices) of comonoids $\{ \comonoid{}{(x)}{} \mid x \mbox{ a type} \}$ are standard contexts (distributive lattices),
and shows why propositions (interpreted as comonoids) and programs (interpreted as terms) are subsumed by a single concept.

A join bisemilattice with one object (universal type) is called a {\em monoidal join semilattice\/}.
A {\em complete Heyting category\/},
abbreviated {\em cHc\/}, 
is the same as a complete join bisemilattice;
that is,
a join bisemilattice {\bf H} whose homsets are complete join semilattices
(arbitrary joins exist) and whose tensor product is join continuous
(completely distributive w.r.t. joins).
Since the homset ${\bf H}[x,z]$ is a complete lattice,
and left tensor product $r \tprod$ is continuous, 
it has (and determines) a right adjoint
$r \tprod s \preceq_{y,z} t \mbox{ iff } s \preceq_{x,z} r \tensorimplysource t$
called {\em left tensor implication\/}.
Similarly,
the right tensor product $\tprod r$ has (and determines) a right adjoint
$t \tprod r \preceq_{z,x} s \mbox{ iff } t \preceq_{z,y} s \tensorimplytarget r$
called {\em right tensor implication\/}.
If {\bf P} is a join bisemilattice,
then the opposite biposet ${\bf P}^{\rm op}$ is also a join bisemilattice.
A {\em morphism of join bisemilattices\/} ${\bf P} \stackrel{H}{\rightarrow} {\bf Q}$
is a functor which preserves homset order and finite homset joins.
An {\em involutive join bisemilattice\/} ${\bf P} = \pair{{\bf P}}{\involution{(\;)}}$
is an involutive biposet {\bf P}
where involution
${\bf P}^{\rm op} \stackrel{\involution{(\,)}}{\longrightarrow} {\bf P}$
is a morphism of join bisemilattices;
that is,
preserves term joins,
so that
$\involution{(r \bsum s)} = \involution{r} \bsum \involution{s}$ for parallel terms $\term{y}{r,s}{x}$, and 
$\involution{0_{y,x}} = 0_{x,y}$.

Any distributive lattice is a monoidal join semilattice,
where tensor product coincides with lattice meet
$s \tprod r \define s \wedge r$.
Any commutative monoid $\triple{M}{\circ}{e}$ is a monoidal join semilattice with the modified ``part (prefix) order''
$m \preceq n$ when there is a $p \memberof M$ and a positive natural number $i \memberof {\bf N}$
such that $m \circ p = n^i$.
The category {\bf JSL} of (finitely-complete) join semilattices
and join-continuous monotonic functions is a large join bisemilattice,
with tensor product $\tprod$ being function composition ``$\cdot$''
and boolean sum $\bsum$ being the join ``$\vee$'' of join-continuous monotonic functions.
{\bf Rel} is a cHc.
A one-object complete Heyting category is called a {\em complete Heyting monoid\/}.
Given an alphabet $A$,
formal $A$-languages $\powerof{A^\ast}$ is a complete Heyting monoid.
More generally,
every biposet {\bf P} has an associated {\em closure subset category\/} $\powerof{{\bf P}}$ which is a cHc:
objects are {\bf P}-types,
arrows are subsets of {\bf P}-terms $\term{y}{R}{x}$ when $R \subseteq {\bf P}[y,x]$,
and homset order is the closed-below order $S \preceq R$ when $S \subseteq \below{R}$.
Since every category {\bf C} is a biposet with the identity order on homsets,
the subset construction $\powerof{{\bf C}}$ is a special case of the closure subset construction.

In subset categories $\powerof{{\bf C}}$ a comonoid of type $x$ 
is either the empty endoterm $\term{x}{\emptyset}{x}$ or the identity singleton $\term{x}{\{x\}}{x}$,
and these can be interpreted as the truth-values {\bf false} and {\bf true},
so that $\comonoid{}{(x)}{}$ is the complete Heyting algebra $\comonoid{}{(x)}{} \cong {\bf 2}$.
In a cHc,
at each type $x$ the distributive lattice of comonoids $\comonoid{}{(x)}{}$
is actually a complete cartesian Heyting monoid;
that is,
a complete Heyting algebra.
Since interiors exist,
for any two comonoids $u,v \memberof \comonoid{}{(x)}{}$
we can define a local {\em standard implication\/} by
$u {\Rightarrow} v \define \interior{(u \tensorimplysource v)}
                         = \interior{(v \tensorimplytarget u)}$.
Standard implication satisfies the fundamental adjointness:
$u \tprod w \preceq v$ iff $w \preceq u {\Rightarrow} v$
In closure subset categories $\powerof{{\bf P}}$
a comonoid $\term{x}{W}{x}$ of type $x$ is a closed-below subset $W \subseteq {\bf P}[x,x]$
of {\bf P}-endoterms $\term{x}{w}{x}$,
which are subparts of the identity $w \preceq x$
and which factor (possibly trivially) $w \preceq v \tprod u$ into two other endoterms $v,u \memberof W$.
Since $\powerof{{\bf P}}$ is a cHc,
the lattice of $\powerof{\bf P}$-comonoids $\comonoid{\powerof{{\rm P}}}{x}{}$ is a complete Heyting algebra.
Any {\bf P}-comonoid $\term{x}{w}{x}$ is embeddable as
the $\powerof{{\bf P}}$-comonoid $\term{x}{\downarrow{w}}{x}$.
So we can regard $\powerof{{\bf P}}$-comonoids as generalized {\bf P}-comonoids
called {\em closure subset {\bf P}-comonoids\/}.
Comonoids in join bisemilattices in general,
but even more strongly in cHc's,
play the role of ``localized truth values''.

\paragraph{Type Sums.}
The closure subset construction $\powerof{{\bf P}}$ does not capture the notion of ``relational structures'' completely.
Although it introduces nondeterminism on the arrows,
it leaves the objects alone.
Type sums introduce distributivity on objects in a constructive fashion.

Assume that we are in a join bisemilattice {\bf P}.
The {\em empty type\/} $0$ is a special type
such that for any type $x$ there are unique terms
between $x$ and $0$ in either direction.
So $0$ is an initial type,
satisfying the condition $\term{0}{r}{x}$ implies $r = 0_{0,x}$;
and $0$ is a terminal type,
satisfying the condition $\term{x}{r}{0}$ implies $r = 0_{x,0}$.
A type that is both initial and terminal is a null type.
The null type $0$ is the ``empty type sum'',
the sum of the empty collection of types.
For any pair of types $y$ and $x$,
the bottom term
$y \stackrel{0_{y,x}}{\rightharpoondown} x$ 
is the composition
$0_{y,x} = 0_{y,0} \tprod 0_{0,x}$.
The empty type
$\term{0}{0_{0,x} \dashv 0_{x,0}}{x}$ is the smallest subtype of any type $x$,
a universal bottom subtype,
and its associated comonoid is the smallest comonoid $0_x \in \comonoid{}{(x)}{}$.

Suppose that {\bf P} is a spannable biposet.
When null types exist
the overlying span operator maps empty {\bf P}-terms to empty functional spans:
$\spano{y}{x}(0_{y,x})
 = (y \stackrel{0_{0,y} \dashv 0_{y,0}}{\leftharpoondown} 0 \stackrel{0_{0,x} \dashv 0_{x,0}}{\rightharpoondown} x)$. 
Two subtypes $\term{z}{i \dashv p}{y}$ and $\term{z'}{i' \dashv p'}{y}$ are {\em disjoint\/}
when their intersection in ${\bf P}^\dashv$ (which must also be a subtype)
is the empty type
$\spano{z}{z'}(i \tprod p')
 = (z \stackrel{0_{0,z} \dashv 0_{z,0}}{\leftharpoondown} 0 \stackrel{0_{0,z'} \dashv 0_{z',0}}{\rightharpoondown} z')$.
This immediately implies the conditions
$i \tprod p' = 0_{z,z'}$ and $i' \tprod p = 0_{z',z}$,
which are equivalent to the fact that the subtype comonoids $p \tprod i$ and $p' \tprod i'$ are disjoint.
In fact,
these conditions are equivalent to disjointness,
since the overlying span operator maps empty {\bf P}-terms to empty functional spans.
In an arbitrary biposet {\bf P}, 
where the intersection of subtypes may not necessarily exist,
we use these conditions to define disjointness:
two subtypes $\term{z}{i \dashv p}{y}$ and $\term{z'}{i' \dashv p'}{y}$ are {\em disjoint\/}
when $i \tprod p' = 0_{z,z'}$ and $i' \tprod p = 0_{z',z}$;
or equivalently,
when the subtype comonoids are disjoint $(p \tprod i) \tprod (p' \tprod i') = 0_y$.

Given two types $y$ and $x$ in a join bisemilattice {\bf P},
the {\em sum} of $y$ and $x$ is a composite type $y \typesum x$
having $y$ and $x$ as disjoint subtypes
$y \stackrel{i_y \dashv p_y}{\rightharpoondown} {y \typesum x} \stackrel{i_x \dashv p_x}{\leftharpoondown} x$ 
which cover $y \typesum x$.
So $y \typesum x$ comes equipped
with two {\em injection terms\/}
$y \stackrel{i_y}{\rightharpoondown} y \typesum x \stackrel{i_x}{\leftharpoondown} x$
and two {\em projection terms\/}
$y \stackrel{p_y}{\leftharpoondown} y \typesum x \stackrel{p_x}{\rightharpoondown} x$
which satisfy the ``comonoid covering equation''
$(p_y \tprod i_y) \bsum (p_x \tprod i_x) =  y \typesum x$
stating that the join of the sum-component subtype comonoids covers the sum type,
and satisfy the ``subtype disjointness equations''
$i_y \tprod p_y = y$,
$i_y \tprod p_x = 0_{y,x}$,
$i_x \tprod p_y = 0_{x,y}$, and
$i_x \tprod p_x = x$,
or the ``comonoid disjointness equation''
$(p_y \tprod i_y) \tprod (p_x \tprod i_x) =  0_{y \scriptsizetypesum x}$.

These conditions defining sum are equivalent to the assertion that
the type $y \typesum x$ is both a coproduct via the injections
and a product via the projections of the types $y$ and $x$.
An object which is both a product and a coproduct of two other objects
is called a {\em biproduct\/}.
So type sums are equivalent to biproducts.
Given any pair of terms
$y \stackrel{t}{\rightharpoondown} z \stackrel{s}{\leftharpoondown} x$
there is a unique term
$y \typesum x \stackrel{\relcopair{t}{s}{}{}}{\rightharpoondown} z$,
called the type sum {\em source pairing\/} of $t$ and $s$,
which satisfies the source pairing conditions
$i_y \tprod \relcopair{t}{s}{}{} = t$ and
$i_x \tprod \relcopair{t}{s}{}{} = s$.
Just define
$\relcopair{t}{s}{}{} \define (p_y \tprod t) \bsum (p_x \tprod s)$.
In particular,
$\relcopair{i_y}{i_x}{}{} = (p_y \tprod i_y) \bsum (p_x \tprod i_x) = y \typesum x$.
These properties say that the sum $y \typesum x$ is a coproduct.
When {\bf P} has type sums,
coproduct copairing operators $\relcopair{\;}{\;}{}{}$
are monotonic in both components and preserve joins:
if $r_1 \preceq r_2$ and $s_1 \preceq s_2$ then $[r_1,s_1] \preceq [r_2,s_2]$;
and $[s_1,r_1] \bsum [s_2,r_2] = [s_1 \bsum s_2,r_1 \bsum r_2]$.
For any term $\term{z}{q}{w}$ it is immediate that
$\relcopair{t}{s}{}{} \tprod q
 = \relcopair{t \tprod q}{s \tprod q}{}{}$.
Dually, given any pair of terms
$y \stackrel{t}{\leftharpoondown} z \stackrel{s}{\rightharpoondown} x$
there is a unique term
$z \stackrel{\relpair{t}{s}{}{}}{\rightharpoondown} y \typesum x$,
called the type sum {\em target pairing\/} of $t$ and $s$,
which satisfies the target pairing conditions
$\relpair{t}{s}{}{} \tprod p_y = t$ and
$\relpair{t}{s}{}{} \tprod p_x = s$.
Just define
$\relpair{t}{s}{}{} \define (t \tprod i_y) \bsum (s \tprod i_x)$.
In particular,
$\relpair{p_y}{p_x}{}{} = (p_y \tprod i_y) \bsum (p_x \tprod i_x) = y \typesum x$.
These properties say that the sum $y \typesum x$
is a product.
Target pairing operators $\relpair{\;}{\;}{}{}$ are monotonic on left and right.
For any term $\term{w}{q}{z}$ it is immediate that
$q \tprod \relpair{t}{s}{}{}
 = \relpair{q \tprod t}{q \tprod s}{}{}$.

A join bisemilattice {\bf P} is said to be {\em summable\/} when it has type sums;
that is,
when all finite type sums exist.
The sum of an two arbitrary {\bf P}-terms $\term{y_2}{s}{y_1}$ and $\term{x_2}{r}{x_1}$
is the term
${y_2 \typesum x_2} \stackrel{{s \scriptsizetypesum r}}{\rightharpoondown} {y_1 \typesum x_1}$
defined by
${s \typesum r}
 = \relcopair{s \tprod i_{y_1}}{r \tprod i_{x_1}}{}{}
 = \relpair{p_{y_2} \tprod s}{p_{x_2} \tprod r}{}{}
 = (p_{y_2} \tprod s \tprod i_{y_1}) \bsum (p_{x_2} \tprod r \tprod i_{x_1})$.
The sum term ${s \typesum r}$ is a kind of ``superposition'' of the terms $s$ and $r$.
We can view a summable bisemilattice {\bf P} as a generalized distributed Petri net \cite{MM}.
In the process interpretation of terms
the sum term-process ${s \typesum r}$ represents the complete parallelism of the term-processes $s$ and $r$.
Three special cases of sum terms are of interest. 
Given any term $r$:
\begin{enumerate}
   \item summing with the zero term
      ${0 \typesum r} = p_{x_2} \tprod r \tprod i_{x_1}$
      embeds a term $\term{x_2}{r}{x_1}$ in a larger type context
      (loosely ${y_2 \typesum x_2} \stackrel{r}{\rightharpoondown} {y_1 \typesum x_1}$);
   \item summing with the identity (nop) term
      ${z \typesum r} = (p_z \tprod i_z) \bsum (p_y \tprod r \tprod i_x)$
      allows us to represent the artificial intelligence notion of ``frame'',
      where the context $z$ outside of the locus of activity of a term-process $\term{y}{r}{x}$
      is maintained as ``status quo''
      (loosely ${z \typesum y} \stackrel{r}{\rightharpoondown} {z \typesum x}$); and
   \item summing with itself
      $\copower{2}{r} \define {r \typesum r} = (p_y^1 \tprod r \tprod i_x^1) \bsum (p_y^0 \tprod r \tprod i_x^0)$
      allows us to define copowers $\term{\copower{2}{y}}{\copower{2}{r}}{\copower{2}{x}}$ of a term $\term{y}{r}{x}$
      where the copower types
      $y \stackrel{i_y^1 \dashv p_y^1}{\rightharpoondown} \copower{2}{y} \stackrel{i_y^0 \dashv p_y^0}{\leftharpoondown} y$ 
      and
      $x \stackrel{i_x^1 \dashv p_x^1}{\rightharpoondown} \copower{2}{x} \stackrel{i_x^0 \dashv p_x^0}{\leftharpoondown} x$
      are types summed with themselves.
\end{enumerate}
The sum operator $\typesum$ is a functor
$\product{{\bf P}}{{\bf P}} \stackrel{\scriptsizetypesum}{\longrightarrow} {\bf P}$.
For any two types $y$ and $x$,
the sum $y \typesum x$ of $y$ and $x$ 
and the sum $x \typesum y$ of $x$ and $y$ 
are isomorphic,
satisfying the ``commutative law'' $y \typesum x \cong x \typesum y$,
with the isomorphism mediated by the mutually inverse term isomorphisms
$x \typesum y \stackrel{\relcopair{i_x}{i_y}{}{}}{\rightharpoondown} y \typesum x$ and
$y \typesum x \stackrel{\relpair{p_x}{p_y}{}{}}{\rightharpoondown} x \typesum y$.
For any three types $z$, $y$ and $x$,
the sum operation satisfies the ``associative law''
$(z \typesum y) \typesum x \cong z \typesum (y \typesum x)$.
For any type $x$,
the sum operation satisfies the ``unit laws''
$ 0 \typesum x \cong x \cong x \typesum 0 $.
So the sum functor $\typesum$ is a symmetric monoidal functor,
and a summable bisemilattice ${\bf P} = \pair{{\bf P}}{\typesum}$ is a symmetric monoidal category.

\begin{Proposition}
   If {\bf P} is a join bisemilattice,
   then {\bf P} has type sums iff ${\bf P}^\dashv$ has coproducts
   which are preserved by the inclusion functor
   ${\bf P}^\dashv \stackrel{{\rm Inc}}{\longrightarrow} {\bf P}$;
   so that $y \typesum_{\rm P} x = y +_{{\rm P}^\dashv} x$.
\end{Proposition}
\begin{proof}
   On the one hand,
   suppose that
   $y \stackrel{i_y \dashv p_y}{\rightharpoondown} y \typesum x \stackrel{i_x \dashv p_x}{\leftharpoondown} x$
   is the coproduct of types $y$ and $x$ in the category of functional terms ${\bf P}^\dashv$.
   Also,
   assume that
   $y \stackrel{i_y}{\rightharpoondown} y \typesum x \stackrel{i_x}{\leftharpoondown} x$
   is the coproduct of $y$ and $x$ in {\bf P}.
   Then
   $y \stackrel{i_y \dashv p_y}{\rightharpoondown} y \typesum x \stackrel{i_x \dashv p_x}{\leftharpoondown} x$
   is a type sum in {\bf P} with product projections
   $p_y = [y,0_{xy}]$ and $p_x = [0_{yx},x]$.
   On the other hand,
   suppose that
   $y \stackrel{i_y \dashv p_y}{\rightharpoondown} y \typesum x \stackrel{i_x \dashv p_x}{\leftharpoondown} x$
   is a type sum in {\bf P}.
   Then it is a coproduct in ${\bf P}^\dashv$ with coproduct copairings defined by
   $[g \dashv g^{\rm op}, f \dashv f^{\rm op}] \define [g,f] \dashv \langle g^{\rm op},f^{\rm op} \rangle$
   for each pair of functional terms
   $\term{y}{g \dashv g^{\rm op}}{z}$ and $\term{x}{f \dashv f^{\rm op}}{z}$.
\end{proof}

\begin{Proposition}
   When the join bisemilattice {\bf P} is a spannable biposet,
   coproducts in ${\bf P}^\dashv$ are pullbacks: given two functional terms
   $\term{y_2}{{g \dashv g^{\rm op}}}{y_1}$ and $\term{x_2}{{f \dashv f^{\rm op}}}{x_1}$
   the coproduct injections
   $y_2 \stackrel{{i_{y_2} \dashv p_{y_2}}}{\rightharpoondown} {y_2 \typesum x_2} \stackrel{{i_{x_2} \dashv p_{x_2}}}{\leftharpoondown} x_2$
   are the pullbacks of the coproduct injections
   $y_1 \stackrel{{i_{y_1} \dashv p_{y_1}}}{\rightharpoondown} {y_1 \typesum x_1} \stackrel{{i_{x_1} \dashv p_{x_1}}}{\leftharpoondown} x_1$
   along the coproduct functional term
   ${y_2 \typesum x_2} \stackrel{{(g \scriptsizetypesum f) \dashv (g^{\rm op} \scriptsizetypesum f^{\rm op})}}{\rightharpoondown} {y_1 \typesum x_1}$.
\end{Proposition}
\begin{proof}
   First of all,
   $i_{x_1} \tprod (g^{\rm op} \typesum f^{\rm op}) = g^{\rm op} \tprod i_{x_2}$.
   Secondly,
   if $l \tprod (g \typesum f) = k \tprod i_{x_1}$ for two functional terms
   $\term{z}{{l \dashv l^{\rm op}}}{{y_2 \typesum x_2}}$ and $\term{z}{{k \dashv k^{\rm op}}}{x_1}$,
   then $(l \tprod p_{x_2}) \dashv (i_{x_2} \tprod l^{\rm op})$ is the unique functional term
   $\term{z}{{h \dashv h^{\rm op}}}{x_2}$ satisfying $h \tprod i_{x_2} = l$ and $h \tprod f = k$.
\end{proof}

When the join bisemilattice {\bf P} is a spannable biposet,
pullbacks in ${\bf P}^\dashv$, as in any exact category {\bf C},
preserve monomorphisms.
Now monomorphisms in ${\bf P}^\dashv$ are identical to subtypes.
So in ${\bf P}^\dashv$ pullbacks of subtypes are again subtypes. 
Also, pullbacks of empty subtypes are again empty subtypes.  
Since pullbacks are right adjoint operators,
pullbacks preserve intersections of subtypes.

\begin{Fact}
   When the join bisemilattice {\bf P} is a spannable biposet,
   pullbacks in ${\bf P}^\dashv$ of disjoint subtypes are again disjoint subtypes. 
\end{Fact}

We want pullbacks in ${\bf P}^\dashv$ to preserve type sums.
A {\em spannable bisemilattice\/} {\bf P} is a join bisemilattice
which satisfies the following axioms:
\begin{enumerate}
   \item {\bf P} is a spannable biposet;
   \item {\bf P} is a summable bisemilattice; and
   \item pullbacks in ${\bf P}^\dashv$ of type sums are type sums.
\end{enumerate}
Axiom 3 means that
if $y_2 \stackrel{{i_2 \dashv p_2}}{\rightharpoondown} {y_2 \typesum y_1} \stackrel{{i_1 \dashv p_1}}{\leftharpoondown} y_1$ is any type sum,
$\term{x}{{f \dashv f^{\rm op}}}{{y_2 \typesum y_1}}$ is any functional term into the sum, and
$\hat{y}_2 \stackrel{{i_{\hat{y}_2} \dashv p_{\hat{y}_2}}}{\rightharpoondown} x$ and
$\hat{y}_1 \stackrel{{i_{\hat{y}_1} \dashv p_{\hat{y}_1}}}{\rightharpoondown} x$
are the pullbacks of the sum subtypes
$y_2 \stackrel{{i_2 \dashv p_2}}{\rightharpoondown} {y_2 \typesum y_1}$ and
$y_2 \stackrel{{i_2 \dashv p_2}}{\rightharpoondown} {y_2 \typesum y_1}$ along $f$,
then $\hat{y}_2 \stackrel{{i_{\hat{y}_2} \dashv p_{\hat{y}_2}}}{\rightharpoondown} x \stackrel{{i_{\hat{y}_1} \dashv p_{\hat{y}_1}}}{\leftharpoondown} \hat{y}_1$ 
is a type sum.

\paragraph{Semiexact Categories.}
We add coproducts to an exact category {\bf C} in order to get boolean sums in $\widehat{{\bf C}}$.
A {\em semiexact category\/} {\bf C} is a category possessing canonical finite limits and canonical finite coproducts.
Since ${\bf C} \cong {\widehat{\bf C}}^\dashv$,
coproducts in {\bf C} are (via Yoneda) coproducts in ${\widehat{\bf C}}^\dashv$.
We want these to be coproducts in the larger category $\widehat{\bf C}$.
So we require that any semiexact category satisfy the following axioms.
\begin{enumerate}
   \item Coproducts are partitions:
      \begin{enumerate}
         \item the initial object is a universal bottom subobject;
         \item coproduct injections are monomorphisms; and
         \item coproduct injections are pairwise disjoint.
      \end{enumerate}
   \item Pullbacks create coproducts:
      \begin{enumerate}
         \item coproducts are pullbacks; and
         \item pullbacks of coproducts are coproducts.
      \end{enumerate}
\end{enumerate}
Axiom 1(a) means that the unique {\bf C}-arrow
(empty coproduct cotupling) $0 \stackrel{0_x}{\rightarrow} x$ 
is a {\bf C}-monomorphism.
Axiom 1(b) is equivalent to the axiom that
the pullback of any coproduct injection along itself is the identity.
Axiom 1(c) means that the pullback of distinct coproduct injections is the empty coproduct.
Axiom 1(c) is equivalent to the ``cancellation axiom'':
$y + x \cong x$ implies $y \cong 0$.
Axiom 2(b) is redundant,
since axiom 2(c) implies 2(b).
Axiom 2 means the following three assertions.
[Nullary case:]
    Pullbacks of the zero coproduct (initial object) are the zero coproduct:
    if $z \stackrel{h}{\rightarrow} 0$ is any {\bf C}-arrow into the empty coproduct,
    then $z = 0$ and $h = 0$.
    This is equivalent to the axiom
    that pullbacks of zero morphisms along {\bf C}-arrows are zero morphisms:
    if $z \stackrel{h}{\rightarrow} x$ is any {\bf C}-arrow,
    then $0 \stackrel{0_z}{\rightarrow} z$
    is the pullback of the empty coproduct cotupling
    $0 \stackrel{0_x}{\rightarrow} x$ along $h$.
[Binary case 2(a):]
    Coproducts are pullbacks: given two {\bf C}-arrows 
    $y_2 \stackrel{g}{\rightarrow} y_1$ and $x_2 \stackrel{f}{\rightarrow} x_1$ 
    the coproduct injections
    $y_2 \stackrel{{\rm in}_{y_2}}{\rightarrow} {y_2 + x_2} \stackrel{{\rm in}_{x_2}}{\leftarrow} x_2$
    are the pullbacks of the coproduct injections
    $y_1 \stackrel{{\rm in}_{y_1}}{\rightarrow} {y_1 + x_1} \stackrel{{\rm in}_{x_1}}{\leftarrow} x_1$
    along the coproduct {\bf C}-arrow
    ${y_2 + x_2} \stackrel{{g + f}}{\longrightarrow} {y_1 + x_1}$.
[Binary case 2(b):]
    Pullbacks of coproducts are coproducts:
    if $z \stackrel{h}{\rightarrow} y + x$ is any {\bf C}-arrow into a binary coproduct,
    $y \stackrel{h_y}{\leftarrow} \hat{y} \stackrel{{\rm in}_{\hat{y}}}{\rightarrow} z$
    is the pullback of the coproduct injection
    $y \stackrel{{\rm in}_y}{\rightarrow} y + x \stackrel{h}{\leftarrow} z$
    along $h$,
    $z \stackrel{{\rm in}_{\hat{x}}}{\leftarrow} \hat{x} \stackrel{h_x}{\rightarrow} x$
    is the pullback of the coproduct injection
    $z \stackrel{h}{\rightarrow} y + x \stackrel{{\rm in}_x}{\leftarrow} x$
    along $h$,
    then
    $z$ is a coproduct
    $\hat{y} \stackrel{{\rm in}_{\hat{y}}}{\rightarrow} z \stackrel{{\rm in}_{\hat{x}}}{\leftarrow} \hat{x}$
    called the {\em pulledback coproduct\/}
    of $\hat{y}$ and $\hat{x}$
    and $h$ is the coproduct ${h_y + h_x}$ of $h_y$ and $h_x$;
    in particular,
    when $h = {\rm in}_y$ the pulledback coproduct is
    $y \stackrel{y}{\rightarrow} y \stackrel{0_{0,y}}{\leftarrow} 0$.

\begin{Fact}
   Axiom 2(b) is equivalent to the axiom
   \begin{center}
      {\em 2(${\rm b}'$) pullbacks of coproduct cotuplings are coproduct cotuplings.}
   \end{center}
\end{Fact}
\begin{proof}
   Axiom 2(${\rm b}'$) means the following.
   Suppose $y \stackrel{g}{\rightarrow} z \stackrel{f}{\leftarrow} x$ is any {\bf C}-opspan
   and $w \stackrel{h}{\rightarrow} z$ is any {\bf C}-arrow into the common target type $z$.
   We can take the pullback of both $g$ and $f$ along $h$:
   suppose that span
   $y \stackrel{\tilde{h}_g}{\leftarrow} \tilde{y} \stackrel{\tilde{g}}{\rightarrow} w$
   is the pullback of the opspan
   $y \stackrel{g}{\rightarrow} z \stackrel{h}{\leftarrow} w$,
   and that span
   $x \stackrel{\tilde{h}_f}{\leftarrow} \tilde{x} \stackrel{\tilde{f}}{\rightarrow} w$
   is the pullback of the opspan
   $x \stackrel{f}{\rightarrow} z \stackrel{h}{\leftarrow} w$.
   Form the coproduct $y + x$ and consider the coproduct copairing
   ${y + x} \stackrel{[g,f]}{\rightarrow} z$.
   Take the pullback of $[g,f]$ along $h$:
   suppose that span
   ${y + x} \stackrel{\tilde{h}}{\leftarrow} \widetilde{y + x} \stackrel{\widetilde{[g,f]}}{\rightarrow} w$
   is the pullback of the opspan
   ${y + x} \stackrel{[g,f]}{\rightarrow} z \stackrel{h}{\leftarrow} w$.
   Then $\widetilde{y + x}$ is the coproduct
   $\widetilde{y + x} = \tilde{y} + \tilde{x}$
   with coproduct injections
   ${\rm in}_{\tilde{y}} = \langle \tilde{g},\tilde{h}_g \cdot {\rm in}_y \rangle$
   and
   ${\rm in}_{\tilde{x}} = \langle \tilde{f},\tilde{h}_f \cdot {\rm in}_x \rangle$,
   and $\widetilde{[g,f]}$ is the coproduct copairing
   $\widetilde{[g,f]} = [\tilde{g},\tilde{f}]$.

   The proof is straightforward.
   First, form the pullback
   ${y + x} \stackrel{\tilde{h}}{\leftarrow} \widetilde{y + x} \stackrel{\widetilde{[g,f]}}{\rightarrow} w$
   of the opspan
   ${y + x} \stackrel{[g,f]}{\rightarrow} z \stackrel{h}{\leftarrow} w$.
   Second, form the pullback
   $y \stackrel{\tilde{h}_g}{\leftarrow} \tilde{y} \stackrel{\widetilde{{\rm in}_y}}{\rightarrow} \widetilde{y + x}$
   of the opspan
   $y \stackrel{{\rm in}_y}{\rightarrow} {y + x} \stackrel{\tilde{h}}{\leftarrow} \widetilde{y + x}$.
   Then,
   by the pasting of pullbacks
   $y \stackrel{\tilde{h}_g}{\leftarrow} \tilde{y} \stackrel{\tilde{g}}{\rightarrow} w$
   is the pullback of the opspan
   $y \stackrel{g}{\rightarrow} z \stackrel{h}{\leftarrow} w$,
   where $\tilde{g}$ is defined to be
   $\tilde{g} \define \tilde{{\rm in}_y} \cdot \widetilde{[g,f]}$.
   Also,
   by axiom 2(b)
   $\widetilde{y + x}$ is the coproduct
   $\widetilde{y + x} = \tilde{y} + \tilde{x}$
   with coproduct injections
   ${\rm in}_{\tilde{y}} = \widetilde{{\rm in}_y}$
   and
   ${\rm in}_{\tilde{x}} = \widetilde{{\rm in}_x}$,
   and $\widetilde{[g,f]}$ is the coproduct copairing
   $\widetilde{[g,f]} = [\tilde{g},\tilde{f}]$.
\end{proof}

\begin{Proposition}
   If {\bf P} is a spannable bisemilattice,
   then ${\bf P}^\dashv$ is a semiexact category.
\end{Proposition}

The limits in {\bf C} are needed for defining the tensor product $\tprod$
in the horizontal aspect of $\widehat{{\bf C}}$,
and the coproducts are needed for defining the boolean sum $\bsum$ in the vertical aspect.
Since any semiexact category is an exact category,
the horizontal aspect of $\widehat{\bf C}$ is already defined.
The vertical aspect of $\widehat{\bf C}$ is defined as follows.
For any parallel pair of spans $\term{y}{\sigma,\rho}{x}$
where $\sigma = (y \stackrel{s_0}{\leftarrow} s \stackrel{s_1}{\rightarrow} x)$
and $\rho = (y \stackrel{r_0}{\leftarrow} r \stackrel{r_1}{\rightarrow} x)$,
the boolean sum is the span $\term{y}{\sigma \bsum \rho}{x}$
where $\sigma \bsum \rho = (y \stackrel{[s_0,r_0]}{\longleftarrow} {s + r} \stackrel{[s_1,r_1]}{\longrightarrow} x)$
with $s + r$ the {\bf C}-coproduct of $s$ and $r$
and $[s_0,r_0]$ and $[s_1,r_1]$ the {\bf C}-coproduct cotuplings of the legs of the spans.
The bottom span $\term{y}{0_{yx}}{x}$ between $y$ and $x$ is the initial span 
$0_{yx} = (y \stackrel{0_y}{\leftarrow} 0 \stackrel{0_x}{\rightarrow} x)$
where $0$ is the initial {\bf C}-object
and $0_y$ and $0_x$ are the unique {\bf C}-arrows to $y$ and $x$, respectively.
In addition,
there is a top span $\term{y}{1_{yx}}{x}$,
which is precisely the product span
$1_{yx} = (y \stackrel{{\rm pr}_y}{\leftarrow} \product{y}{x} \stackrel{{\rm pr}_x}{\rightarrow} x)$.

\begin{Proposition}
   If {\bf C} is a semiexact category,
   then the (horizontal) category of spans $\widehat{\bf C}$ is a join bisemilattice.
\end{Proposition}
\begin{proof} 
   The tensor product of spans distributes over the boolean sum of spans by axiom 2(${\rm b}'$).
\end{proof}

\begin{Lemma}
   If {\bf C} is a semiexact category,
   then ${\widehat{\bf C}}^\dashv$ has coproducts since ${\bf C} \cong {\widehat{\bf C}}^\dashv$
   and coproducts in ${\widehat{{\bf C}}}^\dashv$ are preserved by the inclusion functor
   ${\widehat{{\bf C}}}^\dashv \stackrel{{\rm Inc}}{\longrightarrow} \widehat{{\bf C}}$,
   or equivalently by the Yoneda embedding
   ${\bf C} \stackrel{\yoneda{{\rm C}}{}{1}}{\longrightarrow} \widehat{{\bf C}}$.
   So $\widehat{{\bf C}}$ has type sums with $y \typesum_{\widehat{\rm C}} x = y +_{\rm C} x$.
\end{Lemma}

\begin{Proposition}
   If {\bf C} is a semiexact category,
   then $\widehat{{\bf C}}$ is a spannable bisemilattice.
\end{Proposition}

The category of {\bf C}-spans $\widehat{{\bf C}}$ is important in general topos theory,
and also in the logic of CCS-like languages.
The special case {\bf C} = {\bf Set} is our most fundamental ``spanning'' example of dialectical flow.
This is the cHc of spans of ordinary functions ${\bf Span} = \widehat{\bf Set}$.
Since {\bf Set} is semiexact,
{\bf Span} is a spannable bisemilattice.

\begin{Theorem}
   The spanning construction $\widehat{(\,)}$
   is left adjoint to
   the functionality construction $(\,)^\dashv$
   \begin{center}
      \begin{tabular}{rcl}
         \begin{tabular}{c}
            Semiexact \\ Categories
         \end{tabular}
         &
         $\stackrel{\widehat{(\,)} \dashv (\,)^\dashv}{\longrightarrow}$
         &
         \begin{tabular}{c}
            Spannable \\ Bisemilattices
         \end{tabular}
      \end{tabular}
   \end{center}
   forming a coreflection,
   with $\widehat{(\,)}$ embedding semiexact categories into spannable bisemilattices,
   $(\,)^\dashv$ coreflecting spannable bisemilattices onto semiexact categories.
\end{Theorem} 

\begin{Proposition}
   If {\bf C} is a semiexact category,
   then the Yoneda embedding
   ${\bf C} \stackrel{\yoneda{{\rm C}}{}{1}}{\longrightarrow} \widehat{{\bf C}}$
   preserves finite limits.
\end{Proposition}

\begin{Proposition}
   If {\bf P} is a spannable bisemilattice,
   then $\stackrel{\scriptscriptstyle \rightharpoondown}{\bf P}$
   (the vertical category of term arrows)
   is a semiexact category.
\end{Proposition} 
\begin{proof}
\end{proof}
In the vertical category
$\stackrel{\scriptscriptstyle \rightharpoondown}{\bf P}$
the vertical boolean sum $S \bsumvert R$ of two parallel spans $\twospan{y}{S,R}{x}$
is defined (as usual) by coproduct and coproduct-copairing in
$\stackrel{\scriptscriptstyle \rightharpoondown}{\bf P}$.
The combination of both the horizontal and vertical aspects of 
$\widehat{\stackrel{\scriptscriptstyle \rightharpoondown}{{\bf P}}}$
forms a double bisemilattice, with 
zero-cells (objects) being {\bf P}-types $x$,
one-cells  (arrows)  being {\bf P}-terms $\term{x_0}{x}{x_1}$, and
two-cells  (squares) being vertical spans of {\bf P}-terms $\twospan{y}{R}{x}$.

For a topological spannable bisemilattice {\bf P},
the indexed category
$\comonoid{\stackrel{\scriptscriptstyle \rightharpoondown}{\bf P}}{}{}$
is a full internal/external (2-dimensional) model
for Milner's calculus of concurrent processes with interaction (synchronization).
Milner's calculus can be interpreted in spannable bisemilattices as follows.
\begin{center}
   \begin{tabular}{|lc||lc|}
      \hline
      \multicolumn{2}{|c||}{``Milner's Calculus''} & \multicolumn{2}{c|}{spannable bisemilattices} \\
      \hline
      \hline
      sets of actions                 & $A$            & horizontal monoids                  & $m \type a$   \\
      \multicolumn{2}{|c||}{(synchronization labels)}  &                                     &               \\
      \hline
      typed processes                 & $P \type A$    & typed vertical comonoids            & $U \type m$   \\
                                      &                & \multicolumn{2}{|c|}{(objects of $\comonoid{\stackrel{\scriptscriptstyle \rightharpoondown}{\bf P}}{}{}$)} \\
                                      &                & typed vertical subobjects           & $y \stackrel{F}{\Rightarrow} m$ \\
                                      &                & \multicolumn{2}{|c|}{(objects of $\stackrel{\scriptscriptstyle \rightharpoondown}{\bf P} {\Downarrow} m$)} \\
      \hline
      \hline
      action application              & $a {\bf ;} P$  & horizontal tensor product            & $V \tprod U$  \\
      nondeterministic choice         & $P + Q$        & vertical boolean sum                 & $V \bsumvert U$ \\
      synchronization product         & $P [] Q$       & vertical tensor product              & $V \tprodvert U$ \\
      \hline
      \hline
      restriction                     & $P |_{B}$      & inverse image                 & $\comonoid{I}{(U)}{}
                                                                                         \define \interior{(\yoneda{}{(I)}{1} \tprod U \tprod \yoneda{}{(I)}{0})}$ \\
      \multicolumn{2}{|r||}{where $P \type A$ and $B \subseteq A$} &                   & for monoid monomorphism $n \stackrel{I}{\Rightarrow} m$                   \\
                                      &                & pullback                      & ${\stackrel{\scriptscriptstyle \rightharpoondown}{\bf P}}_{I}(y \stackrel{F}{\Rightarrow} m)
                                                                                         \define \hat{y} \stackrel{\hat{F}}{\Rightarrow} n$ \\
      \hline
      morphism                        & $Q[\phi]$      & direct image                  & $\comonoid{}{(U)}{H} \define (\yoneda{}{(H)}{0} \tprod U \tprod \yoneda{}{(H)}{1})$   \\
      \multicolumn{2}{|r||}{where $Q \type B$ and $B \stackrel{\phi}{\rightarrow} A$}  &                              & for monoid morphism $n \stackrel{H}{\Rightarrow} m$    \\
                                      &                & composition                   & ${\stackrel{\scriptscriptstyle \rightharpoondown}{\bf P}}^{H}(y \stackrel{G}{\Rightarrow} n)
                                                                                         \define y \stackrel{G \cdot H}{\Rightarrow} n$ \\
      \hline
      \hline
      recursion                       & $\vec{x} = \vec{P}(\vec{x})$     & monoidal closure              & $\closure{u}$ \\
      \multicolumn{2}{|c||}{with fixpoint solution ${\bf fix} \; \vec{x}\vec{P}(\vec{x})$} & \multicolumn{2}{c|}{(consideration modality)}                            \\
      \hline
   \end{tabular}
\end{center}

\paragraph{Comonoid Negation.}
Two comonoids $u \type x$ and $u' \type x$ are said to be {\em disjoint\/},
denoted by $u \disjoint{x} u'$,
when $u \tprod u' = u' \tprod u = 0_x$ (term disjointness).
Clearly,
a collection of comonoids is pairwise disjoint
iff the joins of any two disjoint subcollections are disjoint.
Define the {\em negation\/} of a comonoid $u \memberof \comonoid{}{(x)}{}$
to be the largest comonoid (if it exists) disjoint from $u$,
$v \disjoint{x} u$ iff $v \preceq_x \boolneg{u}$.
In this sense,
negation is a local boolean ``complement'' of $u$.
Negation is contravariantly monotonic
$v \preceq_x u$ implies $\boolneg{u} \preceq_x \boolneg{v}$,
and hence is a monotonic function
$\comonoid{}{(x)}{} \stackrel{\boolneg{(\;)}}{\rightarrow} {\comonoid{}{(x)}{}}^{\rm op}$.
In more detail,
since disjointness is a symmetrical notion,
$v \preceq_x \boolneg{u}$
 iff $u \disjoint{x} v$
 iff $u \preceq_x \boolneg{v}$,
 negation is a self-adjoint monotonic function
$\boolneg{(\;)} \dashv {\boolneg{(\;)}}^{\rm op}$.
Since negation $\boolneg{(\;)}$ is self-adjoint,
it maps arbitrary joins to meets
$\boolneg{(\bsum_i u_i)} = \tprod_i (\boolneg{u_i})$,
which in the binary case gives the DeMorgan's law:
$\boolneg{(v \bsum u)} = \boolneg{v} \tprod \boolneg{u}$
and in the nullary case gives the law:
$\boolneg{0_x} =  x$.

Double negation $\boolneg{\boolneg{(\;)}}$ is a local closure operator:
``monotonic''  $u \preceq_x v$ implies $\boolneg{\boolneg{u}} \preceq_x \boolneg{\boolneg{v}}$,
``increasing'' $u \preceq_x \boolneg{\boolneg{u}}$, and
``idempotent'' $\boolneg{\boolneg{(\boolneg{\boolneg{u}})}} = \boolneg{\boolneg{u}}$.
A comonoid $u \memberof \comonoid{}{(x)}{}$ is {\em regular\/} when it is double-negation closed
$u = \boolneg{\boolneg{u}}$;
or equivalently,
when $u = \boolneg{v}$ for some comonoid $v \memberof \comonoid{}{(x)}{}$.
Denote the collection of regular comonoids in $\comonoid{}{(x)}{}$ by $\regular{}{x}{}$.
Then $\regular{}{x}{} = \quintuple{\regular{}{x}{}}{\tprod}{x}{\vee}{\bot}$ is a lattice,
which is a meet-subsemilattice of the lattice $\comonoid{}{(x)}{}$
with meets $\bprod_i u_i$ in $\regular{}{x}{}$ identical to meets in $\comonoid{}{(x)}{}$,
and joins in $\regular{}{x}{}$ defined (following Glivenko) as the double negation closure
$\vee_i u_i = \boolneg{\boolneg{(\bsum_i u_i)}}$ of joins in $\comonoid{}{(x)}{}$.
Double negation 
$\comonoid{}{(x)}{} \stackrel{\boolneg{\boolneg{(\;)}}}{\rightarrow} \regular{}{x}{}$
reflects $\boolneg{\boolneg{(\;)}} \dashv {\rm Inc}$ arbitrary  comonoids into regular comonoids.
The smallest and largest regular comonoids at $x$ are
$\bot_x \define \boolneg{\boolneg{0_x}}
              = \boolneg{x}$ and
$x = \boolneg{0_x}
   = \boolneg{\bot_x}$,
respectively.
\begin{Fact}
   $\regular{}{x}{} = \sextuple{\regular{}{x}{}}{\tprod}{x}{\vee}{\bot}{\boolneg{(\;)}}$ is a Boolean algebra,
   for each {\bf P}-type $x$.
\end{Fact} 
It is these regular subtypes which are the appropriate ``predicates''
used in the classical dialectical flow of predicate transformers
and in precondition/postcondition semantics.

A {\em locally classical join bisemilattice\/} ${\bf P} = \pair{{\bf P}}{\boolneg{(\;)}}$
is a join bisemilattice {\bf P} augmented with a {\em local negation operator\/} $\boolneg{(\;)}$
on the lattice of comonoids $\comonoid{}{(x)}{}$ for each type $x$,
satisfying the axiom $v \disjoint{x} u$ iff $v \preceq_x \boolneg{u}$.
Each local negation defines the Boolean algebra of regular comonoids $\regular{}{x}{} \subseteq \comonoid{}{(x)}{}$.
This Boolean algebra is a Heyting algebra with implication $\imply_x$ defined by
$u \imply_x v \define \boolneg{u} \vee v$.
Then the {\em local implication functor\/}
${\regular{}{x}{}}^{\rm coop} \stackrel{\imply_x}{\longrightarrow} {\bf MSL}$,
defined by $\imply_x(x) \define \regular{}{x}{} = \triple{\regular{}{x}{}}{\tprod}{x}$
and $\imply_x(w) \define \comonoid{}{(x)}{} \stackrel{w \imply (\;)}{\longrightarrow} \comonoid{}{(x)}{}$,
is a contravariant join semilattice functor.
An {\em classical inverse flow category\/} $\pair{{\bf P}}{\inversesquare{(\;)}{}}$
consists of a locally classical join bisemilattice ${\bf P} = \pair{{\bf P}}{\boolneg{(\;)}}$,
and a contravariant meet semilattice functor ${\bf P}^{\rm coop} \stackrel{\inversesquare{(\;)}{}}{\longrightarrow} {\bf MSL}$
into the category of meet semilattices,
which is regular on subtypes.
In more detail,
\begin{enumerate}
   \item $\square{x}{}$ is a meet subsemilattice of regular comonoids
         $\square{x}{} \subseteq \regular{}{x}{} = \triple{\regular{}{x}{}}{\tprod}{x}$
         for each type $x$; 
   \item $\directsquare{y}{} \stackrel{\inversesquare{r}{}}{\leftarrow} \square{x}{}$
         is a morphism of meet semilattices for each term $\term{y}{r}{x}$
         called the {\em inverse flow\/} specified by $r$,
         with $\inversesquare{r}{(x)}= y$
         and $\inversesquare{r}{(u \tprod u')} = \inversesquare{r}{(u)} \tprod \inversesquare{r}{(u')}$;
   \item $\inversesquare{(\;)}{}$ is contravariantly functorial, 
         with $\square{x}{} = {\rm Id}_{\square{x}{}}$, and
         $\inversesquare{s \tprod r}{} = \inversesquare{r}{} \cdot \inversesquare{s}{}$;
   \item $\inversesquare{(\;)}{}$ is a meet semilattice functor,
         (i)   if $s \succeq r$ then $\inversesquare{s}{} \leq \inversesquare{r}{}$,
         (ii)  $\inversesquare{0}{} = \top$, and
         (iii) $\inversesquare{r \bsum s}{} = \inversesquare{r}{} \wedge \inversesquare{s}{}$;
   \item $\inversesquare{(\;)}{}$ is standard on subtypes,
         in that $\inversesquare{(\;)}{}$ restricted to $x$-comonoids is the local implication functor
         $\imply_x = {\rm Inc}_x^{\rm coop} \cdot \directsquare{(\;)}{}$
         that is,
         subtype inverse flow $\square{x}{} \stackrel{\inversesquare{u}{}}{\leftarrow} \square{x}{}$
         is just implication $\inversesquare{u}{(u')} = u \imply u'$
         for each comonoid $u \memberof \square{x}{}$.
\end{enumerate}

\section{Flow Structures}
\subsection{Assertional Categories}

\paragraph{Heyting Categories.}
The full semantics of intuitionistic dialectical logic is defined in terms of Heyting categories \cite{Kent88}.
Concisely speaking,
a {\em Heyting category\/} is a closed bilattice;
that is,
an bilattice {\bf H} whose tensor product has right adjoints on both left and right.
The underlying bilattice represents the structural aspect of a Heyting category,
whereas the closedness property represents the aspect of horizontal dialectical flow.

In more detail,
the flow aspect consists of the following data and axioms.
For any two {\bf H}-terms $\term{y}{r}{x}$ and $\term{z}{s}{x}$ with common target type
there is a composite term $\term{z}{s \tensorimplytarget r}{y}$ between their source types,
defined by the dialectical axiom
$t \tprod r \preceq_{z,x} s \mbox{ iff } t \preceq_{z,y} s \tensorimplytarget r$,
stating that the binary operation $\tensorimplytarget$ of {\em right tensor implication\/},
is right adjoint to tensor product on the right.
Right tensor implication $\tensorimplytarget$,
like all exponentiation or division operators including numerical ones,
is covariantly monotonic on the left and contravariantly monotonic on the right.
This dialectical axiom,
generalizing the deduction theorem of standard logic,
defines the formal semantics of tensor implication $\tensorimplytarget$ in terms of tensor product $\tprod$.
From the dialectical axiom easily follows
the inference rule of right modus ponens
$(s \tensorimplytarget r) \tprod r \preceq s$
and the inference rule
$t \preceq (t \tprod r) \tensorimplytarget r$.
Also immediate from the axioms are
the transitive, reflexive, mixed associative and unital laws:
$(t \tensorimplytarget s) \tprod (s \tensorimplytarget r)
 \preceq (t \tensorimplytarget r)$,
$y \preceq (r \tensorimplytarget r)$,
$t \tensorimplytarget (s \tprod r)
 = (t \tensorimplytarget r) \tensorimplytarget s$,
$(r \tensorimplytarget x) = r$.
Dually,
for any two {\bf H}-terms $\term{y}{r}{x}$ and $\term{y}{t}{z}$ with common source type
there is a composite term $\term{x}{r \tensorimplysource t}{z}$ between their target types,
defined by the dialectical axiom
$r \tprod s \preceq_{y,z} t \mbox{ iff } s \preceq_{x,z} r \tensorimplysource t$,
stating that the binary operation $\tensorimplysource$ of {\em left tensor implication\/},
is right adjoint to tensor product on the left.
Together the left and right implications satisfy the mixed associative law
$s \tensorimplysource (t \tensorimplytarget r)
 = (s \tensorimplysource t) \tensorimplytarget r$.
From both the left and right modus ponens,
we get the derived rules
$(r \tensorimplytarget r) \tensorimplysource r
 = r
 = r \tensorimplytarget (r \tensorimplysource r)$.
Since tensor product is left adjoint on both left and right to tensor implication,
it preserves arbitrary joins
$s \tprod (r \bsum r')
 = (s \tprod r) \bsum (s \tprod r')$,
$s \tprod  0_{y,x} 
 = 0_{z,x}$, 
$(s \bsum s') \tprod r
 = (s \tprod r) \bsum (s' \tprod r)$ and
$0_{z,y} \tprod r
 = 0_{z,x}$.
Since tensor implications are right adjoint to tensor product,
they preserve arbitrary meets
$r \tensorimplysource (t \bprod t')
 = (r \tensorimplysource t) \bprod (r \tensorimplysource t')$,
$r \tensorimplysource 1_{y,z} 
 = 1_{x,z}$, 
$(s \bprod s') \tensorimplytarget r
 = (s \tensorimplytarget r) \bprod (s' \tensorimplytarget r)$ and
$1_{z,x} \tensorimplytarget r
 = 1_{z,y}$.
The two dialectical axioms assert that the bilattice {\bf H} is closed.
Heyting categories are ubiquitous.
Any complete Heyting category is a Heyting category;
in particular,
subset categories and distributer categories (defined below) are Heyting categories.

Any Heyting term $\term{y}{r}{x}$ defines
a join-continuous {\em Heyting direct image\/} monotonic function
${\bf H}[y,y] \stackrel{\scriptbf{H}^r}{\longrightarrow} {\bf H}[x,x]$
defined by ${\bf H}^r(q) \define r \tensorimplysource (q \tprod r)$ for endoterms $\term{y}{q}{y}$,
a join-continuous {\em Heyting inverse image\/} monotonic function
${\bf H}[y,y] \stackrel{\scriptbf{H}_r}{\longleftarrow} {\bf H}[x,x]$
defined by ${\bf H}_r(p) \define (r \tprod p) \tensorimplytarget r$ for endoterms $\term{x}{p}{x}$.
It is easy to check that direct and inverse image is an adjoint pair of monotonic functions
${\bf H}(\term{y}{r}{x}) = {\bf H}[y,y] \stackrel{\scriptbf{H}^r \dashv \scriptbf{H}_r}{\longrightarrow} {\bf H}[x,x]$
for each Heyting term $\term{y}{r}{x}$.
Let {\bf adj} be the category of preorders and adjoint pairs of monotonic functions.
The construction ${\bf H}$,
mapping types to their poset of endoterms
${\bf H}(x) = {\bf H}[x,x]$ and
mapping Heyting terms to their adjoint pair of direct/inverse image adjunction,
is a dialectical base ${\bf H}^\dashv \stackrel{\scriptbf{H}}{\longrightarrow} {\bf adj}$,
mapping Heyting terms into the subcategory of {\bf adj}
consisting of monoidal posets and adjoint pairs of monotonic functions.

In a Heyting category {\bf H},
the precondition/postcondition constraint $v \tprod r \preceq r \tprod u$
is equivalent to
the constraint $v \preceq (r \tprod u) \tensorimplytarget r$.
This suggests that a good definition for inverse subtype flow 
in an affirmation Heyting category {\bf H} would be
$\comonoid{r}{(u)}{} \define \interior{((r \tprod u) \tensorimplytarget r)}$
for each {\bf H}-comonoid $u \memberof \comonoid{}{(x)}{}$.
The monotonic function
$\comonoid{}{(y)}{} \stackrel{\comonoid{r}{}{}}{\longleftarrow} \comonoid{}{(x)}{}$ 
is called the {\em Heyting inverse image\/} monotonic function,
and satisfies the Hoare flow equivalence (see definition below)
$v \tprod r \preceq r \tprod u$ iff $v \preceq \comonoid{r}{(u)}{}$.
For each term $\term{y}{r}{x}$
and each {\bf H}-comonoid $v \memberof \comonoid{}{(y)}{}$,
the endoterm $\term{x}{r \tensorimplysource (v \tprod r)}{x}$ is not necessarily an {\bf H}-comonoid.
So define the {\em Heyting direct image\/} monotonic function by
$\comonoid{}{(v)}{r} \define \interior{(r \tensorimplysource (v \tprod r))}$.
Heyting direct image 
$\comonoid{}{(y)}{} \stackrel{\comonoid{}{}{r}}{\longrightarrow} \comonoid{}{(x)}{}$
is not necessarily left adjoint to Heyting inverse image.
For any functional term $\term{y}{r}{x}$ in a Heyting category,
since left $r^{\rm op}$-implication is equal to left $r$-product $r^{\rm op} \tensorimplysource (\,) = r \tprod (\,)$
and right $r$-implication is equal to right $r^{\rm op}$-product $(\,) \tensorimplytarget r = (\,) \tprod r^{\rm op}$,
we have the adjoint triples
\begin{center}
   $\begin{array}{c@{\;\dashv\;}r@{\;=\;}l@{\;\dashv\;}c}
       r^{\rm op} \tprod (\,) \mbox{  }
          & \mbox{  } r^{\rm op} \tensorimplysource (\,) & r \tprod (\,) \mbox{  }
          & \mbox{  } r \tensorimplysource (\,) \\
       (\,) \tprod r \mbox{  }
          & \mbox{  } (\,) \tensorimplytarget r & (\,) \tprod r^{\rm op} \mbox{  }
          & \mbox{  } (\,) \tensorimplytarget r^{\rm op},
    \end{array}$
\end{center}    
which are involutively equivalent to each other.
These adjoint triples imply that the Heyting direct/inverse image monotonic functions
${\bf H}^r \dashv {\bf H}_r$ and $\comonoid{}{}{r}, \comonoid{r}{}{}$ 
in a affirmation Heyting category {\bf H}
extend the direct/inverse image monotonic functions
${\bf P}^r \dashv {\bf P}_r$ and $\comonoid{}{}{r} \dashv \comonoid{r}{}{}$ 
in an affirmation bisemilattice {\bf P}.

\paragraph{Dialectical Object Flow.}
Let {\bf H} be any Heyting category.
Assume the existence of a special type $1$, called a {\em separator\/} of terms:
for any two parallel terms $\term{y}{s,r}{x}$,
if $\psi \circ s = \psi \circ r$ for all terms $\term{1}{\psi}{y}$ then $s = r$.
A term $\term{1}{\phi}{x}$ is called an {\em object\/} of type $x$.
An important example of objects occurs in relational database theory,
where the Heyting category {\bf H} is the category of monoids and processes \cite{Kent88}
of closed subsets of $\Sigma$-terms for some signature $\Sigma$,
a monoid $m \type x$ represents a constrained database scheme consisting of database scheme $x$ and semantic constraints $m$,
and an $m$-object is a database which satisfies that scheme and those semantic constraints.
Let $\obj{}{(x)}{} = {\bf H}[1,x]$ denote the lattice of all objects of type $x$.
Terms define a dialectical (bidirectional) flow of objects which is expressed in terms of tensor product and implication:
for any term $\term{y}{r}{x}$
let $\obj{}{}{r} = (\,) \tprod r$ denote right tensor product by $r$
defining a {\em direct object flow\/}
$\obj{}{(y)}{} \stackrel{\obj{}{}{r}}{\longrightarrow} \obj{}{(x)}{}$,
and let $\obj{r}{}{} = (\,) \tensorimplytarget r$ denote right tensor implication by $r$
defining an {\em inverse object flow\/}
$\obj{}{(y)}{} \stackrel{\obj{r}{}{}}{\longleftarrow} \obj{}{(x)}{}$.
We identify this dialectical flow of objects as the {\em behavior\/} of the term $r$.
The separator rule states that terms are distinguished (and can be identified) by their direct flow behavior.
Direct object flow $\obj{}{(y)}{} \stackrel{\obj{}{}{r}}{\longrightarrow} \obj{}{(x)}{}$
and inverse object flow $\obj{}{(y)}{} \stackrel{\obj{r}{}{}}{\longleftarrow} \obj{}{(x)}{}$
form an adjoint pair of monotonic functions $\obj{}{}{r} \dashv \obj{r}{}{}$. 
Denoting this adjunction by
$\obj{}{(r)}{} = (\obj{}{}{r} \dashv \obj{r}{}{})$,
the {\em object concept\/} or {\em flow dialectic\/} is represented by the dialectical base
${\bf H} \stackrel{\obj{}{}{}}{\longrightarrow} {\bf adj}$,
mapping types to their object lattice and terms to their dialectical behavior.
This is the sense in which Heyting terms specify the dialectical motion of objects.

In the general theory of dialectics,
there are two natural meanings for ``entities in dialectical motion'':
1. {\em subtypes\/} $u \memberof \comonoid{}{(x)}{} \subseteq {\bf P}[x,x]$; and
2. {\em objects\/}  $\phi \memberof \obj{}{(x)}{} = {\bf H}[1,x]$.
Affirmation bisemilattices naturally specify the vertical dialectical flow of subtypes,
whereas Heyting categories naturally specify the horizontal dialectical flow of objects.
The main purpose of this paper is to discuss a third kind of dialectical flow;
a (horizontal) dialectical flow of subtypes
which is a special case of object flow,
but extends vertical subtype flow from functional terms to arbitrary terms in a bisemilattice.

\paragraph{Hoare Triples.}
For any source and target comonoids
$v \memberof \comonoid{}{(y)}{}$ and $u \memberof \comonoid{}{(x)}{}$
the term $\term{v}{r_{vu}}{u}$ defined by $r_{vu} \define v \tprod r \tprod u$
is called the $(v,u)$-{\em th subterm\/} of $r$.
A {\bf P}-{\em coprocess\/} $\term{v \type y}{r}{u \type x}$
is a {\bf P}-term $\term{y}{r}{x}$
which satisfies the external source constraint $v \tprod r \succeq_{y,x} r$
saying that $r$ restricts to the source comonoid $v \type y$,
and which satisfies the external target constraint $r \preceq_{y,x} r \tprod u$
saying that $r$ corestricts to the target comonoid $u \type x$.
The source/target restriction conditions can be replaced by the two equalities
$v \tprod r = r$ and $r = r \tprod u$;
or by the single equality
$r_{vu} = v \tprod r \tprod u = r$.
Thus,
the notion of coprocess allows comonoids to function as identity arrows,
or objects,
of some category.
To make this precise we define the biposet $\comonoid{}{({\bf P})}{}$,
whose objects are {\bf P}-comonoids and whose arrows are {\bf P}-coprocesses.
The biposet $\comonoid{}{({\bf P})}{}$ has certain completeness properties,
and corresponds to Lawvere's ``Cauchy-completion'' \cite{Lawvere89}.

Given any {\bf P}-term $y \stackrel{r}{\rightarrow} x$,
let $\filtersource{r} \subseteq \comonoid{}{(y)}{}$
denote the collection
$\filtersource{r} \define \{ v \mid v \tprod r \succeq_{y,x} r \}$
of all comonoids at the source type $y$ satisfying source restriction.
Since $\filtersource{r}$ is closed above and closed under finite meets (= tensor products)
it is a filter in the lattice $\comonoid{}{(y)}{}$
called the {\em source filter\/} of $r$.
In particular,
for any comonoid $u \memberof \comonoid{}{(x)}{}$
the source filter is $\filtersource{u} = \principalfilter{u}$,
the principal filter generated by $u$.
Similarly,
the {\em target filter\/} $\filtertarget{r}$ of $r$ is the collection
$\filtertarget{r} \define \{ u \mid r \preceq_{y,x} r \tprod u \}
                  \subseteq \comonoid{}{(x)}{}$
of all comonoids at $x$ satisfying target corestriction,
and for any comonoid $u \memberof \comonoid{}{(x)}{}$
the target filter is the principal filter $\filtertarget{u} = \principalfilter{u}$.
Given two comonoids $v \type y$ and $u \type x$,
a term $\term{y}{r}{x}$ is a coprocess $\term{v \type y}{r}{u \type x}$
iff $v \memberof \filtersource{r}$ and $u \memberof \filtertarget{r}$.

For the fundamental case {\bf P} = {\bf Rel} of sets and relations,
a comonoid $U \type X$ is a subset $U \subseteq X$ and hence occurs as a type itself,
and an ordinary relation satisfies the source/target constraints
$V \tprod R \supseteq R$ and $R \subseteq R \tprod U$
iff $R \subseteq \product{V}{U}$.
So $\term{V \type Y}{R}{U \type X}$ iff $\term{V}{R}{U}$.
Then for functionality,
$\term{V \type Y}{R \dashv S}{U \type X}$ in $\comonoid{}{({\bf Rel})}{}$
iff
$\term{V}{R \dashv S}{U}$ in {\bf Rel}
iff
$R = {\rm graph}(f)$ for some unique function $V \stackrel{f}{\rightarrow} U$ and $S = R^{\rm op}$.
So ${\comonoid{}{({\bf Rel})}{}}^{\dashv} = {\bf Pfn}$ the category of sets and {\em partial functions\/}.
In a general join bisemilatice {\bf P},
a functional term in $\comonoid{}{({\bf P})}{}$ is called a {\em partial functional term\/} in {\bf P},
and the category ${\comonoid{}{({\bf P})}{}}^{\dashv}$ is called the category of partial functional {\bf P}-terms.

Unfortunately,
the category $\comonoid{}{({\bf P})}{}$ is not as useful as one might desire;
in particular,
there is no canonical functor to the underlying category {\bf P} of types and terms
since identities are not preserved.
But by suitably weakening the constraint $v \tprod r = r = r \tprod u$
we get a very useful and interesting category.
A {\em Hoare triple\/} or {\em Hoare assertion\/} $\term{v \type y}{r}{u \type x}$,
denoted traditionally although imprecisely by $\{v\}r\{u\}$,
consists of a ``flow specifying'' {\bf P}-term $\term{y}{r}{x}$ and two {\bf P}-comonoids,
a ``precondition'' or source comonoid $v \memberof \comonoid{}{(y)}{}$
and a ``postcondition'' or target comonoid $u \memberof \comonoid{}{(x)}{}$,
which satisfy the ``precondition/postcondition constraint''
$v \tprod r \preceq r \tprod u$.
Composition of Hoare triples $\{w\}s\{v\} \tprod \{v\}r\{u\} = \{w\}(s \tprod r)\{u\}$ is well-defined
and $\{u\}x\{u\}$ is the identity Hoare triple at the comonoid $u \type x$.
Also,
there is a zero triple $\{v\}0_{y,x}\{u\}$ for any precondition $v \memberof \comonoid{}{(y)}{}$ and postcondition $u \memberof \comonoid{}{(x)}{}$,
and if $\{v\}r\{u\}$ and $\{v\}s\{u\}$ are two triples with the same precondition and postcondition
then $\{v\}(r \bsum s)\{u\}$ is also a triple.
So typed comonoids as objects and Hoare triples as arrows form a join bisemilattice $\Hoare{{\bf P}}$
called the {\em Hoare assertional category\/} over {\bf P}.
There is an obvious underlying type/term functor
$\Hoare{{\bf P}} \stackrel{T_P}{\longrightarrow} {\bf P}$
which is a morphism of join bisemilattices.
For each type $x$ in {\bf P},
the {\em fiber\/} over $x$ is the subcategory
$T_P^{\rm -1}(x) \subseteq \Hoare{{\bf P}}$
of all comonoids and triples which map to $x$.
The objects in $T_P^{\rm -1}(x)$ are the comonoids of type $x$
and the triples in $T_P^{\rm -1}(x)$ are of the form $\{u'\}x\{u\}$,
pairs of comonoids of type $x$ satisfying $u' \preceq u$.
Hence,
the fiber over $x$ is just the join semilattice (actually, lattice) of comonoids
$T_P^{\rm -1}(x) = \comonoid{}{(x)}{}$.

When {\bf P} is an affirmation bisemilattice,
the fact that the (vertical flow) comonoid construction $\Omega$ is an indexed adjointness
(dialectical base)
${\bf P}^\dashv \stackrel{\Omega}{\longrightarrow} {\bf adj}$
can equivalently be expressed by the fact that
there is an indexed adjointness
$\Omega_P \stackrel{T^{\dashv}}{\longrightarrow} {\bf P}^{\dashv}$:
objects of $\Omega_P$ are pairs $u \type x$ or typed comonoids $u \memberof \comonoid{}{(x)}{}$;
arrows $\term{v \type y}{f}{u \type x}$ of $\Omega_P$
are functional terms $\term{y}{f \dashv f^{\rm op}}{x}$ in {\bf P}
satisfying $\comonoid{}{u}{f} \preceq u$ iff $v \tprod f \preceq f \tprod u$ iff $v \preceq \comonoid{f}{u}{}$.
The indexed category
$\Omega_P \stackrel{T^{\dashv}}{\longrightarrow} {\bf P}^{\dashv}$
is a subcategory of $\Hoare{{\bf P}}$ satisfying
${\rm Inc} \cdot T_P = T^{\dashv} \cdot {\rm Inc}$.
The important additional axiom,
which states that $\triple{\Hoare{{\bf P}}}{T_P}{{\bf P}}$ is an indexed category
(asserting certain optimality conditions on the fibers of $T_P$),
and defines the notion of a ``flow category'',
is discussed below.

Just as we can replace the category of comonoids and coprocesses $\comonoid{}{({\bf P})}{}$
by the Hoare assertional category $\Hoare{{\bf P}}$ to get better mathematical properties,
so also we can replace the category of partial functional {\bf P}-terms ${\comonoid{}{({\bf P})}{}}^{\dashv}$
by the category ${\Hoare{{\bf P}}}^{\dashv}$ of relative partial functional {\bf P}-terms.
An arrow in ${\Hoare{{\bf P}}}^{\dashv}$,
a relative partial functional {\bf P}-term,
is a functional term in $\Hoare{{\bf P}}$
and satisfies the conditions:
{\bf functionality} $y \preceq f \tprod f^{\rm op}$ and $f^{\rm op} \tprod f \preceq x$,
and {\bf partialness} $v \tprod f \preceq f \tprod u$ and $u \tprod f^{\rm op} \preceq f^{\rm op} \tprod v$.
(when an involution exists, these two partialness conditions are equivalent to the single condition $v \tprod f = f \tprod u$;
 but in general $v \tprod f \neq f$).
It is easy to check that,
if $\term{v \type y}{f}{u \type x}$ is a functional term in $\Hoare{{\bf P}}$,
then
$\term{v \type y}{v \tprod f \tprod u}{u \type x}$ is a partial functional term in ${\bf P}$.

The {\em source ideal\/} of $r$ is the collection
$\idealsource{r} \define \{ v \memberof \comonoid{}{(y)}{} \mid v \tprod r = 0 \}
                 \subseteq \comonoid{}{(y)}{}$
of all preconditions which are ``disjoint'' from $r$,
and
the {\em target ideal\/} of $r$ is the collection
$\idealtarget{r} \define \{ u \memberof \comonoid{}{(x)}{} \mid 0 = r \tprod u \}
                 \subseteq \comonoid{}{(x)}{}$
of all postconditions which are disjoint from $r$.
For any {\bf P}-term $\term{y}{r}{x}$,
define the $r$-{\em flow relation\/} $\{\}r\{\} \subseteq \product{\comonoid{}{(y)}{}}{\comonoid{}{(x)}{}}$
by $v\{\}r\{\}u$ when $\{v\}r\{u\}$.
The flow relation $\{\}r\{\}$ has a {\em direct image\/} monotonic function
$\comonoid{}{(y)}{} \stackrel{\{(\;)\}r\{\}}{\longrightarrow} \filter{\comonoid{}{(x)}{}}$
where $\{v\}r\{\}$ is the $v$-{\em th target filter\/} of $r$
defined by $\{v\}r\{\} \define \{ u \memberof \comonoid{}{(x)}{} \mid \{v\}r\{u\} \}$
for any precondition $v \memberof \comonoid{}{(y)}{}$.
Then $\{v\}r\{\}$ is a filter in the lattice $\comonoid{}{(x)}{}$ having the same meets as $\comonoid{}{(x)}{}$,
since $\{v\}r\{u \tprod u'\}$ iff $\{v\}r\{u\}$ and $\{v\}r\{u'\}$.
The flow relation $\{\}r\{\}$ has an {\em inverse image\/} monotonic function
$\ideal{\comonoid{}{(y)}{}} \stackrel{\{\}r\{(\:)\}}{\longleftarrow} \comonoid{}{(x)}{}$
where $\{\}r\{u\}$ is the $u$-{\em th source ideal\/} of $r$
defined by $\{\}r\{u\} \define \{ v \memberof \comonoid{}{(x)}{} \mid \{v\}r\{u\} \}$
for any postcondition $u \memberof \comonoid{}{(x)}{}$.
Then $\{\}r\{u\}$ is an ideal in the lattice $\comonoid{}{(y)}{}$ having the same joins as $\comonoid{}{(y)}{}$,
since $\{v \bsum v'\}r\{u\}$ iff $\{v\}r\{u\}$ and $\{v\}r\{u'\}$.

A {\em Hoare cotriple\/} or {\em dual Hoare assertion\/} $\term{v \type y}{r}{u \type x}$,
denoted also by $\}v\{r\}u\{$,
consists of a flow specifier $\term{y}{r}{x}$,
a precondition $v \memberof \comonoid{}{(y)}{}$ and a postcondition $u \memberof \comonoid{}{(x)}{}$,
which satisfy the ``dual precondition/postcondition constraint''
$v \tprod r \succeq r \tprod u$.
Typed comonoids as objects and Hoare cotriples as arrows form a dual join bisemilattice $\dualHoare{{\bf P}}$
called the {\em dual Hoare assertional category\/} over {\bf P}.
All of the properties of Hoare triples hold also for Hoare cotriples
with the words ``ideal'' and ``filter'' interchanged.
In particular,
$\dualHoare{{\bf P}}$ is a join bisemilattice,
and there is an obvious underlying type/term functor
$\dualHoare{{\bf P}} \stackrel{T_P}{\longrightarrow} {\bf P}$
which is a morphism of join bisemilattices.
For each type $x$ in {\bf P},
the fiber over $x$ is $T_P^{\rm -1}(x) = {\comonoid{}{(x)}{}}^{\rm op}$,
the opposite of the lattice of comonoids,
since triples in $T_P^{\rm -1}(x)$ are of the form $\}u'\{x\}u\{$,
pairs of comonoids of type $x$ satisfying $u' \succeq u$.
An involution ${\bf P}^{\rm op} \stackrel{\involution{(\;)}}{\longrightarrow} {\bf P}$
extends to an ``involution on assertions''
${\dualHoare{{\bf P}}}^{\rm op} \stackrel{\involution{(\;)}}{\longrightarrow} \Hoare{{\bf P}}$;
a pair of inverse morphisms of join bisemilattices defining an isomorphism ${\dualHoare{{\bf P}}}^{\rm op} \equiv \Hoare{{\bf P}}$
and satisfying $\involution{(\,)} \cdot T_P = {T_P}^{\rm op} \cdot \involution{(\,)}$.

For any {\bf P}-term $\term{y}{r}{x}$,
define the {\em dual\/} $r$-{\em flow relation\/} ${\}\{r\}\{} \subseteq \product{\comonoid{}{(y)}{}}{\comonoid{}{(x)}{}}$
by $v\}\{r\}\{u$ when $\}v\{r\}u\{$.
The flow relation $\}\{r\}\{$ has a {\em direct image\/} monotonic function
$\filter{\comonoid{}{(y)}{}} \stackrel{\}\{r\}(\:)\{}{\longleftarrow} \comonoid{}{(x)}{}$
where $\}\{r\}u\{$ is the $u$-{\em th source filter\/}
defined by ${\}\{r\}u\{} \define \{ v \memberof \comonoid{}{(x)}{} \mid \; \}v\{r\}u\{ \; \}$
for any postcondition $u \memberof \comonoid{}{(x)}{}$.
Then $\}\{r\}u\{$ is a filter in the lattice $\comonoid{}{(y)}{}$ having the same meets as $\comonoid{}{(y)}{}$.
The flow relation $\}\{r\}\{$ has an {\em inverse image\/} monotonic function
$\comonoid{}{(y)}{} \stackrel{\}(\;)\{r\}\{}{\longrightarrow} \ideal{\comonoid{}{(x)}{}}$
where $\}v\{r\}\{$ is the $v$-{\em th target ideal\/} of $r$
defined by $\}v\{r\}\{ \; \define \{ u \memberof \comonoid{}{(x)}{} \mid \; \}v\{r\}u\{ \; \}$
for any precondition $v \memberof \comonoid{}{(y)}{}$.
Then $\}v\{r\}\{$ is an ideal in the lattice $\comonoid{}{(x)}{}$ having the same joins as $\comonoid{}{(x)}{}$.
\begin{Facts}
   The source and target filters and the source and target ideals 
   are the dual $x$-th source filter, the $y$-th target filter,
   the $0$-th source ideal, and the dual $0$-th target ideal,
   respectively:
   \begin{center}
      $\begin{array}{r@{\;=\;}lcr@{\;=\;}l}
          \filtersource{r} & \}\{r\}x\{
          && 
          \filtertarget{r} & \{y\}r\{\} \\
          \idealsource{r}  & \{\}r\{0\}
          && 
          \idealtarget{r}  & \}0\{r\}\{
       \end{array}$
   \end{center}
   The other four possible combinations are trivial:
   $\}\{r\}0\{ = \comonoid{}{(y)}{} = \{\}r\{x\}$, and
   $\{0\}r\{\} = \comonoid{}{(x)}{} = \}y\{r\}\{$.
   As we shall see later,
   we can axiomatize the notions of domain, range, kernel and cokernel
   via existence of the following meets of filters and joins of ideals:
   \begin{center}
      $\begin{array}{rr@{\;=\;}lcrr@{\;=\;}l} 
          \mbox{domain}   & \domain{(r)}   & \bigwedge \filtersource{r}
          && 
          \mbox{range}    & \range{(r)}    & \bigwedge \filtertarget{r} \\
          \mbox{kernel}   & \kernel{(r)}   & \bigvee \idealsource{r}
          && 
          \mbox{cokernel} & \cokernel{(r)} & \bigvee \idealtarget{r}
       \end{array}$
   \end{center}
\end{Facts}

\subsection{Flow Categories}

\paragraph{Direct Flow Categories.}
For each type $x$ in {\bf P},
the lattice of comonoids $\comonoid{}{(x)}{}$ is a (one object) join sub-bisemilattice of {\bf P},
and the inclusion functor $\comonoid{}{(x)}{} \stackrel{{\rm Inc}_x}{\longrightarrow} {\bf P}$ is a morphism of join bisemilattices.
Tensor product,
which is lattice meet in $\comonoid{}{(x)}{}$,
forms a {\em local conjunction functor\/}
$\comonoid{}{(x)}{} \stackrel{\tprod_x}{\longrightarrow} {\bf JSL}$
into the category of join semilattices,
defined by $\tprod_x(x) \define \comonoid{}{(x)}{} = \triple{\comonoid{}{(x)}{}}{\bsum}{0}$
and $\tprod_x(u) \define \comonoid{}{(x)}{} \stackrel{(\;) \tprod u}{\longrightarrow} \comonoid{}{(x)}{}$.
Conjunction is a morphism of join bisemilattices.
This example is a special case of the following construct.
An {\em indexed join semilattice\/} $\pair{{\bf P}}{\directsquare{(\;)}{}}$ consists of:
   1. a join bisemilattice {\bf P}, and
   2. a morphism of join bisemilattices ${\bf P} \stackrel{\directsquare{(\;)}{}}{\longrightarrow} {\bf JSL}$
      to the special join bisemilattice {\bf JSL}.
In more detail,
            (a) $\directsquare{x}{}$ is a join semilattice for each type $x$; 
            (b) $\directsquare{y}{} \stackrel{\directsquare{r}{}}{\rightarrow} \directsquare{x}{}$
                  is a morphism of join semilattices for each term $\term{y}{r}{x}$
                  called the {\em direct flow\/} specified by $r$,
                  with $\directsquare{r}{(0_y)} = 0_x$
                  and $\directsquare{r}{(v \bsum v')} = \directsquare{r}{(v)} \bsum \directsquare{r}{(v')}$;
            (c) $\directsquare{(\;)}{}$ is functorial, 
                  with $\directsquare{x}{} = {\rm Id}_{\directsquare{x}{}}$, and
                  $\directsquare{s \tprod r}{} = \directsquare{s}{} \cdot \directsquare{r}{}$;
            (d) $\directsquare{(\;)}{}$ is a join semilattice functor,
                  (i)   if $r \preceq s$ then $\directsquare{r}{} \leq \directsquare{s}{}$,
                  (ii)  $\directsquare{0}{} = \bot$, and
                  (iii) $\directsquare{r \bsum s}{} = \directsquare{r}{} \vee \directsquare{s}{}$.
Equivalently,
an indexed join semilattice
is a join bisemilattice morphism ${\bf H} \stackrel{T}{\rightarrow} {\bf P}$ 
from some join bisemilattice {\bf H}
called the {\em underlying type/term functor\/},
which as a functor is an indexed category (an opfibration).
This underlying type/term functor has as left adjoint $0_P \dashv T_P$ 
the {\em bottom functor\/} (join bisemilattice morphism)
${\bf P} \stackrel{0_P}{\longrightarrow} \Hoare{{\bf P}}$
mapping types as $0_P(x) \define 0 \type x$
and terms as $0_P(\term{y}{r}{x}) \define \term{0 \type y}{r}{0 \type x}$.
This is a coreflective pair of functors
with $0_P$ embedding {\bf P} into $\Hoare{{\bf P}}$
and $T_P$ coreflecting $\Hoare{{\bf P}}$ onto {\bf P}.
A {\em morphism of indexed join semilattices\/}
$\pair{{\bf P}}{\directsquare{P,(\;)}{}} \stackrel{H}{\longrightarrow} \pair{{\bf Q}}{\directsquare{Q,(\;)}{}}$
is a morphism of join bisemilattices ${\bf P} \stackrel{H}{\longrightarrow} {\bf Q}$
which preserves flow $H \cdot \directsquare{Q,(\;)}{} = \directsquare{P,(\;)}{}$.
Equivalently,
a morphism of indexed join semilattices is a morphism of join bisemilattices
${\bf P} \stackrel{H}{\longrightarrow} {\bf Q}$,
which commutes with the underlying type/term functors $T_P \cdot H = T_Q$.

A {\em direct flow category\/} $\pair{{\bf P}}{\directsquare{(\;)}{}}$ is an indexed join semilattice,
which is (3) standard on subtypes:
            (a) $\directsquare{x}{}$ is a join subsemilattice of comonoids
                  $\directsquare{x}{} \subseteq \comonoid{}{(x)}{} = \triple{\comonoid{}{(x)}{}}{\bsum}{0}$
                  for each type $x$; 
            (b) $\directsquare{(\;)}{}$ restricted to $x$-comonoids is the local conjunction functor
                  ${\rm Inc}_x \cdot \directsquare{(\;)}{} = \tprod_x$;
                  that is,
                  subtype direct flow $\directsquare{x}{} \stackrel{\directsquare{u}{}}{\rightarrow} \directsquare{x}{}$
                  is just conjunction $\directsquare{u}{(u')} = u' \tprod u$
                  for each comonoid $u \memberof \directsquare{x}{}$.
More precisely,
the inclusion $\directsquare{x}{} \subseteq \comonoid{}{(x)}{}$
should be replace by a coreflective pair of monotonic functions
which embed $\directsquare{x}{}$ into $\comonoid{}{(x)}{}$
and coreflect $\comonoid{}{(x)}{}$ back to $\directsquare{x}{}$.
For any functional term $\term{y}{f \dashv f^{\rm op}}{x}$
direct flow along $f$ is called {\em existential quantification\/} along $f$
and denoted by $\exists_f \define \directsquare{f}{}$,
whereas direct flow along $f^{\rm op}$ is called either {\em substitution\/} or {\em inverse image\/} along $f$
and denoted by ${\rm sub}_f \define \directsquare{f^{\rm op}}{}$.
Thus,
there is a dialectical base
${\bf P}^\dashv \stackrel{\directsquare{(\,)}{}}{\longrightarrow} {\bf adj}$
defined on functional terms $\term{y}{f \dashv f^{\rm op}}{x}$ by
$\directsquare{f}{} \define (\exists_f \dashv {\rm sub}_f)$.
Comonoids in $\directsquare{x}{}$ and conjunction form a direct flow category $\pair{\directsquare{x}{}}{\tprod_x}$ for each type $x$.
A {\em morphism of direct flow categories\/}
$\pair{{\bf P}}{\directsquare{P,(\;)}{}} \stackrel{H}{\longrightarrow} \pair{{\bf Q}}{\directsquare{Q,(\;)}{}}$
is a morphism of indexed join semilattices.
In particular,
inclusion $\pair{\directsquare{x}{}}{\tprod_x} \stackrel{{\rm Inc}_x}{\longrightarrow} \pair{{\bf P}}{\directsquare{(\;)}{}}$
is a morphism of direct flow categories.

A {\em contravariant direct flow category\/} $\pair{{\bf P}}{\directdiamond{(\;)}{}}$
consists of a join bisemilattice {\bf P}, and
a join semilattice functor ${\bf P}^{\rm op} \stackrel{\directdiamond{(\;)}{}}{\longrightarrow} {\bf JSL}$,
such that $\pair{{\bf P}^{\rm op}}{\directdiamond{(\;)}{}}$ is a covariant direct flow category.
An {\em involutive direct flow category\/} ${\bf P} = \triple{{\bf P}}{\directdiamond{(\;)}{}}{\directsquare{(\;)}{}}$
consists of
an involutive join bisemilattice ${\bf P} = \pair{{\bf P}}{\involution{(\;)}}$,
a contravariant direct flow category ${\bf P} = \pair{{\bf P}}{\directdiamond{(\;)}{}}$, and
a covariant direct flow category     ${\bf P} = \pair{{\bf P}}{\directsquare{(\;)}{}}$,
where involution
$\pair{{\bf P}^{\rm op}}{\directdiamond{(\;)}{}} \stackrel{\involution{(\;)}}{\longrightarrow} \pair{{\bf P}}{\directsquare{(\;)}{}}$
is a morphism of direct flow categories,
so that involution satisfies the condition:
$\directsquare{\involution{r}}{} = \directdiamond{r}{}$ for all terms $\term{y}{r}{x}$. 
In particular,
$\directsquare{p}{} = \directdiamond{p}{}$ for all self-involutive endoterms $\term{x}{r}{x}$,
which includes all subobjects and {\em a fortiriori\/} all comonoids.

A join bisemilattice {\bf P} has {\em direct Hoare flow\/}
when for any term $\term{y}{r}{x}$ and any precondition $v \memberof \comonoid{}{(y)}{}$,
there is a postcondition $\directsquare{r}{(v)} \memberof \comonoid{}{(x)}{}$
called the {\em strongest postcondition\/} of $r$
which satisfies the axiom
\begin{center}
   $\begin{array}{ccccc}
       \directsquare{r}{(v)} \preceq u
       & \mbox{iff} &
       v \tprod r \preceq r \tprod u
    \end{array}$
\end{center}
iff $(\term{v \type y}{r}{u \type x}) \memberof {\rm Ar}(\Hoare{{\bf P}})$
for any postcondition $u \memberof \comonoid{}{(x)}{}$.
So,
direct Hoare flow can be expressed as the meet
$\directsquare{r}{(v)} = \bigwedge \{ u \in \comonoid{}{(x)}{} \mid v \tprod r \preceq r \tprod u \}$.
This axiom states that
$\directsquare{r}{(v)} = \bigwedge_x \{v\}r\{\}$ for any postcondition $v \memberof \comonoid{}{(y)}{}$,
or equivalently that the $v$-th target filter of $r$ is the principal filter
$\{v\}r\{\} = {\uparrow}_x (\directsquare{r}{(v)})$.
So the direct image function factors
$\{(\;)\}r\{\} = {\directsquare{r}{\,}}^{\rm op} \cdot {\uparrow}_x$
as direct flow followed by principal filter,
and that direct flow factors as direct image followed by meet
(when meets exist for all filters of comonoids),
since the meet operator
$\filter{\comonoid{}{(x)}{}} \stackrel{\wedge_x}{\longrightarrow} {\comonoid{}{(x)}{}}^{\rm op}$
forms a reflective pair $\bigwedge_x \dashv {\uparrow}_x$ with the principal filter operator
${\comonoid{}{(x)}{}}^{\rm op} \stackrel{\uparrow_x}{\longrightarrow} \filter{\comonoid{}{(x)}{}}$.
Then for any two composable {\bf P}-terms $\term{z}{s}{y}$ and $\term{y}{r}{x}$
the inequality
$\directsquare{s \tprod r}{(w)} \preceq \directsquare{r}{(\directsquare{s}{(w)})}$
holds for any comonoid $w \memberof \comonoid{}{(z)}{}$.
Since we want this to be equality,
we must also assume that the axiom
\begin{center}
   $\directsquare{r}{(\directsquare{s}{(w)})} \preceq \directsquare{s \tprod r}{(w)}$
\end{center}
holds for any comonoid $w \memberof \comonoid{}{(z)}{}$.
We also assume that the direct Hoare flow operator $\directsquare{(\,)}{}$ is monotonic:
if $r \preceq s$ then $\directsquare{r}{} \preceq \directsquare{s}{}$.
Some identities for the direct flow operator $\directsquare{(\;)}{}$ are:
$\directsquare{u}{(u')} = u' \tprod u$ for all comonoids $u \memberof \comonoid{}{(x)}{}$,
in particular $\directsquare{x}{} = {\rm Id}$ for any type $x$;
$\directsquare{s \tprod r}{(w)} = \directsquare{r}{(\directsquare{s}{(w)})}$
for two composable {\bf P}-terms $\term{z}{s}{y}$ and $\term{y}{r}{x}$;
$\directsquare{r}{(v)} = 0$ iff $v \tprod r = 0$ iff $v \memberof \idealsource{r}$,
or equivalently
$\idealsource{\;} = \directsquare{(\;)}{} \cdot {\rm ker\/}$,
stating that the source ideal $\idealsource{r}$ is the kernel of the direct flow function $\directsquare{r}{}$.
So, amongst other things, $\triple{\Hoare{{\bf P}}}{T_P}{{\bf P}}$ is an indexed category.

A join bisemilattice {\bf P} has {\em ranges\/}
when for any {\bf P}-term $\term{y}{r}{x}$
there is a {\em range postcondition\/} $\range{(r)} \memberof \comonoid{}{(x)}{}$
which satisfies the axioms
\begin{center}
   \begin{tabular}{cl}
      $\range{(r)} \preceq u$ iff $r \preceq r \tprod u$
         & for any postcondition $u \memberof \comonoid{}{(x)}{}$ \\
      $\range{(s \tprod r)} = \range{(\range{(s)} \tprod r)}$
         & for any composable {\bf P}-term $\term{z}{s}{y}$       \\
      if $r \preceq s$ then $\range{(r)} \preceq \range{(s)}$
         & for any parallel {\bf P}-term $\term{y}{s}{x}$
   \end{tabular}
\end{center}
The first axiom is equivalent to
$\range{(r)} = \bigwedge \{ u \in \comonoid{}{(x)}{} \mid r \preceq r \tprod u \}$.
It immediately implies that
$\range{(s \tprod r)} \preceq \range{(r)}$ and $r \preceq r \tprod \range{(r)}$;
hence also implies
$\range{(s \tprod r)} \preceq \range{(s \tprod \range{(s)} \tprod r)}
                      \preceq \range{(\range{(s)} \tprod r)}$,
which is half of the second axiom.
The first axiom states that
$\range{(r)} = \bigwedge \filtertarget{r} \in \filtertarget{r}$
is the bottom of the target filter of $r$,
or equivalently that the target filter of $r$ is the principal filter
$\filtertarget{r} = \abov{\range{(r)}}$.
Some identities for the range operator $\range{}$ are:
``subtypes are their own range''
$\range{(u)} = u$ for any comonoid $u \memberof \comonoid{}{(x)}{}$;
``the range of a subterm is the subterm of the range''
$\range{(r \tprod u)} = \range{(r)} \tprod u$ for any term $\term{y}{r}{x}$ and any postcondition $u \memberof \comonoid{}{(x)}{}$; and
``only zero has empty range''
$\range{(r)} = 0_x$ iff $r = 0_{y,x}$ for any term $\term{y}{r}{x}$.
If {\bf P} has direct Hoare flow $\directsquare{(\;)}{}$,
then it has ranges $\range{}$
defined to be the direct flow of the top (identity) precondition
$\range{(r)} \define \directsquare{r}{(y)}$
for any term $\term{y}{r}{x}$.
Conversely,
if {\bf P} has ranges,
then it has direct Hoare flow
defined to be the range of the tensor product (guarded term)
$\directsquare{r}{(v)} \define \range{(v \tprod r)}$;
which states that
direct flow is the target readout of the interaction of a source condition process $v$ with the general process $r$.
A {\em direct Hoare flow category\/} is a join bisemilattice which has direct Hoare flow,
or equivalently, ranges.
\begin{Proposition}
   A join bisemilattice {\bf P} is a direct Hoare flow category
   iff
   the associated functor $\Hoare{{\bf P}} \stackrel{T_P}{\longrightarrow} {\bf P}$
   is an indexed join semilattice $\triple{\Hoare{{\bf P}}}{T_P}{{\bf P}}$. 
   In fact,
   any direct Hoare flow category is a direct flow category.
\end{Proposition}

A join bisemilattice {\bf P} has {\em dual direct Hoare flow\/}
when for any term $\term{y}{r}{x}$ and any postcondition $u \memberof \comonoid{}{(x)}{}$,
there is a precondition $\directdiamond{r}{(u)} \memberof \comonoid{}{(y)}{}$
called the {\em weakest dual precondition\/} of $r$
which satisfies the axiom
\begin{center}
   $\begin{array}{ccccc}
       v \succeq \directdiamond{r}{(u)}
       & \mbox{iff} &
       v \tprod r \succeq r \tprod u
    \end{array}$
\end{center}
iff $(\term{v \type y}{r}{u \type x}) \memberof {\rm Ar}(\dualHoare{{\bf P}})$,
for any precondition $v \memberof \comonoid{}{(y)}{}$,
or $\directdiamond{r}{(u)} = \bigwedge \{ v \in \comonoid{}{(y)}{} \mid v \tprod r \succeq r \tprod u \}$.
This axiom states that
$\directdiamond{r}{(u)} = \bigwedge_y \}\{r\}u\{$ for any postcondition $u \memberof \comonoid{}{(x)}{}$,
or equivalently that the $u$-th source filter of $r$ is the principal filter
${\}\{r\}u\{} = {\uparrow}_y (\directdiamond{r}{(u)})$.
So the dual direct image function factors
${\}\{r\}(\;)\{} = {\directdiamond{r}{\,}}^{\rm op} \cdot \uparrow_y$
as direct flow followed by principal filter,
and that direct flow factors as direct image followed by meet
(when meets exist for all filters of comonoids).
Then for any two composable {\bf P}-terms $\term{z}{s}{y}$ and $\term{y}{r}{x}$
the inequality
$\directdiamond{s \tprod r}{(u)} \preceq \directdiamond{s}{(\directdiamond{r}{(u)})}$
holds for any comonoid $u \memberof \comonoid{}{(x)}{}$.
Since we want this to be equality,
we must also assume that the axiom
\begin{center}
   $\begin{array}{c}
       \directdiamond{s}{(\directdiamond{r}{(u)})} \preceq \directdiamond{s \tprod r}{(u)}
    \end{array}$
\end{center}
holds for any comonoid $u \memberof \comonoid{}{(x)}{}$.

A join bisemilattice {\bf P} has {\em domains\/}
when for any {\bf P}-term $\term{y}{r}{x}$
there is a {\em domain precondition\/} $\domain{(r)} \memberof \comonoid{}{(y)}{}$
which satisfies the axioms
\begin{center}
   $\begin{array}{ccccc}
       v \succeq \domain{(r)}
       & \mbox{iff} &
       v \tprod r \succeq r
       & \mbox{iff} &
       v \memberof \filtersource{r} \\
       \multicolumn{3}{c}{\domain{(s \tprod r)} = \domain{(s \tprod \domain{(r)})}}
       &&
    \end{array}$
\end{center}
for any precondition $v \memberof \comonoid{}{(y)}{}$,
and any composable {\bf P}-term $\term{z}{s}{y}$.
The first axiom states that
$\domain{(r)} = \bigwedge \filtersource{r} \in \filtersource{r}$
is the bottom of the source filter of $r$,
or equivalently that the source filter of $r$ is the principal filter
$\filtersource{r} = \abov{\domain{(r)}}$.
Some identities for the domain operator $\domain{}$ are:
``subtypes are their own domain''
$\domain{(u)} = u$ for any comonoid $u \memberof \comonoid{}{(x)}{}$;
``the domain of a subterm is the subterm of the domain''
$\domain{(v \tprod r)} = v \tprod \domain{(r)}$ for any term $\term{y}{r}{x}$ and any precondition $v \memberof \comonoid{}{(y)}{}$; and
``only zero has empty domain''
$\domain{(r)} = 0_y$ iff $r = 0_{y,x}$ for any term $\term{y}{r}{x}$.
A term $\term{y}{r}{x}$ is {\em total\/}
when its domain is the entire source type $\domain{(r)} = y$.
Any functional term $\term{y}{f \dashv f^{\rm op}}{x}$ is total,
since the counit inequality $y \preceq f \tprod f^{\rm op}$ implies
$y = \domain{(y)}
 \preceq \domain{(f \tprod f^{\rm op})}
       = \domain{(f \tprod \domain{(f^{\rm op})})}
 \preceq \domain{(f \tprod x)}
       = \domain{(f \tprod x)}
       = \domain{(f)}
 \preceq y$
by the composite term axiom above.
Given any two total terms $\term{z}{s}{y}$ and $\term{y}{r}{x}$,
the composite term $\term{z}{s \tprod r}{x}$ is also total,
since 
$\domain{(s \tprod r)}
 = \domain{(s \tprod \domain{(r)})}
 = \domain{(s \tprod y)}
 = \domain{(s)}
 = z$.
Total terms are close above w.r.t. term entailment order.
Since functional terms (in particular, identity terms) are total,
and the composite of total terms are also total,
total terms form a biposet ${\bf P}^\dagger$,
a subbiposet of ${\bf P}$,
${\bf P}^\dashv \subseteq {\bf P}^\dagger \subseteq {\bf P}$,
which is the homset order closure of ${\bf P}^\dashv$.
The empty term $\bot_{y,x}$ is never total for $y \neq 0$.
So ${\bf P}^\dagger$ is a subbiposet ${\bf P}$,
which preserves homset joins but usually does not have a bottom.
If {\bf P} has dual direct Hoare flow $\directdiamond{(\;)}{}$,
then it has domains $\domain{}$
defined to be the dual direct flow of the top (identity) postcondition
$\domain{(r)} \define \directdiamond{r}{(x)}$
for any term $\term{y}{r}{x}$.
Conversely,
if {\bf P} has domains,
then it has dual direct Hoare flow
defined to be the domain of the tensor product
$\directdiamond{r}{(u)} \define \domain{(r \tprod u)}$.
A {\em contravariant direct Hoare flow category\/} is a join bisemilattice which has dual direct Hoare flow,
or equivalently, domains.

An {\em involutive direct Hoare flow category\/} ${\bf P} = \triple{{\bf P}}{\involution{(\;)}}{\square{}{}}$
consists of
an involutive join bisemilattice ${\bf P} = \pair{{\bf P}}{\involution{(\;)}}$,
and a covariant direct Hoare flow category ${\bf P} = \pair{{\bf P}}{\square{(\;)}{}}$.
There is an equivalent contravariant direct Hoare flow category
${\bf P} = \pair{{\bf P}}{\diamond{}{}}$
defined by $\diamond{}{} \define \involution{(\,)} \cdot \square{}{}$,
so that involution satisfies either of the equivalent conditions:
$\directsquare{\involution{r}}{} = \directdiamond{r}{}$
or
$\range{(\involution{r})} = \domain{(r)}$
for all terms $\term{y}{r}{x}$.
\begin{Proposition}
   Any involutive direct Hoare flow category is an involutive direct flow category.
\end{Proposition}

\paragraph{Inverse Flow Categories.}
A {\em locally cartesian closed join bisemilattice\/} ({\em lcc join bisemilattice\/}) ${\bf P} = \pair{{\bf P}}{\imply}$
is a join bisemilattice {\bf P} augmented with a {\em local implication operator\/} $\imply$
which makes each lattice of comonoids $\comonoid{}{(x)}{}$ into a Heyting algebra for each type $x$,
by satisfying the dialectical axiom $v \tprod w \preceq_x u$ iff $v \preceq_x w \imply u$.
The {\em local implication functor\/} ${\comonoid{}{(x)}{}}^{\rm coop} \stackrel{\imply_x}{\longrightarrow} {\bf MSL}$,
defined by $\imply_x(x) \define \comonoid{}{(x)}{} = \triple{\comonoid{}{(x)}{}}{\tprod}{x}$
and $\imply_x(w) \define \comonoid{}{(x)}{} \stackrel{w \imply (\;)}{\longrightarrow} \comonoid{}{(x)}{}$,
is a contravariant join semilattice functor, since
(i)   $\imply_x(x) = {\rm Id}_{\comonoid{}{(x)}{}}$,
(ii)  $\imply_x(v \tprod u) \;=\; \imply_x(u) \;\cdot\; \imply_x(v)$,
(iii) if $v \succeq u$ then $\imply_x(v) \leq\; \imply_x(u)$,
(iv)  $\imply_x(0) = \top$, and
(v)   $\imply_x(u \bsum v) \;=\; \imply_x(u) \;\wedge\; \imply_x(v)$.
An {\em inverse flow category\/} $\pair{{\bf P}}{\inversesquare{(\;)}{}}$
consists of a lcc join bisemilattice ${\bf P} = \pair{{\bf P}}{\imply}$, and
a contravariant meet semilattice functor ${\bf P}^{\rm coop} \stackrel{\inversesquare{(\;)}{}}{\longrightarrow} {\bf MSL}$
into the category of meet semilattices,
which is standard on subtypes.
In more detail,
\begin{enumerate}
   \item $\inversesquare{x}{}$ is a meet subsemilattice of comonoids
         $\inversesquare{x}{} \subseteq \comonoid{}{(x)}{} = \triple{\comonoid{}{(x)}{}}{\tprod}{x}$
         for each type $x$; 
   \item $\inversesquare{y}{} \stackrel{\inversesquare{r}{}}{\leftarrow} \inversesquare{x}{}$
         is a morphism of meet semilattices for each term $\term{y}{r}{x}$
         called the {\em inverse flow\/} specified by $r$,
         with $\inversesquare{r}{(x)} = y$
         and $\inversesquare{r}{(u \tprod u')} = \inversesquare{r}{(u)} \tprod \inversesquare{r}{(u')}$;
   \item $\inversesquare{(\;)}{}$ is contravariantly functorial, 
         with $\inversesquare{x}{} = {\rm Id}_{\inversesquare{x}{}}$, and
         $\inversesquare{s \tprod r}{} = \inversesquare{r}{} \cdot \inversesquare{s}{}$;
   \item $\inversesquare{(\;)}{}$ is a meet semilattice functor,
         (i)   if $s \succeq r$ then $\inversesquare{s}{} \leq \inversesquare{r}{}$,
         (ii)  $\inversesquare{0}{} = \top$, and
         (iii) $\inversesquare{r \bsum s}{} = \inversesquare{r}{} \wedge \inversesquare{s}{}$;
   \item $\inversesquare{(\;)}{}$ is standard on subtypes,
         in that $\inversesquare{(\;)}{}$ restricted to $x$-comonoids is the local implication functor
         $\imply_x = {\rm Inc}_x^{\rm coop} \cdot \inversesquare{(\;)}{}$
         that is,
         subtype inverse flow $\inversesquare{x}{} \stackrel{\inversesquare{u}{}}{\leftarrow} \inversesquare{x}{}$
         is just implication $\inversesquare{u}{(u')} = u \imply u'$
         for each comonoid $u \memberof \inversesquare{x}{}$.
\end{enumerate}
For any functional term $\term{y}{f \dashv f^{\rm op}}{x}$
inverse flow along $f$ is called either {\em substitution\/} or {\em inverse image\/} along $f$
and denoted by ${\rm sub}_f \define \inversesquare{f}{}$.
whereas inverse flow along $f^{\rm op}$ is called {\em universal quantification\/} along $f$
and denoted by $\forall_f \define \inversesquare{f^{\rm op}}{}$.
Thus,
there is a dialectical base
$({\bf P}^\dashv)^{\rm op} \stackrel{\inversesquare{(\,)}{}}{\longrightarrow} {\bf adj}$
defined on functional terms $\term{y}{f \dashv f^{\rm op}}{x}$ by
$\inversesquare{f}{} \define ({\rm sub}_f \dashv \forall_f)$.
Comonoids and implication form an inverse flow category $\pair{\inversesquare{x}{}}{\imply_x}$ for each type $x$.
A {\em morphism of inverse flow categories\/}
$\pair{{\bf P}}{\inversesquare{P,(\;)}{}} \stackrel{H}{\longrightarrow} \pair{{\bf Q}}{\inversesquare{Q,(\;)}{}}$
is a morphism of join bisemilattices ${\bf P} \stackrel{H}{\longrightarrow} {\bf Q}$
which preserves flow $H^{\rm coop} \cdot \inversesquare{Q,(\;)}{} = \inversesquare{P,(\;)}{}$.
So inclusion $\pair{\inversesquare{x}{}}{\imply_x} \stackrel{{\rm Inc}_x^{\rm coop}}{\longrightarrow} \pair{{\bf P}}{\inversesquare{(\;)}{}}$
is a morphism of inverse flow categories.
A {\em contravariant inverse flow category\/} $\pair{{\bf P}}{\inversediamond{(\;)}{}}$
consists of a join bisemilattice {\bf P}, and
a meet semilattice functor ${\bf P}^{\rm co} \stackrel{\inversediamond{(\;)}{}}{\longrightarrow} {\bf MSL}$,
such that $\pair{{\bf P}^{\rm co}}{\inversediamond{(\;)}{}}$ is a covariant inverse flow category.

An lcc join bisemilattice {\bf P} has {\em inverse Hoare flow\/}
when for any {\bf P}-term $\term{y}{r}{x}$ and any postcondition (target comonoid) $u \memberof \comonoid{}{(x)}{}$,
there is a precondition (source comonoid) $\inversesquare{r}{(v)} \memberof \comonoid{}{(y)}{}$
called the {\em weakest (liberal) precondition\/} of $r$
which satisfies the axioms (dual to the direct flow case)
\begin{center}
   $\begin{array}{ccccc}
       v \preceq \inversesquare{r}{(u)} & \mbox{iff} & v \tprod r \preceq r \tprod u
       & \mbox{iff} & (v \type y \stackrel{r}{\rightarrow} u \type x) \memberof {\rm Ar}(\Hoare{{\bf P}}) \\
       \multicolumn{3}{c}{\inversesquare{s \tprod r}{(u)} \preceq \inversesquare{s}{(\inversesquare{r}{(u)})}}
       &&
    \end{array}$
\end{center}
for any comonoid $v \memberof \comonoid{}{(y)}{}$,
and any composable {\bf P}-term $\term{z}{s}{y}$.
The first axiom states that
$\inversesquare{r}{(u)} = \bigvee \{ v \in \comonoid{}{(y)}{} \mid v \tprod r \preceq r \tprod u \}$.
In addition,
any inverse Hoare flow must satisfy the closure compatibility axiom
\begin{center}
   $\begin{array}{c}
       \inversesquare{r}{(\complement{}{\complement{}{u}})} = \inversesquare{r}{(u)}
    \end{array}$
\end{center}
for any comonoid $u \memberof \comonoid{}{(x)}{}$,
where $\complement{}{u}$ is the boolean complement
$\complement{}{u} \define u \imply 0_x$.

Some identities for $\inversesquare{(\;)}{}$ are:
$\inversesquare{u}{(u')} = u \imply u'$ for all comonoids $u \memberof \comonoid{}{(x)}{}$;
$\inversesquare{s \tprod r}{(u)} = \inversesquare{s}{(\inversesquare{r}{(u)})}$
for two composable {\bf P}-terms $\term{z}{s}{y}$ and $\term{y}{r}{x}$;
$\inversesquare{r}{(u)} = y$ iff $r \preceq r \tprod u$ iff $u \memberof \filtertarget{r}$.

A lcc join bisemilattice {\bf P} has {\em kernels\/}
when for any {\bf P}-term $\term{y}{r}{x}$
there is a {\em kernel precondition\/} $\kernel{(r)} \memberof \comonoid{}{(y)}{}$
which satisfies the axioms
\begin{center}
   $\begin{array}{ccccc}
       v \preceq \kernel{(r)}
       & \mbox{iff} &
       v \tprod r = 0
       & \mbox{iff} &
       v \memberof \idealsource{r} \\
       \multicolumn{3}{c}{\kernel{(s \tprod r)} = \kernel{(s \tprod \complement{}{\kernel{(r)}})}}
       &&
    \end{array}$
\end{center}
for any precondition $v \memberof \comonoid{}{(y)}{}$,
and any composable {\bf P}-term $\term{z}{s}{y}$.

The first axiom states that
$\kernel{(r)} = \bigvee \idealsource{r} \in \idealsource{r}$,
or equivalently that the source ideal of $r$ is the principal ideal
$\idealsource{r} = \below{\kernel{(r)}}$.

So the source ideal operator factors 
$\kernel{(r)} \cdot \uparrow_y = \idealsource{} = \inversesquare{(\;)}{} \cdot {\rm ker}$
as kernel followed by principal ideal
$\{(\;)\}r\{\} = {\inversesquare{r}{}}^{\rm op} \cdot \uparrow_x$
and that kernel factors as source ideal followed by join 
$\kernel{} = \idealsource{} \cdot \bigvee_y$
(when joins exist for all ideals of comonoids),
since the join operator
$\ideal{\comonoid{}{(y)}{}} \stackrel{\vee_y}{\longrightarrow} \comonoid{}{(y)}{}$
forms a reflective pair $\bigvee_y \dashv \downarrow_y$ with the principal ideal operator
$\comonoid{}{(y)}{} \stackrel{\downarrow_y}{\longrightarrow} \ideal{\comonoid{}{(y)}{}}$.
(so the kernel of $\directsquare{r}{}$ is the source ideal $\idealsource{r}$,
 which matches the fact that the kernel of $r$ is the join of the source ideal $\bigvee_y \idealsource{r}$).

Some identities for the kernel operator $\kernel{}$ are:
``kernel is standard (negation) on subtypes''
$\kernel{(u)} = \complement{}{u}$ for any comonoid $u \memberof \comonoid{}{(x)}{}$;
``kernel translates guards to implications''
$\kernel{(v \tprod r)} = v \imply \kernel{(r)}$ for any guarded term $\term{y}{v \tprod r}{x}$; and
``only zero has top kernel''
$\kernel{(r)} = y$ iff $r = 0_{y,x}$ for any term $\term{y}{r}{x}$.
A term $\term{y}{r}{x}$ is {\em weakly total\/}
when its kernel is the empty comonoid $\kernel{(r)} = 0_y$.
If {\bf P} has inverse Hoare flow $\inversesquare{(\;)}{}$,
then it has kernels $\kernel{}$
defined to be the inverse flow of the bottom (zero) postcondition
$\kernel{(r)} \define \inversesquare{r}{(0_x)}$
for any term $\term{y}{r}{x}$.
Conversely,
if {\bf P} has kernels,
then it has inverse Hoare flow
defined to be the kernel of the tensor product
$\inversesquare{r}{(u)} \define \kernel{(r \tprod \complement{}{u})}$.
A {\em inverse Hoare flow category\/} is a lcc join bisemilattice which has inverse Hoare flow,
or equivalently, kernels.
\begin{Proposition}
   Any inverse Hoare flow category {\bf P} is a inverse flow category $\pair{{\bf P}}{\inversesquare{(\;)}{}}$.
\end{Proposition}

\paragraph{Dialectical Flow Categories.}
A {\em covariant dialectical flow category\/} $\pair{{\bf P}}{\square{}}$,
or {\em dialectical category\/} for short,
consists of an affirmation bisemilattice ${\bf P}$,
and a dialectical base ${\bf P} \stackrel{\square{}}{\longrightarrow} {\bf adj}$ of comonoids,
such that the direct flow aspect $\pair{{\bf P}}{\directsquare{(\;)}{}}$ is a direct flow category.
The inverse flow aspect $\pair{{\bf P}}{\inversesquare{(\;)}{}}$ is an inverse flow category
with local implication defined to be interior of inverse flow
$v \imply u \define \interior{\inversesquare{v}{(u)}}$
for comonoids $v,u \memberof \comonoid{}{(x)}{}$,
since
$w \tprod v \preceq u$
iff $\directsquare{v}{(w)} \preceq u$
iff $w \preceq \inversesquare{v}{(u)}$
iff $w \preceq \interior{\inversesquare{v}{(u)}}$.
Note: the inverse flow $\inversesquare{v}{(u)}$ alone satisfies
$\inversesquare{v}{(u)} \preceq x$ and
$\inversesquare{v}{(u)} = \inversesquare{v}{(u)} \wedge \inversesquare{v}{(u)}$
for comonoids $v,u \memberof \comonoid{}{(x)}{}$.
A {\em morphism of dialectical categories\/}
$\pair{{\bf P}}{\square{}_P} \stackrel{H}{\longrightarrow} \pair{{\bf Q}}{\square{}_Q}$
is a morphism of join bisemilattices ${\bf P} \stackrel{H}{\rightarrow} {\bf Q}$
such that $H$ is
both a morphism of direct flow categories
$\pair{{\bf P}}{\directsquare{P,(\;)}{}} \stackrel{H}{\longrightarrow} \pair{{\bf Q}}{\directsquare{Q,(\;)}{}}$,
and a morphism of inverse flow categories
$\pair{{\bf P}}{\inversesquare{P,(\;)}{}} \stackrel{H}{\longrightarrow} \pair{{\bf Q}}{\inversesquare{Q,(\;)}{}}$.
Comonoids and conjunction/implication form a one-object dialectical category
$\pair{\square{(x)}}{{\tprod_x \dashv \imply_x}}$
for each type $x$.
So inclusion
$\pair{\square{(x)}}{{\tprod_x \dashv \imply_x}} \stackrel{{\rm Inc}_x}{\longrightarrow} \pair{{\bf P}}{\square{}}$
is a morphism of dialectical categories for each type $x$ in {\bf P}.

A {\em dialectical Hoare flow category\/} is an affirmation bisemilattice
which has direct/inverse Hoare flow,
or equivalently, ranges and kernels.
\begin{Proposition}
   Any dialectical Hoare flow category is a dialectical flow category.
\end{Proposition}
When a dialectical Hoare flow category has an involution,
it has domains, kernels, ranges and cokernels for any terms $\term{y}{r}{x}$.
The domain and kernel are disjoint for any term $\term{y}{r}{x}$,
since
$\kernel{(r)} \tprod r = 0_{y,x}$
implies
$\domain{(\kernel{(r)} \tprod \domain{(r)})}
 = \domain{(\kernel{(r)} \tprod r)}
 = \domain{(0_{y,x})}
 = 0_y$
implies
$\kernel{(r)} \tprod \domain{(r)} = 0_y$. 
However,
they usually do {\em not\/} cover the source type $\kernel{(r)} \bsum \domain{(r)} \neq y$.
Disjointness of domain and kernel implies that,
if $r$ is total then $r$ is weakly total.
Furthermore,
if $\term{y}{r}{x}$ is weakly total then
$s \tprod r = 0_{z,x}$ implies $s = 0_{z,y}$ for all terms $\term{z}{s}{y}$
(this is often an alternate definition of ``totalness'').

A {\em contravariant dialectical flow category\/} $\pair{{\bf P}}{\diamond{}}$
consists of an affirmation bisemilattice ${\bf P}$,
and a dialectical base ${\bf P}^{\rm op} \stackrel{\diamond{}}{\longrightarrow} {\bf adj}$,
such that $\pair{{\bf P}^{\rm op}}{\diamond{}}$ is a covariant dialectical category.
An {\em involutive dialectical flow category\/} ${\bf P} = \triple{{\bf P}}{\involution{(\;)}}{\square{}}$
consists of
an involutive affirmation bisemilattice ${\bf P} = \pair{{\bf P}}{\involution{(\;)}}$,
and a covariant dialectical flow category     ${\bf P} = \pair{{\bf P}}{\square{}}$.
Defining $\diamond{}$ by $\diamond{} \define \involution{(\,)} \cdot \square{}$
makes ${\bf P} = \pair{{\bf P}}{\diamond{}}$ into a contravariant dialectical flow category,
where involution
$\pair{{\bf P}^{\rm op}}{\diamond{}} \stackrel{\involution{(\;)}}{\longrightarrow} \pair{{\bf P}}{\square{}}$
is a morphism of dialectical flow categories;
so that involution satisfies the condition
$\directsquare{\involution{r}}{} = \directdiamond{r}{}$,
or the equivalent condition
$\inversesquare{\involution{r}}{} = \inversediamond{r}{}$,
for all terms $\term{y}{r}{x}$. 
\begin{Facts}
   Let $\term{y}{f \dashv f^{\rm op}}{x}$ be any functional term in a dialectical flow category {\bf P}.
   \begin{enumerate} 
      \item When ${\bf P} = \pair{{\bf P}}{\square{}}$ is a covariant dialectical flow category,
            inverse $f$-flow is equal to direct $f^{\rm op}$-flow $\inversesquare{f}{} = \directsquare{f^{\rm op}}{}$,
            and we have the ``square'' adjoint triple
            \begin{center}
               $\begin{array}{c@{\;\dashv\;}r@{\;=\;}l@{\;\dashv\;}c}
                   \directsquare{f}{} \mbox{  }
                      & \mbox{  } \inversesquare{f}{} & \directsquare{f^{\rm op}}{} \mbox{  }
                      & \mbox{  } \inversesquare{f^{\rm op}}{}
                \end{array}$
            \end{center}    
      \item When ${\bf P} = \pair{{\bf P}}{\diamond{}}$ is a contravariant dialectical flow category,
            inverse $f^{\rm op}$-flow is equal to direct $f$-flow $\inversediamond{f^{\rm op}}{} = \directdiamond{f}{}$,
            and we have the ``diamond'' adjoint triple
            \begin{center}
               $\begin{array}{c@{\;\dashv\;}r@{\;=\;}l@{\;\dashv\;}c}
                   \directdiamond{f^{\rm op}}{} \mbox{  }
                      & \mbox{  } \inversediamond{f^{\rm op}}{} & \directdiamond{f}{} \mbox{  }
                      & \mbox{  } \inversediamond{f}{}
                \end{array}$
            \end{center}    
      \item When ${\bf P} = \triple{{\bf P}}{\involution{(\;)}}{\square{}}$
            is an involutive dialectical flow category
            with $\diamond{} = \involution{(\,)} \cdot \square{}$,
            the ``square'' and ``diamond'' adjoint triples above are identical.
            Either of these three cases justifies the common notation
            \begin{center}
               $\begin{array}{c@{\;\dashv\;}c@{\;\dashv\;}c}
                   \exists_f \mbox{  } & \mbox{  } {\rm sub}_f \mbox{  } & \mbox{  } \forall_f
                \end{array}$
            \end{center}    
   \end{enumerate} 
\end{Facts}

\paragraph{Hyperdoctrines.}
We discuss here the very important example of ``dialectical functional spans (hyperdoctrines)''.

The ``action'' of a term-process $\term{y}{r}{x}$ is concentrated in and localized to two ``loci of activity'',
a source subtype called the domain(-of-definition) of $r$ and a target subtype called the range of $r$.
These loci are polar ways to compute the ``effect'' or ``read-out'' of $r$,
and define dialectically opposed predicate transformations.
So term-processes $r$ become dialectical predicate transformers.

We assume that we are in a spannable join bisemilattice {\bf P}.
For any type $x$ let $\widehat{\Omega^\dashv}(x)$ denote $x$-comonoids
in the join bisemilattice $\widehat{{\bf P}^\dashv}$.
These are diagonal functional spans of the form
$\phi = (x \stackrel{f \dashv f^{\rm op}}{\leftharpoondown} y \stackrel{f \dashv f^{\rm op}}{\rightharpoondown} x)$,
and can be identified with functional terms $\term{y}{f \dashv f^{\rm op}}{x}$ into $x$.
By definition of $\widehat{\Omega^\dashv}(x)$,
the underlying {\bf P}-term operator
$\widehat{{\rm P}^\dashv}[x,x] \stackrel{\uterm{x}{x}}{\longrightarrow} {\bf P}[x,x]$
restricts to comonoids:
for any diagonal functional span $\phi$ as above,
the underlying {\bf P}-term is
$\flat_x(\phi) =  \uterm{x}{x}(\phi) = f^{\rm op} \tprod f$,
the comonoid associated with the functional term
$\term{y}{f \dashv f^{\rm op}}{x}$.  
So $\flat_x$ is the {\em associated comonoid\/} operator.
Analogous to the term/span reflective pair $\utermspano{x}{x} = (\uterm{x}{x} \dashv \spano{x}{x})$,
we assume the existence of a right adjoint left inverse
$\widehat{\Omega^\dashv}(x) \stackrel{\sharp_x}{\longleftarrow} \comonoid{}{(x)}{}$
to the associated comonoid operator $\flat_x$.
This means existence of a reflective pair 
$\widehat{\Omega^\dashv}(x) \stackrel{\natural_x}{\longrightarrow} \comonoid{}{(x)}{}$
of monotonic functions $\natural_x \define (\flat_x \dashv \sharp_x)$.
We assume that the reflective pair $\natural_x$ and the term/span reflective pair $\utermspano{x}{x}$
commute with the inclusion/interior coreflective pairs,
forming a commuting square
\begin{center}
   $\begin{array}{ccc}
       \widehat{\Omega^\dashv}(x)
       & 
       \makebox{\rule{.1in}{0in}}
       \makebox[0in]{\raisebox{.12in}[0in][0in]{\scriptsize$\natural_x$}} \makebox[0in]{$\longrightarrow$}
       \makebox{\rule{.1in}{0in}}
       &
       \comonoid{}{(x)}{}
       \\
       \makebox[0in][r]{\scriptsize${\rm Inc} \dashv \interior{(\,)}$} \downarrow
       &
       & \downarrow \makebox[0in][l]{\scriptsize${\rm Inc} \dashv \interior{(\,)}$}
       \\
       \widehat{{\rm P}^\dashv}[x,x]
       &
       \makebox[0in]{\raisebox{-.12in}[0in][0in]{\scriptsize$\utermspano{x}{x}$}} \makebox[0in]{$\longrightarrow$}
       &
       {\bf P}[x,x]
    \end{array}$
\end{center}
for any type $x$.
This implies that $\sharp_x$ must be defined by
$\sharp_x = {\rm Inc} \cdot \spano{x}{x} \cdot \interior{(\,)}$,
and must satisfy the equality
$\sharp_x \cdot \flat_x = {\rm Id}$;
so that for any $x$-subobject $\term{y}{f \dashv f^{\rm op}}{x}$ 
and any $x$-comonoid $u \memberof \comonoid{}{(x)}{}$,
we have the equivalence
$\flat_x(\term{y}{f \dashv f^{\rm op}}{x}) \preceq u$
iff
$(\term{y}{f \dashv f^{\rm op}}{x}) \preceq (\term{x}{e \dashv e^{\rm op}}{x})$
where $\sharp_x(u) = \term{z}{e \dashv e^{\rm op}}{x}$ and $e^{\rm op} \tprod e = u$
iff
there is a functional term
$\term{y}{h \dashv h^{\rm op}}{z}$
such that $f = h \tprod e$ and $e^{\rm op} \tprod h^{\rm op} = f^{\rm op}$.
So we can interpret $\sharp_x(u)$ to be the ``largest'' $x$-subobject
whose associated comonoid is $u$.
Interpreting in terms of graphs,
for any {\bf P}-comonoid $u \memberof \comonoid{}{(x)}{}$
with overlying span (${\bf P}^\dashv$-graph)
$\spano{x}{x}(u)
 = (x \stackrel{p_0 \dashv p_0^{\rm op}}{\leftharpoondown} y \stackrel{p_1 \dashv p_1^{\rm op}}{\rightharpoondown} x)$
we must have
$\sharp_x(u)
 = \term{z}{e \dashv e^{\rm op}}{x}
 = (x \stackrel{e \dashv e^{\rm op}}{\leftharpoondown} z \stackrel{e \dashv e^{\rm op}}{\rightharpoondown} x)$ 
the equalizer (self-loops) of this ${\bf P}^\dashv$-graph.
Moreover,
$e^{\rm op} \tprod e = p_0^{\rm op} \tprod p_1 = u$.
So $u$ specifies self-loops for ${\bf P}^\dashv$-graphs.


Let {\bf C} be a semiexact category that has epi-mono factorizations
which are preserved by pullback.
A {\bf C}-span
$\tau = (y \stackrel{t_0}{\leftarrow} t \stackrel{t_1}{\rightarrow} x)$
is {\em total\/} when the source leg $t \stackrel{t_0}{\rightarrow} y$ is an epimorphism.
Total spans are closed under identities and tensor products,
and hence form a subcategory of spans.
{\bf C}-arrows, as {\bf C}-spans via Yoneda, are total.
The {\em domain\/} of any span
$\rho = (y \stackrel{r_0}{\leftarrow} r \stackrel{r_1}{\rightarrow} x)$
is the {\bf C}-monomorphism $d_r \stackrel{m_0}{\rightarrow} y$,
or more precisely the subtype $\term{d_r}{\yoneda{{\rm C}}{(m_0)}{1}}{y}$,
where $r_0 = e_0 \cdot m_0$ is the epi-mono factorization of $r_0$ through the image {\bf C}-object $d_r$.
The {\em totalization\/} of any span
$\rho = (y \stackrel{r_0}{\leftarrow} r \stackrel{r_1}{\rightarrow} x)$
is the span
$\total{\rho} = (d_r \stackrel{e_0}{\leftarrow} r \stackrel{r_1}{\rightarrow} x)$.
Clearly,
$\yoneda{{\rm C}}{(m_0)}{0} \tprod \total{\rho}
 = {\yoneda{{\rm C}}{(m_0)}{1}}^{\rm op} \tprod \total{\rho} 
 = \rho$
and
$\yoneda{{\rm C}}{(m_0)}{1} \tprod \rho
 = \total{\rho}$.
Total spans have the following property:
if $\sigma \tprod \rho = 0_{zx}$ for some span $\term{z}{\sigma}{y}$,
then $\sigma = 0_{zy}$.

Our approach regards the notion of domain-of-definition as fundamental,
and defines totalness as a derived notion.
The {\em domain subtype\/} of any term $\term{y}{r}{x}$
is the source subtype $\domain{(r)} = \term{d_r}{i_r \dashv p_r}{y}$
which satisfies the axioms:
(1) ``minimality''
    $z \succeq \domain{(r)}$ iff $p \tprod i \tprod r = r$
    for any source subtype $\term{z}{i \dashv p}{y}$;
(2) ``composition''
    $\domain{(s \tprod r)} = \domain{(s \tprod p_r)}$
    for any composable term $\term{z}{s}{y}$; and
(3) ``monotonicity''
    $r \preceq r'$ implies $\domain{(r)} \preceq \domain{(r')}$
    for any parallel term $\term{y}{r'}{x}$.
Define the {\em totalization\/} of $r$ to be the $r$-subterm $\total{r} \define i_r \tprod r$.
A term $\term{y}{r}{x}$ is {\em total\/}
when its domain is the largest source subtype,
the entire source type $\domain{(r)} = y$.
Some identities for the domain operator $\domain{}$ are:
types are their own domain
$\domain{(x)} = x$;
the totalization is total, since
$\domain{(\total{r})}
 = \domain{(i_r \tprod r)}
 = \domain{(i_r \tprod p_r)}
 = \domain{(d_r)}
 = d_r$;
functional terms $\term{y}{f \dashv f^{\rm op}}{x}$ are total,
since the counit inequality $y \preceq f \tprod f^{\rm op}$ implies
$y = \domain{(y)}
 \preceq \domain{(f \tprod f^{\rm op})}
       = \domain{(f \tprod p_{f^{\rm op}})}
 \preceq \domain{(f \tprod x)}
       = \domain{(f)}
 \preceq y$;
in particular,
subtypes are total
$\domain{(\term{y}{i \dashv p}{x})} = y$;
domain subtypes are their own domain, since
$\domain{(p_r)}
 = \domain{(p_r \tprod d_r)}
 = \domain{(p_r \tprod \total{r})}
 = \domain{(r)}
 = d_r$;
only zero has empty domain
$\domain{(r)} = \term{0}{\bot_{0,y} \dashv \bot_{y,0}}{y}$
iff $r = 0_{y,x}$ for any term $\term{y}{r}{x}$; and
given any two total terms $\term{z}{s}{y}$ and $\term{y}{r}{x}$,
the composite term $\term{z}{s \tprod r}{x}$ is also total,
since 
$\domain{(s \tprod r)}
 = \domain{(s \tprod p_r)}
 = \domain{(s \tprod y)}
 = \domain{(s)}
 = z$.

The {\em domain subobject\/} of any term $\term{y}{t}{x}$
is the source subobject
$d_t \stackrel{i_d \dashv p_d}{\rightharpoondown} y$
where the term $t$ has non-nil action. 
The domain subobject (if it exists) is the ``smallest'' source subobject
such that the term $t$ is recoverable from the associated subterm by the identity
$t = (p_d \tprod i_d) \tprod t$;
so that, 
$t \preceq v \tprod t$
iff $\flat_y(d_t \stackrel{i_d \dashv p_d}{\rightharpoondown} y) \preceq v$
iff $(d_t \stackrel{i_d \dashv p_d}{\rightharpoondown} y) \preceq \sharp_y(v)$
for any source comonoid $v \memberof \comonoid{}{(y)}{}$.
The associated $t$-subterm
$t^\dagger = i_d \tprod t$ 
is called the {\em totalization\/} of $t$.
The term $t$ and its totalization $t^\dagger$ are equivalent by the identities
$\left\{\begin{array}{r@{\;=\;}c@{\;=\;}l}
           t         & p_d \tprod t^\dagger & {\bf P}^{[i_d,x]}(t^\dagger) \\
           t^\dagger & i_d \tprod t         & {\bf P}_{[i_d,x]}(t)
        \end{array} \right\}$.
A term $t$ is {\em total\/} when its domain subobject is the total source type $d_t = y$;
and then $t^\dagger = t$.

Total terms are close above w.r.t. term entailment order.
Since functional terms (in particular, identity terms) are total,
and the composite of total terms are also total,
total terms form a biposet $\total{{\bf P}}$,
a subbiposet of ${\bf P}$,
${\bf P}^\dashv \subseteq \total{{\bf P}} \subseteq {\bf P}$,
which is the homset order closure of ${\bf P}^\dashv$.
So $\total{{\bf P}}$ is a subbiposet ${\bf P}$,
which preserves homset joins but usually does not have a bottom.
Total terms in Heyting categories have been suggested \cite{Hoare} (although not by that name)
as good models for programs
(brief discussion in the section on Heyting categories).


Concurrent with the development of this paper,
an algebraic theory for the ``laws of progamming'' has been advocated \cite{Hoare},
whose axioms are essentially those for Heyting categories;
or more precisely,
Heyting categories (in particular, cHc) with affirmation/consideration modalities and domain subtypes.
But most of the ``laws of programming'' can be interpreted in
the category of spans $\widehat{{\bf C}}$ of a semiexact category {\bf C}.
In the program interpretation,
arbitrary {\bf C}-spans represent progam specifications,
total {\bf C}-spans represent programs,
and either arbitrary or monomorphic subtypes (either {\bf C}-arrows or {\bf C}-monomorphisms) represent conditions.
Types represent local contexts for local states of the system.
Span entailment order is interpreted as a measure of ``nondeterminism''
with $\rho \preceq \sigma$ asserting that $\rho$ is more deterministic than $\sigma$.
The top span $\term{y}{1_{yx}}{x}$,
which is the product span,
represents the worst (most nondeterministic) program,
and {\bf C}-arrows (as spans via Yoneda) represent fully deterministic (minimally nondeterministic) programs.
The bottom (initial) span $\term{y}{0_{y,x}}{x}$, although deterministic, is not a program since its domain-of-definition is empty.
The totalization $\term{d_r}{\total{\rho}}{x}$ of a span $\term{y}{\rho}{x}$
is the least deterministic program (on the domain-of-definition) of that specification.
In summary,
the ``Laws of Programming'' can be interpreted in categories of spans as follows.
\begin{center}
   \begin{tabular}{|lc||lc|}
      \hline
      \multicolumn{2}{|c||}{``Laws of Programming''} & \multicolumn{2}{c|}{categories of spans} \\
      \hline
      \hline
      program specifications          & $S$               & spans                       & $\term{y}{\rho}{x}$               \\
      programs                        & $P$               & total spans                 & $\term{y}{\tau}{x}$               \\
      conditions                      & $b$               & comonoids                   & $\phi \memberof \comonoid{}{(x)}{}$ \\
                                      &                   & subtypes                    & $y \stackrel{f}{\rightarrow} x$   \\
      \hline
      nondeterminism order            & $P \subseteq Q$   & span entailment order       & $\rho \preceq \sigma$             \\
      sequential composition          & $P {\bf ;} Q$     & tensor product              & $\sigma \tprod \rho$              \\
      nondeterministic choice         & $P \bigcup Q$     & boolean sum                 & $\sigma \bsum \rho$               \\
      \verb:SKIP:, the nop            & {\rm II}          & identity (objects-as-spans) & $\term{x}{x}{x}$                  \\ 
      \verb:ABORT:, the worst program & $\bot$            & top span                    & $\term{y}{1_{yx}}{x}$     \\
      \hline
      conditional or branch           & $P {\triangleleft} b {\triangleright} Q$ & derived expression & $(v \tprod \rho) \bsum (\boolneg{v} \tprod \sigma)$  \\
                                      & {\bf if} $b$ {\bf then} $P$ {\bf else} $Q$      & \multicolumn{2}{c|}{where $\boolneg{v} \define (v \imply \bot_y)$} \\
      iteration or {\bf while}-loop   & $b \ast P$        & derived expression          & $\closure{(u \tprod \rho)} \tprod \boolneg{u}$                     \\
                                      & {\bf while} $b$ {\bf do} $P$                    & \multicolumn{2}{c|}{where $\closure{(\,)}$ is the consideration modality} \\
      \hline
   \end{tabular}
\end{center}
In this paper these laws (concerning structure and flow in categories of spans)
are connected with the older program semantics which uses Hoare triples.

Given any {\bf C}-arrow $y \stackrel{h}{\rightarrow} x$,
composition defines a direct image monotonic function
${\bf Sub}(y) \stackrel{\exists^\ast_h}{\longrightarrow} {\bf Sub}(x)$
where $\exists^\ast_h(z \stackrel{g}{\rightarrow} y) = g \cdot_C h$,
and an inverse image monotonic function
${\bf Sub}(y) \stackrel{{\rm sub}^\ast_h}{\longleftarrow} {\bf Sub}(x)$
where ${\rm sub}^\ast_h(w \stackrel{f}{\rightarrow} x) = \hat{w} \stackrel{\hat{f}}{\rightarrow} y$
the pullback of $f$ along $h$.

It is easy to check that $\widehat{\bf C}$ 
has both domains and ranges,
with $\domain{(\rho)} = (y \stackrel{r_0}{\leftarrow} r \stackrel{r_0}{\rightarrow} y)$
and  $\range{(\rho)}  = (x \stackrel{r_1}{\leftarrow} r \stackrel{r_1}{\rightarrow} x)$
for any span $\rho = (y \stackrel{r_0}{\leftarrow} r \stackrel{r_1}{\rightarrow} x)$;
so it has direct Hoare flow $\directsquare{\rho}{}$ defined by
$\directsquare{\rho}{(\psi)}
 \define  \range{(\psi \tprod \rho)}
       =  \exists^\ast_{r_1}({\rm sub}^\ast_{r_0}(g))$
for any $y$-comonoid $\psi = (y \stackrel{g}{\leftarrow} z \stackrel{g}{\rightarrow} y)$.
With the obvious involution $\widehat{\bf C}$ is an involutive direct Hoare flow category;
so it has dual direct Hoare flow $\directdiamond{\rho}{}$ defined by
$\directdiamond{\rho}{(\psi)}
 = \directsquare{{\rho}^{\rm op}}{(\psi)}
 = \domain{(\rho \tprod \phi)}
 = \exists^\ast_{r_0}({\rm sub}^\ast_{r_1}(f))$
for any $x$-comonoid $\phi = (x \stackrel{f}{\leftarrow} w \stackrel{f}{\rightarrow} x)$.
By assuming local cartesian closure,
we can prove \cite{Freyd} that $\widehat{\bf C}$ is standard cartesian.

When $\widehat{\bf C} = \pair{\widehat{\bf C}}{\directsquare{(\,)}{}}$ is a direct flow category,
the direct flow along any term $\term{y}{\rho}{x}$ decomposes as
$\directsquare{\rho}{}
 = \directsquare{r_0^{\rm op} \tprod r_1}{}
 = \directsquare{r_0^{\rm op}}{} \cdot \directsquare{r_1}{}
 = {\rm sub}_{r_0} \cdot \exists_{r_1}$, 
and satisfies the ``Beck condition''
$\exists_f \cdot {\rm sub}_g
 = \directsquare{f}{} \cdot \directsquare{g^{\rm op}}{}
 = \directsquare{f \tprod g^{\rm op}}{}
 = \directsquare{\hat{g}^{\rm op} \tprod \hat{f}}{}
 = \directsquare{\hat{g}^{\rm op}}{} \cdot \directsquare{\hat{f}}{}
 = {\rm sub}_{\hat{g}} \cdot \exists_{\hat{f}}$
for the pullback span
$y \stackrel{\hat{g}}{\leftarrow} \hat{x} \stackrel{\hat{g}}{\rightarrow} z$
of any opspan
$y \stackrel{f}{\rightarrow} x \stackrel{g}{\leftarrow} z$,
and when $\widehat{\bf C} = \pair{\widehat{\bf C}}{\inversesquare{(\,)}{}}$ is a inverse flow category,
the inverse flow along any term $\term{y}{\rho}{x}$ decomposes as
$\inversesquare{\rho}{}
 = \inversesquare{r_0^{\rm op} \tprod r_1}{}
 = \inversesquare{r_1}{} \cdot \inversesquare{r_0^{\rm op}}{}
 = {\rm sub}_{r_1} \cdot \forall_{r_0}$,
and satisfies the ``Beck condition''
$\forall_g \cdot {\rm sub}_f
 = \inversesquare{g^{\rm op}}{} \cdot \inversesquare{f}{}
 = \inversesquare{f \tprod g^{\rm op}}{}
 = \inversesquare{\hat{g}^{\rm op} \tprod \hat{f}}{}
 = \inversesquare{\hat{f}}{} \cdot \inversesquare{\hat{g}^{\rm op}}{}
 = {\rm sub}_{\hat{f}} \cdot \forall_{\hat{g}}$.

Generalizing from this,
a {\em hyperdoctrine (of comonoids)\/} $\quadruple{{\bf C}}{\exists}{{\rm sub}}{\forall}$
consists of
(1) a semiexact category {\bf C}, and
(2) two connected dialectical bases
    ${\bf C} \stackrel{{\rm C}^{(\,)}}{\longrightarrow} {\bf adj}$ and
    ${\bf C}^{\rm op} \stackrel{{\rm C}_{(\,)}}{\longrightarrow} {\bf adj}$,
where ${\bf C}^f = {\bf Sub}(y) \stackrel{\exists_f \dashv {\rm sub}_f}{\longrightarrow} {\bf Sub}(x)$
and   ${\bf C}_f = {\bf Sub}(x) \stackrel{{\rm sub}_f \dashv \forall_f}{\longrightarrow} {\bf Sub}(y)$
for any {\bf C}-arrow $y \stackrel{f}{\rightarrow} x$,
which satisfy either of the equivalent ``Beck conditions'' above.

Suppose that {\bf P} is a spannable dialectical flow category.
Flow along arbitrary {\bf P}-terms factors into flow along functional {\bf P}-terms:
$\directsquare{r}{}
       = \directsquare{r_0^{\rm op} \tprod r_1}{}
       = \directsquare{r_0^{\rm op}}{} \cdot \directsquare{r_1}{}
       = {\rm sub}_{r_0} \cdot \exists_{r_1}$
for any term $\term{y}{r}{x}$ with overlying span of functional terms
$\spano{y}{x}(r)
 = (y \stackrel{r_0 \dashv r_0^{\rm op}}{\leftharpoondown} r_{01} \stackrel{r_1 \dashv r_1^{\rm op}}{\rightharpoondown} x)$,
since each {\bf P}-term factors through its associated functional span
$r = r_0^{\rm op} \tprod r_1$.
This decomposition of flow
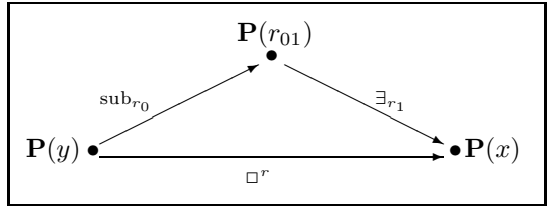
\begin{center}
\fbox{
\begin{picture}(190,70)(-28,-15)
   \put(0,5){\vector(2,1){60}}
   \put(70,35){\vector(2,-1){60}}
   \put(0,0){\vector(1,0){130}}
   \put(-5,0){$\bullet$}
   \put(132,0){$\bullet$}
   \put(62.5,36){$\bullet$}
   \put(0,20){\mbox{\scriptsize${\rm sub}_{r_0}$}}
   \put(104,20){\mbox{\scriptsize$\exists_{r_1}$}}
   \put(55,-10){\mbox{\scriptsize$\directsquare{r}{}$}}
   \put(-28,0){${\bf P}(y)$}
   \put(138,0){${\bf P}(x)$}
   \put(52,44){${\bf P}(r_{01})$}
\end{picture}
}
\end{center}
corresponds to the following diagram of \cite{Lawvere89}
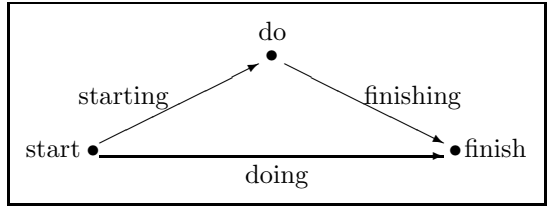
\begin{center}
\fbox{
\begin{picture}(190,70)(-28,-15)
   \put(0,5){\vector(2,1){60}}
   \put(70,35){\vector(2,-1){60}}
   \put(0,0){\vector(1,0){130}}
   \put(-5,0){$\bullet$}
   \put(132,0){$\bullet$}
   \put(62.5,36){$\bullet$}
   \put(-8,20){starting}
   \put(100,20){finishing}
   \put(55,-10){doing}
   \put(-28,0){start}
   \put(138,0){finish}
   \put(60,44){do}
\end{picture}
}
\end{center}
We automatically satisfy a generalized ``Beck condition''.

\begin{Proposition}
   \begin{enumerate}
      \item If {\bf P} is a spannable dialectical flow category,
         then ${\bf P}^\dashv$ is a hyperdoctrine of comonoids.
      \item If {\bf C} is a hyperdoctrine of comonoids,
         then $\widehat{{\bf C}}$ is a spannable dialectical flow category.
   \end{enumerate}
\end{Proposition}

\begin{Theorem}
   The spanning construction $\widehat{(\,)}$
   is left adjoint to
   the functionality construction $(\,)^\dashv$
   \begin{center}
      \begin{tabular}{rcl}
         \begin{tabular}{c}
            Hyperdoctrines \\ of Comonoids
         \end{tabular}
         &
         $\stackrel{\widehat{(\,)} \dashv (\,)^\dashv}{\longrightarrow}$
         &
         \begin{tabular}{c}
            Spannable Dialectical \\ Flow Categories
         \end{tabular}
      \end{tabular}
   \end{center}
   forming a coreflection,
   with $\widehat{(\,)}$ embedding hyperdoctrines into spannable dialectical flow categories,
   $(\,)^\dashv$ coreflecting spannable dialectical flow categories onto hyperdoctrines.
\end{Theorem}

\appendix
\section{Matrices}

There is a cHc with biproducts $\mat{{\bf R}}$ associated with the closed poset of reals 
${\bf R} = \sextuple{[0,\infty]}{\geq}{+}{0}{\wedge}{\infty}$,
whose objects are sets $X,Y,Z, \cdots$,
whose morphisms $\term{Y}{\phi}{X}$ are $\product{Y}{X}$-indexed collections of reals
$\phi = \{ \phi_{yx} \mid y \memberof Y, x \memberof X \}$
(that is, real-valued characteristic functions $\product{Y}{X} \stackrel{\phi}{\rightarrow} [0,\infty]$),
whose composition $\term{Z}{\psi \circ \phi}{X}$ for morphisms $\term{Z}{\psi}{Y}$ and $\term{Y}{\phi}{X}$ is pointwise addition
$(\psi \circ \phi)_{zx} \define \bigwedge_{y \in Y} [\psi_{zy} + \phi_{yx}]$,
and whose identity $\term{X}{X}{X}$ at $X$ is defined by
$X_{x'x} = 0 \mbox{ if } x'=x,
         = \infty \mbox{ otherwise}$.
Terms $\term{Y}{\phi}{X}$ can be viewed as {\em fuzzy relations\/},
where $\phi_{yx}$ measures the degree of membership in $\phi$,
with $\phi_{yx} = 0$ asserting full (crisp) membership $(y,x) \memberof \phi$
and  $\phi_{yx} = \infty$ asserting full nonmembership $(y,x) \not\in \phi$.
For the complete (cartesian) closed poset of boolean values
${\bf 2} = \sextuple{\{0,1\}}{\leq}{\wedge}{1}{\vee}{0}$
the associated cHc with biproducts is $\mat{{\bf 2}} = {\bf Rel}$ the category of ordinary relations.

More generally,
every cHc {\bf H} has an associated {\em matrix category\/} $\mat{{\bf H}}$,
whose objects are {\em {\bf H}-vectors\/} ${\cal X} = \pair{X}{|\;|_{\cal X}}$
where $X$ is an indexing (node) set and $X \stackrel{|\;|_{\cal X}}{\rightarrow} {\rm Obj}({\bf H})$ is a (typing) function,
whose arrows $\term{{\cal Y}}{R}{{\cal X}}$ are {\em {\bf H}-matrices\/}
where $R$ is a $\product{Y}{X}$-indexed collection of {\bf H}-terms
$R = \left( \term{|y|_{\cal Y}}{r_{yx}}{|x|_{\cal X}} \mid y \memberof Y, x \memberof X \right)$
(in other words, a generalized ${\rm Ar}({\bf H})$-valued characteristic functions
 $\product{Y}{X} \stackrel{r}{\rightarrow} {\rm Ar}({\bf H})$
 compatible with source and target),
whose homset order is pointwise order
$(s_{yx}) \preceq (r_{yx})$ when $s_{yx} \preceq r_{yx}$
for all $y \memberof Y$ and $x \memberof X$,
whose composition is matrix tensor product
$(S \circ R)_{zx} = S_{zY} \circ R_{Yx}
                  = \bigvee_{y \in Y} (s_{zy} \circ r_{yx})$
                  ``{\em matrix tensor product}''
for composable matrices $\term{{\cal Z}}{S}{{\cal Y}}$ and $\term{{\cal Y}}{R}{{\cal X}}$,
whose identity at ${\cal X}$ is the diagonal matrix $\term{{\cal X}}{{\cal X}}{{\cal X}}$
defined as identity {\bf H}-terms
${\cal X}_{xx} = \term{|x|_{\cal X}}{|x|_{\cal X}}{|x|_{\cal X}}$
on the diagonal
and zero (bottom) {\bf H}-terms
${\cal X}_{x'x} = \term{|x'|_{\cal X}}{0}{|x|_{\cal X}}$
off the diagonal,
and whose matrix tensor implications are
$(S \tensorimplytarget R)_{zy} = S_{zX} \tensorimplytarget R_{yX}
                               = \bigwedge_{x \in X} (s_{zx} \tensorimplytarget r_{yx})$
                               ``{\em right matrix tensor implication}'' and
$(R \tensorimplysource T)_{xz} = R_{Yx} \tensorimplysource T_{Yz}
                               = \bigwedge_{y \in Y} (r_{yx} \tensorimplysource t_{yz})$
                               ``{\em left matrix tensor implication}''.
Matrices $\term{Y}{R}{X}$ can be viewed as {\em fuzzy {\bf H}-relations\/}.
For any cHc {\bf H},
the matrix category $\mat{{\bf H}}$ is a complete Heyting category
for which biproducts (type sums) exist.

The sum of an arbitrary indexed collection of {\bf H}-types
is precisely the biproduct of an {\bf H}-vector.
Given any {\bf H}-vector ${\cal X} = \pair{X}{|\,|_{\cal X}}$,
the {\em sum type\/} of ${\cal X}$ is the composite {\bf H}-type $\sumtype{\cal X}$ having
$\{ \term{|x|_{\cal X}}{\iota_x \dashv \pi_x}{\sumtype{\cal X}} \mid x \memberof X \}$
as a pairwise disjoint collection of subtypes which cover $\sumtype{\cal X}$.
So the type $\sumtype{\cal X}$,
and its component injections $\iota_x$ and projections $\pi_x$,
satisfy
the ``comonoid covering axiom''
$\bigvee_{x \in X} (\pi_x \circ \iota_x) = \sumtype{\cal X}$,
and the ``subtype disjointness axioms''
$\iota_x \circ \pi_{x'} = x \mbox{ if } x = x', = 0_{x,x'} \mbox{ otherwise}$,
or the ``comonoid pairwise disjointness axioms''
$(\pi_{x'} \circ \iota_{x'}) \wedge (\pi_x \circ \iota_x) = 0_{\scriptsizesumtype{\cal X}}$ for $x' \neq x$.
Assume {\bf H} is a cHc with biproducts.
The type sum operator is a {\em sum functor\/}
$\mat{\bf H} \stackrel{\scriptsizesumtype}{\rightarrow} {\bf H}$,
which maps each {\bf H}-vector ${\cal X} = \pair{X}{|\,|_{\cal X}}$ to its underlying {\em sum type\/} $\sumtype{\cal X}$,
and maps each {\bf H}-matrix $\term{{\cal Y}}{R}{{\cal X}}$
where $R = \left( r_{yx} \mid y \memberof Y, x \memberof X \right)$ to its {\em sum term\/} $\term{\sumtype{\cal Y}}{\scriptsizesumtype{R}}{\sumtype{\cal X}}$
defined by $\sumtype R \define \bigvee_{y \in Y,x \in X} (\pi_y \circ r_{yx} \circ \iota_x)$
the sum of all the internal subterms of (internalized term entries in) $R$.
The $({\cal Y},{\cal X})$-th component of the sum functor $\sumtype$ is a join-continuous monotonic function
$\mat{{\bf H}}[{\cal Y},{\cal X}] \stackrel{\scriptsizesumtype_{\cal Y,X}}{\longrightarrow} {\bf H}[\sumtype{\cal Y},\sumtype{\cal X}]$.
The category ${\bf H}$ can be embedded into the category of matrices $\mat{{\bf H}}$
by the {\em singleton functor\/}
${\bf H} \stackrel{\singleton{}}{\longrightarrow} \mat{{\bf H}}$,
which embeds scalar objects ({\bf H}-types) as {\bf H}-vectors $x \mapsto \singleton{x} = \pair{{\bf 1}}{x}$
and embeds scalar arrows ({\bf H}-terms) as {\bf H}-matrices $r \mapsto \singleton{r} = \singleton{\term{y}{r}{x}}$.
This functor is clearly fully-faithful,
since for two fixed types $y$ and $x$,
there is a bijection ${\bf H}[y,x] \cong \mat{{\bf H}}[\singleton{y},\singleton{x}]$.
Also,
the composition of singleton with sum is the identity functor
$\singleton{} \cdot \sumtype = {\rm Id}_H$.
This implies that the sum functor is surjective on objects.

Let ${\cal Y}$ and ${\cal X}$ be any two {\bf H}-vectors,
and let $\term{\sumtype{\cal Y}}{r}{\sumtype{\cal X}}$ be any {\bf H}-term.
The matrix $\term{{\cal Y}}{\decomposition{r}{{\cal Y}}{{\cal X}}}{{\cal X}}$ defined by
$\decomposition{r}{{\cal Y}}{{\cal X}} \define \{ \term{|y|_{\cal Y}}{r_{yx}}{|x|_{\cal X}} \mid y \memberof Y, x \memberof X \}$,
where $r_{yx} \define \iota_y \circ r \circ \pi_x$ is the $(y,x)$-th external subterm of $r$,
is called the {\em decomposition matrix\/} of $r$.
(The external $r$-subterms $\term{|y|_{\cal Y}}{r_{yx}}{|x|_{\cal X}}$
are equivalent and in bijection with
the internal $r$-subterms $\term{|y|_{\cal Y}}{r'_{yx}}{|x|_{\cal X}}$
defined by
$r'_{yx} \define \pi_y \circ r_{yx} \circ \iota_x
              = (\pi_y \circ \iota_y) \circ r \circ (\pi_x \circ \iota_x)$
for $y \memberof Y$ and $x \memberof X$,
and $r$ is the join $r = \bigvee_{y \in Y} \bigvee_{x \in X} r'_{yx}$.
These internal subterms are predominant in topomatrices (next section).
Such decompositions,
especially w.r.t. topological bases of comonoids (next section),
give an internal representation of cHc's as distributor categories.)
This defines a {\em partition\/} join-continuous monotonic function
${\bf H}[\sumtype{\cal Y},\sumtype{\cal X}] \stackrel{\#_{\cal Y,X}}{\longrightarrow} \mat{{\bf H}}[{\cal Y},{\cal X}]$,
where $\#_{\cal Y,X}(r) \define \decomposition{r}{{\cal Y}}{{\cal X}}$.
Moreover,
by the comonoid covering axioms for biproducts,
any {\bf H}-term $\term{\sumtype{\cal Y}}{r}{\sumtype{\cal X}}$ is recoverable from its decomposition matrix $\decomposition{r}{{\cal Y}}{{\cal X}}$ by applying the sum functor
$\sumtype_{{\cal Y},{\cal X}}(\#_{\cal Y,X}(r)) = \sumtype_{{\cal Y},{\cal X}}(\decomposition{r}{{\cal Y}}{{\cal X}}) = \bigvee_{y \in Y, x \in X} r_{y,x} = r$.
This means that the sum functor is full (surjective on arrows).
Conversely,
by the subtype disjointness axioms for biproducts,
an {\bf H}-matrix $\term{{\cal Y}}{R}{{\cal X}}$ is recoverable from its sum term $\sumtype R$ by applying the partition function
$\#_{\cal Y,X}(\sumtype_{\cal Y,X}(R)) = R$.
This means that the sum functor is faithful (injective on arrows).
So for two fixed vectors ${\cal Y}$ and ${\cal X}$, 
the partition and sum monotonic functions are inverse to each other,
and define an isomorphism
${\bf H}[\sumtype{\cal Y,X}] \cong \mat{{\bf H}}[{\cal Y},{\cal X}]$. 
\begin{Lemma}
   The sum functor $\mat{{\bf H}} \stackrel{\scriptsizesumtype}{\rightarrow} {\bf H}$ is fully-faithful and surjective on objects.
\end{Lemma}

A matrix $\term{{\cal Y}}{R}{\singleton{x}}$ is called a {\em column {\bf H}-vector\/}.
If $\term{\sumtype{\cal Y}}{r}{x}$ is any term,
then the ${\cal Y}$-{\em source decomposition\/} of $r$ is the column vector
$\term{{\cal Y}}{\source{r}{{\cal Y}}}{\singleton{x}}$
defined by
$\source{r}{{\cal Y}} \define \left( \term{|y|_{\cal Y}}{r_{yx}}{x} \mid r_{yx} = \iota_y \circ r, y \memberof Y \right)$. 
The ${\cal Y}$-{\em source cotupling\/} of a column vector $\term{{\cal Y}}{R}{\singleton{x}}$,
where $R$ is the $Y$-indexed collection of terms $\left( \term{|y|_{\cal Y}}{r_{yx}}{x} \mid y \memberof Y \right)$,
is the {\bf H}-term $\term{\sumtype{\cal Y}}{\cotuple{R}{{\cal Y}}}{x}$
defined by $\cotuple{R}{{\cal Y}} \define \bigvee_{y \in Y} (\pi_y \circ r_{yx})$. 
The source decomposition and cotupling operations are inverse to each other,
with $\cotuple{\source{r}{{\cal Y}}}{{\cal Y}} = r$ and $\source{\cotuple{R}{{\cal Y}}}{{\cal Y}} = R$.
Dually,
a matrix $\term{\singleton{y}}{R}{{\cal X}}$ is called a {\em row {\bf H}-vector\/}.
If $\term{y}{r}{\sumtype{\cal X}}$ is any term,
then the ${\cal X}$-{\em target decomposition\/} of $r$ is the row vector
$\term{\singleton{y}}{\target{r}{{\cal X}}}{{\cal X}}$
defined by
$\target{r}{{\cal X}} \define \left( \term{y}{r_{yx}}{|x|_{\cal X}} \mid r_{yx} = r \circ \pi_x, x \memberof X \right)$.
The ${\cal X}$-{\em target tupling\/} of a row vector $\term{\singleton{y}}{R}{{\cal X}}$,
where $R$ is the $X$-indexed collection of terms $\left( \term{y}{r_{yx}}{|x|_{\cal X}} \mid x \memberof X \right)$,
is the {\bf H}-term $\term{y}{\tuple{R}{{\cal X}}}{\sumtype{\cal X}}$
defined by $\tuple{R}{{\cal X}} \define \bigvee_{x \in X} (r_{yx} \circ \iota_x)$.
The target decomposition and tupling operations are inverse to each other,
with $\tuple{\target{r}{{\cal X}}}{{\cal X}} = r$ and $\target{\tuple{R}{{\cal X}}}{{\cal X}} = R$.

Any {\bf H}-vector ${\cal X}$
decomposes the identity term $\term{\sumtype{\cal X}}{\scriptsizesumtype{\cal X}}{\sumtype{\cal X}}$ in either of two ways:
as the source decomposition column vector  
$\term{{\cal X}}{\iota_{\cal X}}{\singleton{\sumtype{\cal X}}}$
defined by $\iota_{\cal X} \define \source{x}{{\cal X}} = \left( \term{|x|_{\cal X}}{\iota_x}{\sumtype{\cal X}} \mid x \memberof X \right)$,
or as the target decomposition row vector  
$\term{\singleton{\sumtype{\cal X}}}{\pi_{\cal X}}{{\cal X}}$
defined by $\pi_{\cal X} \define \target{x}{{\cal X}} = \left( \term{\sumtype{\cal X}}{\pi_x}{|x|_{\cal X}} \mid x \memberof X \right)$.
Moreover,
the identity matrix at ${\cal X}$ decomposes as $\iota_{\cal X} \circ \pi_{\cal X}$,
and the identity matrix at $\singleton{\sumtype{\cal X}}$ decomposes as $\pi_{\cal X} \circ \iota_{\cal X}$,
so that $\term{{\cal X}}{\iota_{\cal X}}{\singleton{\sumtype{\cal X}}}$
and $\term{\singleton{\sumtype{\cal X}}}{\pi_{\cal X}}{{\cal X}}$ are inverse matrices.
Since $\iota_{\cal X}$ and $\pi_{\cal X}$ are inverse pairs,
they are adjoint pairs in both directions
$\term{{\cal X}}{\iota_{\cal X} \dashv \pi_{\cal X}}{\singleton{\sumtype{\cal X}}}$
and
$\term{\singleton{\sumtype{\cal X}}}{\pi_{\cal X} \dashv \iota_{\cal X}}{{\cal X}}$.
So,
given any term $\term{\sumtype{\cal Y}}{r}{x}$,
(1) the term $r$ and its source decomposition $\source{r}{{\cal Y}}$ are expressible
in terms of each other via the direct and inverse left flow expressions
$\source{r}{{\cal Y}} = \iota_{\cal Y} \circ \singleton{r}
               = \pi_{\cal Y} \tensorimplysource \singleton{r}$
and
$\singleton{r} = \pi_{\cal Y} \circ \source{r}{{\cal Y}}
               = \iota_{\cal Y} \tensorimplysource \source{r}{{\cal Y}}$,
and
(2) the term $r$ and its target decomposition $\target{r}{{\cal X}}$ are expressible
in terms of each other via the direct and inverse right flow expressions
$\target{r}{{\cal X}} = \singleton{r} \circ \pi_{\cal X}
               = \singleton{r} \tensorimplytarget \iota_{\cal X}$
and
$\singleton{r} = \target{r}{{\cal X}} \circ \iota_{\cal X}
               = \target{r}{{\cal X}} \tensorimplytarget \pi_{\cal X}$.
Furthermore,
given any two {\bf H}-vectors ${\cal Y}$ and ${\cal X}$,
(1) a term $\term{\sumtype{\cal Y}}{r}{\sumtype{\cal X}}$ and its decomposition matrix $\#_{\cal Y,X}(r) = \decomposition{r}{{\cal Y}}{{\cal X}}$
are expressible in terms of each other via the direct flow expressions
$r = \pi_{\cal Y} \circ \#_{\cal Y,X}(r) \circ \iota_{\cal X}$
and
$\#_{\cal Y,X}(r) = \iota_{\cal Y} \circ \singleton{r} \circ \pi_{\cal X}$,
and
(2) an {\bf H}-matrix $\term{{\cal Y}}{R}{{\cal X}}$ and its sum term $\term{\sumtype{\cal Y}}{\scriptsizesumtype R}{\sumtype{\cal X}}$
are expressible in terms of each other via the direct flow expressions 
$R = \iota_{\cal Y} \circ \singleton{\sumtype R} \circ \pi_{\cal X}$
and
$\sumtype R = \pi_{\cal Y} \circ R \circ \iota_{\cal X}$.

For each vector ${\cal X}$
the matrix isomorphism $\term{\singleton{\sumtype{\cal X}}}{\pi_{\cal X}}{{\cal X}}$ 
is the ${\cal X}$-th component of a ``counit'' natural isomorphism
$\pi \type \sumtype \cdot \singleton{} \Longrightarrow {\rm Id}_{\mat{H}}$,
since $\singleton{\sumtype R} \circ \pi_{\cal X} = \pi_{\cal Y} \circ R$.
\begin{Theorem}
   For every cHc {\bf H} with biproducts,
   the singleton and sum functors form a categorical equivalence
   $\singleton{} \dashv \sumtype$
   between {\bf H} and its category of matrices $\mat{{\bf H}}$,
   with identity unit ${\rm Id}_H = \singleton{} \cdot \sumtype$
   and natural isomorphism counit
   $\pi \type \sumtype \cdot \singleton{} \Longrightarrow {\rm Id}_{\mat{H}}$.
\end{Theorem}

For any term $\term{y}{t}{x}$,
since the source type $y$ is the direct sum
$y = d_t \sumtype k_t$
and the target type $x$ is the direct sum
$x = r_t \sumtype c_t$,
the term $t$ can be expressed as the $\product{2}{2}$ matrix
$t = \left( \begin{array}{cc}
                      t^\bullet_\bullet & 0_{d_t,c_t} \\
                      0_{k_t,r_t} & 0_{k_t,c_t}
                   \end{array} \right)$
where $t^\bullet_\bullet$ is both total and cototal.

Every category {\bf C} has an associated {\em distributor category\/} $\distrib{{\bf C}}$
defined by $\distrib{{\bf C}} \define \mat{\powerof{{\bf C}}}$.
In more detail,
$\distrib{{\bf C}}$ is the category,
whose objects are {\em distributed {\bf C}-objects\/} or {\bf C}-{\em vectors\/} ${\cal X} = \pair{X}{|\;|_{\cal X}}$ as above,
whose arrows $\term{{\cal Y}}{R}{{\cal X}}$ are {\em distributed {\bf C}-arrows\/} or {\bf C}-{\em distributors}
where $R \subseteq \triproduct{Y}{{\rm Ar}({\bf C})}{X}$ is a digraph between the underlying node sets
consisting of compatible triples:
if $(y,r,x) \memberof R$ then $|y|_{\cal Y} \stackrel{r}{\rightarrow} |x|_{\cal X}$ is a {\bf C}-arrow,
whose tensor product is defined pointwise as
$(S \circ R)_{z,x} \define \bigcup_{y \in Y} [S_{zy} \circ R_{yx}]$,
and whose identity at ${\cal X}$ is the {\bf C}-distributor
${\cal X} \define \{ (x,|x|_{\cal X},x) \mid x \memberof X \}
          \subseteq \triproduct{X}{{\rm Ar}({\bf C})}{X}$
consisting (on the diagonal) of all the {\bf C}-identities indexed by ${\cal X}$.
The $(y,x)$-th fiber of a $\distrib{{\bf C}}$-term $\term{{\cal Y}}{R}{{\cal X}}$,
defined by $R_{yx} \define \{ \term{y}{r}{x} \mid r \memberof R \}$,
is a $\powerof{{\bf C}}$-term $\term{y}{R_{yx}}{x}$,
and $R$ is the disjoint union
$R = \coprod_{y \in {\cal Y}, x \in {\cal X}} R_{yx}$
of its $\powerof{{\bf C}}$-term fibers.
For any category {\bf C},
the distributor category $\distrib{{\bf C}}$ is a complete Heyting category
for which biproducts (type sums) exist.
The distributor category generalizes the ``state space construction'' from automaton theory.
In distributor categories $\distrib{{\bf C}}$ a comonoid $W$ of type $X$
is essentially a subobject (subset) $W \subseteq X$,
and so $\comonoid{}{X}{} \cong \powerof{X}$.
More generally,
every biposet {\bf P} has an associated {\em closure distributor category\/} $\distrib{{\bf P}} \define \mat{\powerof{{\bf P}}}$,
whose objects, arrows, tensor product and identities are as above,
and whose homset order is the pointwise closed-below order.
Given any set of attributes or sorts $A$,
a signature
$\Sigma = \{ \Sigma_{y,a} \mid y \memberof \mbox{multiset}(A), a \memberof A \}$ 
over $A$
determines a term category ${\bf T}_\Sigma$,
the initial algebraic theory over $\Sigma$,
whose objects are multisubsets of $A$ 
(arities, tuplings, etc.)
and whose arrows are tuples of $\Sigma$-terms.
A parallel pair of arrows
$\term{{\cal Y}}{S,R}{{\cal X}}$
in the distributor category $\distrib{{\bf T}_\Sigma^{\rm op}}$
is a Horn clause logic program,
whose predicate names are ${\cal X}$-nodes,
whose clause names are ${\cal Y}$-nodes, 
whose clause-head atoms are (w.l.o.g.) collected together as $S$,
whose clause-body atoms are collected together as $R$,
and whose associated fixpoint operator is the inverse/direct flow composite
$((\,) \tensorimplytarget R) \circ S$
defined on Herbrand interpretations with database scheme ${\cal X}$.
In much of the logic of dialectical processes
(in particular, for Girard's completeness theorem)
closure subset categories suffice.
However,
for the constraint dialectic,
the full nondeterminism and parallelism of distributor categories is essential.

\section*{Summary}

The most important improvement made by dialectical logic over dynamic logic is in the correct and rigorous treatment of subtypes.
It is a serious conceptual error \cite{Kozen} to view dynamic logic as a two-sorted structure:
one sort being programs and the other sort being propositions.
The central viewpoint of dialectical logic is that predicates (here called subtypes, or more precisely, {\em comonoids\/})
are special local idempotent kinds of programs (here called {\em terms\/} or {\em processes\/}),
which by their idempotent and coreflexive nature form the standard logical structure of Heyting algebra in the intuitionistic case
or Boolean algebra in the classical case.
The two dynamic logic operations of program sequencing and predicate conjunction
are combined into the one (horizontal) dialectical logic operation of {\em tensor product\/} of terms, 
and the two dynamic logic operations of program summing and predicate disjunction
are combined into the one (vertical) dialectical logic operation of {\em boolean sum\/}.
Now,
tensor product and boolean sum are global operations on terms.
In addition,
dialectical logic has complement operations called 
{\em tensor implications\/} and {\em tensor negation\/} \cite{Kent88},
which are also global.
In contrast to these,
dialectical program semantics,
introduces local complement operations called {\em standard implication\/} and {\em standard negation\/}.
The global operation of tensor implication (negation) is replaced by
the local standard implication (negation) and direct/inverse flow.

Global products and coproducts of precondition/postcondition assertions
are defined in terms of {\em biproducts\/} in the indexing category underlying a dialectical flow category. 
Biproducts model the semantic notion of ``type sum''.
Completely general axioms for {\em domains-of-definition\/} and {\em ranges\/},
and their negation duals {\em kernels\/} and {\em cokernels\/},
can be given,
which are equivalent to predicate transformer axioms,
and do {\em not\/} require the notion of type sum.
A nice program semantics has already been given \cite{Manes}
which is based upon the notions of {\em sums\/} and {\em bikernels\/},
but one of the purposes of this paper is to show that dialectical program semantics,
the standard logical semantics of ``relational structures'',
does not require sums and only indirectly requires bikernels.
{\em Iterates\/},
the dialectical logic rendition of the ``consideration modality'' of linear logic \cite{Girard},
are defined as freely generated monoids,
and dialectical categories with consideration modality are introduced to ensure the existence of iterates.
The important doctrine of linear logic,
paraphrased by the statement that
``the familiar connective of standard negation factors into two operations:
  {\em linear negation\/}, which is the purely negative part of negation;
  and the modality {\tt of}\_{\tt course}, which has the meaning of reaffirmation'',
is verified in dialectical logic,
since the local operation of standard implication (standard negation) of subtypes factors
into the global operation of tensor implication  (tensor negation) followed by comonoidal {\em support\/},
the dialectical logic rendition of the ``affirmation modality'' of linear logic.
Term hom-set completeness defines the notion of {\em topology of subtypes\/},
thereby making further contact with the affirmation modality.
In such complete semantics,
topologized matrices of terms are defined
and shown to be (categorically) equivalent to single terms
via the inverse operations of ``partitioning'' and ``summing''.
With the introduction of type sums a nontopological matrix theory is developed,
where ordinary matrices of terms are defined
and shown to be (categorically) equivalent to terms with biproducts.

In summary,
with dialectical program semantics
we hope to unify small-scale and large-scale program semantics
by giving a concrete foundation for the observation that
``precondition/postcondition assertions are similar in structure to relational database constraints''.
I am now exploring the close connection
between the functional aspect of dialectical program semantics and Martin-L\"{o}f type theory
given via locally cartesian closed categories \cite{Seeley}.
Furthermore,
there is a strong connection between dialectical program semantics
and algebraic and temporal logic models of regulation in feedback control systems \cite{Wonham}.

   \tableofcontents

\end{document}